# Geochemical and planetary dynamical views on the origin of Earth's atmosphere and oceans

Nicolas Dauphas[1,*] & Alessandro Morbidelli[2]

[1]Origins Laboratory, Department of the Geophysical Sciences and Enrico Fermi Institute, The University of Chicago, 5734 South Ellis Avenue, Chicago IL 60637, USA (dauphas@uchicago.edu)

[2]Université de Nice Sophia Antipolis, CNRS, Observatoire de la Côte d'Azur, Laboratoire Cassiopée, Boulevard de l'Observatoire, B.P. 4229, 06304 Nice Cedex 4, France (Alessandro.MORBIDELLI@obs-nice.fr)

[*]Visiting scientist at UJF-Grenoble 1/CNRS-INSU, Institut de Planétologie et d'Astrophysique de Grenoble (IPAG), UMR 5274, Grenoble, F-38041, France

**Abstract.** Earth's volatile elements (H, C, and N) are essential to maintaining habitable conditions for metazoans and simpler life forms. However, identifying the sources (comets, meteorites, and trapped nebular gas) that supplied volatiles to Earth is not straightforward because secondary processes like mantle degassing, crustal recycling, and escape to space modified the composition of the atmosphere. Here, we review two complementary approaches to investigate the origin of Earth's atmosphere and oceans. The geochemical approach uses volatile element abundances and isotopic compositions to identify the possible contributors to the atmosphere and to disentangle the processes that shaped it. In that respect, noble gases (He, Ne, Ar, Kr, and Xe), elements that are chemically inert and possess several isotopes produced by radioactivity, play a critical role. The dynamical approach uses our knowledge of planetary dynamics to track volatile delivery to the Earth, starting with dust transport in the disk to planet-building processes. The main conclusion is that Earth acquired most of its major volatile elements by accretion of planetesimals or embryos akin to volatile-rich meteorites. At the same time, solar/meteoritic noble gases were captured by embryos and some gases were lost to space, by hydrodynamic escape and large impacts. Comets did not contribute much H, C, and N but may have delivered significant noble gases, which could represent the only fingerprints of the bombardment of our planet with icy bodies. The processes that governed the delivery of volatile elements to the Earth are thought to be relatively common and it is likely that Earth-like planets covered with oceans exist in extra-solar systems.

## 1. Introduction

The oldest water-laid chemical sediments identified in the geologic record are found in southern West Greenland and Northern Québec and are dated at ~3.8 Ga[*], indicating that liquid water was present at that time (CATES and MOJZSIS, 2007; DAUPHAS et al., 2007; DAUPHAS et al., 2004b; MOORBATH et al., 1973; NUTMAN et al., 1997; O'NEIL et al., 2007). This age corresponds to the transition between the Hadean and the Archean and marks the time when the rock record starts. The Hadean has also been referred to as the dark age of the Earth because, for lack of samples, very little is known about that period. However, there are direct witnesses of the Hadean in the form of zircons ($ZrSiO_4$), which are tiny minerals resistant to chemical alteration and abrasion that are readily datable by the U-Pb method (HARRISON, 2009). The oldest zircons are found in detrital sediments at Jacks Hill (Australia) and some of these have been dated at 4.3 Ga. Oxygen isotope ratios indicate that the zircons were crystallized from a magma that was produced by melting of a clay-rich sediment produced by water-rock interaction at Earth's surface (MOJZSIS et al., 2001; WILDE et al., 2001). This suggests that liquid water was present on Earth as early as 4.3 Ga.

In undifferentiated meteorites (*i.e.*, chondrites), water is present in the structure of hydrated minerals as well as in fluid inclusions. In comets, water is in the form of ice. However, largely anhydrous bodies also exist. Isotopically, the meteorites that best match the composition of the Earth are enstatite chondrites (*i.e.*, they have identical O, Ti, Cr, and Mo isotopic ratios to terrestrial rocks, Clayton, 1993; Dauphas et al., 2002; Trinquier et al., 2007; Trinquier et al., 2009; Fig. 1), leading some to suggest that the Earth formed from these meteorites (Javoy et al., 2010). A notable difficulty with this idea is that enstatite chondrites have a high Si/Mg ratio, so that unrealistic amounts of Si in the core would be needed to explain the lower Si/Mg ratio of Earth's mantle. Javoy et al. (2010) solved this conundrum by proposing that the Earth was made of material isotopically similar to enstatite chondrites, yet chemically distinct from these meteorites. There is no reason why any of the surviving meteorite groups should match the main source population for Earth. A planet formed solely from such highly reduced bodies would undoubtedly be sterile, posing the question of how and when Earth acquired its water.

---

[*]We use annum (as in Ga) for absolute time relative to present and year (as in Gyr) for relative time and duration.

Any model that attempts to explain the origin of the oceans and atmosphere also must be able to explain the elemental and isotopic compositions of noble gases in air. These are trace gases that play no biological role but their inertness makes them particularly powerful to unravel the sources and processes that have shaped the atmosphere (Ozima and Podosek, 2001). In addition, some noble gas isotopes are part of radioactive decay schemes (*e.g.*, $^{40}$K-$^{40}$Ar, $^{129}$I-$^{129}$Xe) that can be used to date processes like mantle degassing or atmosphere loss to space.

In this contribution, we examine how geochemical and planetary dynamical considerations can provide constraints on the formation of Earth's atmosphere. Marty & Yokochi (2006), and Pepin (2006) presented detailed reviews on that topic. The question of the origin of Earth's atmosphere covers two aspects: how volatile elements were delivered to the Earth and how secondary processes modified the chemical and isotopic compositions of the atmosphere. This history must be understood in the framework of the formation of the Earth itself, which is discussed in Sect. 2 (reviewed by Dauphas and Chaussidon, 2011; Raymond, 2010).

## 2. Making terrestrial planets

We start this review by discussing the process of terrestrial planet formation from the astrophysical point of view. The purpose is to build a framework that can be useful to interpret the volatile record on the Earth. Terrestrial planet formation is thought to proceed in three steps. In step I, the first planetesimals are formed, from a disk of gas and dust. In step II, the collision evolution of the planetesimal population gives birth to a new class of objects, called planetary embryos, which represent an intermediate stage between planetesimals and planets. At the same time, giant planets are formed. In step III, after the disappearance of gas from the proto-planetary disk, the embryos become unstable, and their mutual collisions give birth to a small number of massive objects, known as the terrestrial planets.

**Step I: from dust to planetesimals.**

In the proto-planetary disk, the most refractory materials condense first, gradually followed (while the local temperature drops) by more and more volatile elements. Coexisting with the condensate grains thus formed are dust particles inherited from circumstellar/interstellar chemistry that escaped vaporization in the solar nebula. Collisions helped by electrostatic and magnetic forces stick the grains together, forming fractal aggregates. Other collisions then rearrange the aggregates, and compact them. When the grains reach a size of about a millimeter to a centimeter, they begin to rapidly sediment onto the median plane of the disk in a time of the order of $10^3$ yr. This timescale, however, can be longer if the nebula is strongly turbulent.

The growth from these grains to kilometer-size planetesimals is still quite a mystery. In principle, one could expect that grains stick to each other to form progressively bigger and bigger objects, in an ordered-growth process. However, particles of cm-size are too small for gravity to be effective in particle-particle collisions, but they are too big to stick through electrostatic forces (although sticking may be possible for particles with large size ratios). Moreover, grains are subject to gas drag, which makes them drift towards the central star (Weideschilling, 1977). The drift speed is size dependent; thus, particles of different sizes must collide with non-negligible relative velocities (of the order m/s). At these velocities particles should break, rather than coagulate (Blum and Wurm, 2008). Because the drift speed towards the central star is maximal for meter-size boulders, this issue is known as the "meter-size barrier problem", but it is likely that this bottleneck for accretion starts already at much smaller sizes (cm or dm).

A new alternative to this ordered-growth process is that planetesimals form thanks to the collective gravity of massive swarms of small particles, concentrated at some locations (vortices or inter-vortex regions, depending on particle sizes) by the turbulence of the disk (Johansen et al., 2007; Cuzzi et al., 2008). This model can explain the formation of large planetesimals (100 km or larger) without passing through intermediate small sizes, so that the meter-size barrier problem is circumvented. Thus, in these gravito-turbulent models, planetesimals form big. The size distribution of objects in the asteroid belt and in the Kuiper belt, where most of the mass is concentrated in 100 km objects, supports this scenario (Morbidelli et al., 2009). The existence and the properties of Kuiper belt binary objects also are best explained by the gravitational collapse of massive swarms of small particles that have too large angular momentum to form a single object (Nesvorny et al., 2010). This new view of planetesimal formation is rapidly gaining support. Although more work is needed to explore all its facets we can start to discuss, in a broad sense, its implications.

Once enough small particles are concentrated at some location, the formation of a planetesimal is extremely rapid (Johansen et al., 2007; Cuzzi et al., 2008). However, the formation of self-gravitating clumps of small particles is sporadic (Cuzzi et al., 2010; Chambers, 2010). Therefore, planetesimal formation can proceed over a long time interval. The planetesimals that form first are rich in the short-lived radionuclide $^{26}$Al and therefore they can melt and differentiate in a core-mantle-crust structure. Those that form after a couple of millions of years probably can escape differentiation because the majority of short-lived radioactive elements have already decayed. This can explain the co-existence of differentiated and undifferentiated planetesimals that we deduce from the meteorite collections (iron and basaltic meteorites being fragments of differentiated objects and chondritic meteorites being representative of undifferentiated ones). However, it is not necessarily true that each region of the proto-planetary disk had to form both differentiated and undifferentiated planetesimals. The reason is that sufficient clumping of small particles to form planetesimals is possible only if the solid/gas density ratio is larger than some threshold value (Johansen et al., 2009). For instance in the innermost part of the disk, this condition might have been met very early, thus leading to a first generation of planetesimals rich enough in radioactive elements to melt. But in other regions of the disk, this condition might have been met only later, after the removal by photo-evaporation of a substantial fraction of the gas, or the diffusion into the considered region of debris from elsewhere in the disk. Thus, in these regions, all planetesimals formed would remain undifferentiated. This might have been the case of the asteroid belt, in which the solid/gas ratio might have become large enough only after the formation of chondrules (Scott, 2006).

At a given time, the temperature in the disk decreases with the distance from the Sun. Thus, close to the Sun only refractory elements can be in solid form, whereas further away more volatile elements are locked up in solids, following the classical condensation sequence. Thus, if all planetesimals had formed at the same time, we would expect two basic properties to be manifested: (a) the planetesimal disk should be characterized by a clear radial gradient of chemical properties; (b) the abundance pattern in each planetesimal should be characterized by a very sharp transition between elements that were refractory enough to condense and those that were too volatile and stayed in the gas. However, this is not the pattern of abundance that is observed in chondrites, which show instead a relatively smooth pattern of depletion as a function of element volatility. Moreover, asteroid belt objects of different compositions have partially overlapping distributions in orbital semi major axis (Gradie and Tedesco, 1982). Why neither (a) nor (b) are true can be understood by the following reasons:

(i) According to the gravito-turbulent models, planetesimals form sporadically so that, even at the same location they do not form at the same time (Johansen et al., 2007; Cuzzi et al., 2008). This is consistent with existing chronological constraints on meteorite formation, showing that it spanned ~3-4 Myr. With time passing, the temperature decreases and more volatile elements can condense in solids. Thus, this can explain why (a) is not true.

(ii) Dynamical evolution after planetesimal formation can partially mix planetesimals originally born at distinct locations (see below). This also can explain why (a) is not true.

(iii) Dust transport in the disk partially obliterated the strong heliocentric gradient in temperature. The most blatant evidence for that is the presence of highly refractory dust (*i.e.*, calcium-aluminum-rich inclusion) formed at a temperature of 1500 K in comet Wild 2 that coexists with water ice condensed at a temperature of presumably less than ~70 K. This is also seen in meteorites where CAIs coexist with carbonaceous material formed by cold chemistry in the interstellar medium or in outer regions of the disk. This mechanism can explain why (b) is not true (Cassen, 2001 and references herein).

In the absence of clear predictions from the formation models, we can turn to observational constraints to deduce properties of the planetesimal disk. There are basically three classes of chondritic meteorites: enstatite, ordinary and carbonaceous. Their chemistry and mineralogy suggest that they formed at overall decreasing temperatures. For instance, water is essentially absent on enstatite meteorites, and quite abundant in (some subclasses of) carbonaceous chondrites, while the water-content in ordinary chondrites is between the two (Robert, 2003). Spectroscopic observations link these three classes of meteorites to asteroids of different taxonomic type: enstatite chondrites can be linked with E-type asteroids (Fornasier et al., 2008), which are predominant in the Hungaria region at 1.8 astronomical units (AU; 1 AU is the mean Sun-Earth distance); ordinary chondrites are linked to S-type asteroids (Binzel et al., 1996; Nakamura et al., 2011), which are predominant in the inner belt (2.1-2.8 AU); carbonaceous chondrites are linked to C-type asteroids (Burbine, 2000), which are predominant in the outer belt (beyond 2.8 AU).

Comets are representative of the planetesimal disk that formed at larger distances than the asteroid belt, *i.e.* in between the giant planet orbits and beyond. The classical view is that, while the parent bodies of carbonaceous chondrites are rich in hydrated minerals, comets are rich in water ice, presumably because they formed in a colder environment. The difference between carbonaceous chondrites (or C-type asteroids) and comets, though, is becoming less well defined, with new discoveries. The close flyby images of comets (*e.g.,* comet Borrelly) show very little surface ice and small active regions (Sunshine et al., 2006). The Stardust samples turned out to be quite similar to meteoritic samples (Zolensky et al., 2006). Modeling work on the origin of the dust that produces the zodiacal light (Nesvorny et al., 2010) predicts that at least 50 % of the micro-meteorites collected on Earth are cometary; however, we see no clear separation of micro-meteorites into two categories, which could be traced to asteroidal and cometary dust (Levison et al., 2009). Water-ice has been found on the C-type asteroid Themis (Campins et al., 2010; Rivkin et al., 2010) and some C-type asteroids in the main belt show cometary activity (Hsieh and Jewitt, 2006). The possibility of a continuum in physical and chemical properties between carbonaceous asteroids and comets is well described in Gounelle et al. (2008). The Rosetta encounter with comet 67P/Churyumov-Gerasimenko may shed some light on the question of the relationship between cometary dust and carbonaceous chondrites.

Putting all this information together in a coherent picture is not a simple task. However at the very least we can say that there is evidence that a radial gradient in temperature existed in the disk at the time(s) when planetesimals formed, although this gradient has probably been smeared by the processes (i) and (ii) described above. In particular, planetesimals in the inner disk (in the inner asteroid belt region and presumably also in the terrestrial planet region) appear dry and volatile poor. As proposed by Albarède (2009), probably the gas was removed from the system before that the temperature decreased enough to allow the condensation of the volatiles in the inner solar system and hence primitive objects (*i.e.,* objects with near-solar compositions for moderately volatile elements) could not form there.

**Step II: from planetesimals to planetary embryos**

Once the proto-planetary disk contains a substantial population of planetesimals, the second stage of planet formation can start. The dynamics of accretion is initially dominated by the effect of the gravitational attraction between pairs of planetesimals. A *runaway growth* phase starts, during which the big bodies grow faster than the small ones, hence increasing their relative difference in mass (Greenberg et al, 1978). This process can be summarized by the equation:

$$d/dt\,(M_1/M_2) > 0, \qquad (1)$$

where $M_1$ and $M_2$ are respectively the characteristic masses of the "big" and of the "small" bodies. The reasons for runaway growth can be explained as follows. At the beginning of the growth phase, the largest planetesimals represent only a small fraction of the total mass. Hence the dynamics is governed by the small bodies, in the sense that the relative velocities among the bodies is of order of the escape velocity of the small bodies $V_{esc(2)}$. This velocity is independent of the mass $M_1$ of the big bodies and is smaller than the escape velocity of the large bodies $V_{esc(1)}$. For a given body, the collisional cross-section is enhanced with respect to the geometrical cross-section by the so-called *gravitational focusing* factor $F_g$, so that:

$$dM/dt \sim R^2\,F_g \qquad (2)$$

The gravitational focussing factor is given by (Greenzweig and Lissauer, 1992):

$$F_g = 1 + V_{esc}^2/V_{rel}^2 \qquad (3)$$

where $V_{esc}$ is the body's escape velocity and $V_{rel}$ is the relative velocity of the other particles in its environment. Because $V_{rel} \sim V_{esc(2)}$, the gravitational focusing factor of the small bodies ($V_{esc}=V_{esc(2)}$) is $\sim 2$, while that of the large bodies ($V_{esc}=V_{esc(1)} \gg V_{esc(2)}$) is much larger, of order $V_{esc(1)}^2/V_{rel}^2$. In this situation, remembering that both $V_{esc(1)}$ and the geometrical cross section are proportional to $M_1^{2/3}$, the mass growth of a big body is described by the equation

$$1/M_1\,dM_1/dt \sim M_1^{1/3}\,V_{rel}^{-2} \qquad (4)$$

(Ida and Makino, 1993). Therefore, the relative growth rate is an increasing function of the body's mass, which is the condition for the runaway growth (Fig. 2). Runaway growth stops when the masses of the large bodies become important (Ida and

Makino, 1993) and the latter start to govern the dynamics. The condition for this to occur is:

$$n_1 M_1^2 > n_2 M_2^2, \qquad (5)$$

where $n_1$ (resp. $n_2$) is the number of big bodies (resp. small bodies). In this case, $V_{rel} \sim V_{esc(1)}$, so that $F_g \sim 2$ ; hence $(1/M_1)(dM_1/dt) \sim M_1^{-1/3}$. The growth rate of the embryos gets slower and slower as the bodies grow and the relative differences in mass among the embryos also slowly decreases. In principle, one could expect that the small bodies catch up, narrowing their mass difference with the embryos. But in reality, the now large relative velocities prevent the small bodies from accreting with each other due to collisional fragmentation. The small bodies can only participate to the growth of the embryos. This phase is called 'oligarchic growth'.

The runaway growth phase happens with timescales that depend on the local dynamical time (keplerian time) and on the local density of available solid material. The density also determine the maximum size of the embryos when the runaway growth ends (Lissauer, 1987). Assuming a reasonable surface density of solid materials, the runaway growth process forms planetary embryos of lunar to martian mass at 1 AU in $10^5$-$10^6$y, separated by a few $10^{-2}$ AU (Kokubo and Ida, 1998). Thus, the planetary embryos are not yet the final terrestrial planets. They are not massive enough, they are too numerous and they are closely packed relative to the terrestrial planets that we know. Moreover they form too quickly, compared to the timescale of several $10^7$y suggested for the Earth by radioactive chronometers (Kleine et al., 2010).

Because runaway growth is a local process, the embryos form essentially from the local planetesimals in their neighbourhoods. Little radial mixing is expected at this stage. Thus, if the planetesimal disk is characterized by a radial gradient of chemical properties, such a gradient is expected to be reflected in the embryos distribution. Nevertheless, embryos can undergo internal physical modifications. In view of their large mass and their rapid formation timescale, they can undergo differentiation. We stress, though, that embryo formation cannot be faster than planetesimal formation, because a massive planetesimal population is needed to trigger the runaway growth of the embryos. [It is not correct to say that planetesimals formed late. Iron meteorites formed very rapidly and the meteorite collections could give us a biased sampling of the objects that were present when embryos formed] Embryos formed early would have incorporated enough $^{26}$Al to melt (Dauphas and Pourmand 2011) while embryos formed late could have escaped global differentiation, similarly to what is invoked for Callisto (Canup and Ward, 2002) or Titan (Sotin et al., 2010). The lack of differentiation could have helped the embryos formed in the outer asteroid belt to preserve the water inherited from the local carbonaceous chondrite-like planetesimals. Nevertheless, even if differentiation occurred, water would not have been necessarily lost; water ice could have formed a mantle around a rocky interior, possibly differentiated itself into a metallic core and a silicate outer layer, like on Europa and Ganymede.

The formation of the giant planets is intimately related to the runaway/oligarchic growth of embryos. Beyond the so-called *snow line* at about 4

AU, where condensation of water ice occurred thanks to the low temperature, enhancing the surface density of solid material, the embryos would have been bigger, probably approaching an Earth mass. The formation of the massive cores of the giant planets (of about 10 Earth masses each) is still not well understood. It has been proposed that convergent migration processes would have brought these embryos together, favoring their rapid mutual accretion (Morbidelli et al., 2008; Lyra et al., 2009; Zandor et al., 2010). Once formed, the cores started to accrete massive atmospheres of hydrogen and helium from the proto-planetay disk, thus becoming the giant planets that we know.

**Step III: from embryos to terrestrial planets**

At the disappearance of the gas from the proto-planetary disk, the solar system should have had the following structure: a disk of planetesimals and planetary embryos, roughly of equal total masses, in the inner part; the system of the giant planets already fully formed, in the central part; another disk of planetesimals beyond the orbits of the giant planets. The orbits of the giant planets were inherited from their previous dynamical evolution, dominated by their gravitational interactions with the gas disk. They were therefore likely different from the current ones. We will come back to this important issue below.

Nebular gas was present in the inner part of the disk probably for 3-5 Myr after solar system birth (Dauphas and Chaussidon 2011).The gas has a stabilizing effect on the system of embryos and planetesimals because it continuously damps their orbital eccentricities, through tidal and drag effects, respectively. Thus, when the gas is removed by accretion onto the central star and by photoevaporative loss (Alexander 2008, Hollenbach et al. 2000), the embryos rapidly become unstable and their orbits begin to intersect; collisions can take place (Chambers and Wetherill, 1998). Numerical simulations (Chambers and Wetherill, 2001; Raymond et al., 2004, 2005; O'brien et al., 2006) show that the dynamical evolution is very different in the terrestrial planet region and in the asteroid belt. In the terrestrial planet region, where the perturbations exerted by Jupiter are weak, the embryos' eccentricities remain relatively small, and the embryos can accrete each other in low velocity collisions. Instead, the asteroid belt is crossed by several powerful resonances with Jupiter, which excite the eccentricities of the resonant objects. The orbits of embryos and planetesimals change continuously due to mutual encounters; every time that one of them temporarily falls into a resonance, its eccentricity is rapidly enhanced. Thus, most of the original population eventually leaves the asteroid belt region by acquiring orbits that are so eccentric to cross the terrestrial planet region. Ultimately, the majority of the population originally in the asteroid belt is dynamically removed by collisions with the Sun or ejections on hyperbolic orbits, but a fraction of it can also be accreted by the growing planets inside of 2 AU (see Fig. 3 for an illustration of this process). The typical result of this highly chaotic phase --simulated with several numerical N-body integrations-- is the elimination of all the embryos originally situated in the asteroid belt and the formation a small number of terrestrial planets on stable orbits in the 0.5--2 AU region in a timescale of several tens of millions of years (Chambers and Wetherill, 2001; O'Brien et al., 2006).

This scenario has several strong points:

(i) Typically, 2 to 4 planets are formed on well-separated and stable orbits. If the initial disk of embryos and planetesimals contains about 4-5 Earth masses of solid material, typically the two largest planets are about one Earth mass each. Moreover, in the most modern simulations, accounting for the dynamical interaction between embryos and planetesimals (O'Brien et al., 2006), the final eccentricities and inclinations of the synthetic terrestrial planets are comparable or even smaller than those of the real planets.

(ii) Quasi-tangent collisions of Mars-mass embryos onto the proto-planets are quite frequent (Agnor et al., 1999; Morishima et al., 2009). These collisions are expected to generate a disk of ejecta around the proto-planets (Canup and Asphaug, 2002), from which a satellite is likely to accrete (Canup and Esposito, 1996). This is the standard, generally accepted, scenario for the formation of the Moon.

(iii) The accretion timescale of the terrestrial planets in the simulations is ~30-100 Myr. This is in gross agreement with the timescale of Earth accretion deduced from radioactive chronometers (whose estimates change from one study to another over a comparable range; Yin et al., 2002; Touboul et al., 2007; Kleine et al. 2009; Allègre et al. 1995).

(iv) No embryos and only a small fraction of the original planetesimals typically remain in the asteroid belt on stable orbits at the end of the process of terrestrial planet formation (Petit et al, 2001; O'Brien et al., 2007). This explains well the current mass deficit of the asteroid belt. The orbital eccentricities and inclinations of these surviving particles compare relatively well with those of the largest asteroids in the current belt. Moreover, because of the scattering suffered from the embryos, the surviving particles are randomly displaced in semi major axis, relative to their original position, by about half of an AU. This can explain the partial mixing of asteroids of different taxonomic types, discussed above.

There is a clear dependence of the final outcomes of the simulations on the orbital architecture assumed for the giant planets. As said above, the giant planets should be fully formed by the time the gas is removed from the disk, *i.e.* well before the formation of the terrestrial planets. The simulations with larger orbital eccentricities of the giant planets (up to 2 times the current eccentricities) form terrestrial planets on a shorter timescale and on more circular final orbits than those with more circular giant planets; moreover, the terrestrial planets accrete fewer material from the asteroid belt and the synthetic planet produced at the place of Mars is smaller (Raymond et al., 2009). All these properties are related to each other. In fact, eccentric giant planets deplete more violently the asteroid belt: embryos and planetesimals originally in the belt are removed by collisions with the Sun or ejection on hyperbolic orbit before they have a significant chance to interact with the growing terrestrial planets inside of 2 AU. Thus, terrestrial planet formation proceeds as in a "close" system, with little material (or none) coming into the game from outside of 1.5-2.0 AU. Accretion proceeds faster because the material that builds the planets is more radially confined. On the other hand, faster accretion timescales imply that more planetesimals are still in the system at the end of the terrestrial planet accretion process, so that the orbital eccentricities of the terrestrial planets can be damped more efficiently by planet-planetesimal interactions. Finally, Mars forms smaller because it is close to the edge of the radial distribution of the mass that participates to the construction of the terrestrial planets (Raymond et al., 2009).

To understand which giant planet orbital architecture was more likely at the time of terrestrial planet formation, one has to examine the dynamical evolution that the giant planets should have had in the disk of gas in which they formed. It is well known that, by interacting gravitationally with the gas-disk, the orbits of the giant planets migrate (see Ward, 1997, for a review). Eventually the planets tend to achieve a multi-resonance configuration, in which the period of each object is in integer ratio with that of its neighbour (Morbidelli et al., 2007). The interaction with the disk also damps the planets' orbital eccentricities. Thus, at the disappearance of the gas disk, the giant planets should have been closer to each other, on resonant and quasi-circular orbits (the giant planets could have achieved the current orbits at a much later time, corresponding to the so-called Late Heavy Bombardment; see Morbidelli, 2011 and Sect. 7). Unfortunately, this kind of orbital configuration of the giant planets systematically leads to synthetic planets at ~1.5 AU that are much more massive than the real Mars (Raymond et al., 2009).

Hansen (2009) convincingly showed that the key parameter for obtaining a small Mars is the radial distribution of the solid material in the disk. If the outer edge of the disk of embryos and planetesimals is at about 1 AU, with no solid material outside of this distance, even simulations with giant planets on circular orbits achieve systematically a small Mars (together with a big Earth). This scenario would also imply a short accretion timescale for Mars, of the order of a few My, consistent with estimates obtained from Hf-W systematics (Dauphas and Pourmand 2011). The issue is then how to justify the existence of such an outer edge and how to explain its compatibility with the existence of the asteroid belt, between 2 and 4 AU. The asteroid belt as a whole has a small mass (about $6 \times 10^{-4}$ Earth masses; Krasinsky et al., 2002), but it is well known that it had to contain at least a thousand times more solid material when the asteroids formed (Wetherill, 1989).

The result by Hansen motivated Walsh et al. (2011) to look in more details at the possible orbital history of the giant planets and their ability to sculpt the disk in the inner solar system. For the first time, the giant planets were not assumed to be on static orbits (even if different from the current ones); instead Walsh et al. studied the co-evolution of the orbits of the giant planets and of the precursors of the terrestrial planets. Walsh et al. built their model on previous hydro-dynamical simulations showing that the migration of Jupiter can be in two regimes: when Jupiter is the only giant planet in the disk, it migrates inwards (Lin and Papaloizou, 1986); when it is neighboured by Saturn, both planets typically migrate outward, locked in a 2:3 mean motion resonance (where the orbital period of Saturn is 3/2 of that of Jupiter; Masset and Snellgrove, 2001; Morbidelli and Crida, 2007). Thus, assuming that Saturn formed later than Jupiter, Walsh et al. envisioned the following scenario: first, Jupiter migrated inwards while Saturn was still growing; then, when Saturn reached a mass close to its current one, it started to migrate inwards more rapidly than Jupiter, until it captured Jupiter in the 3/2 resonance (Masset and Snellgrove, 2001; Pierens and Nelson, 2008) when the latter was at ~1.5 AU; finally the two planets migrated outwards until Jupiter reached ~5.5 AU at the complete disappearance of the disk of gas. The reversal of Jupiter's migration at 1.5 AU provides a natural explanation for the existence of the outer edge at 1 AU of the inner disk of embryos and planetesimals, required to produce a small Mars (Fig. 4,5). Because of the prominent inward-then-outward migration of Jupiter that it assumes, Walsh et al. scenario is nicknamed "Grand Tack".

Several giant extra-solar planets have been discovered orbiting their star at a distance of 1-2 AU, so the idea that Jupiter was sometime in the past at 1.5 AU from the Sun is not shocking by itself. A crucial diagnostic of this scenario, though, is the survival of the asteroid belt. Given that Jupiter should have migrated through the asteroid belt region twice, first inwards, then outwards, one could expect that the asteroid belt should now be totally empty. However, the numerical simulations by Walsh et al. show that the asteroid belt is first fully depleted by the passage of the giant planets, but then, while Jupiter leaves the region for the last time, it is re-populated by a small fraction of the planetesimals scattered by the giant planets during their migration. In particular, the inner asteroid belt is dominantly repopulated by planetesimals that were originally inside the orbit on which Jupiter formed, while the outer part of the asteroid belt is dominantly repopulated by planetesimals originally in between and beyond the orbits of the giant planets. Assuming that Jupiter accreted at the location of the snow line, it is then tempting to identify the planetesimals originally closer to the Sun with the largely anhydrous asteroids of E- and S-type and those originally in between and beyond the orbits of the giant planets with the "primitive" C-type asteroids. With this assumption, the Grand Tack scenario explains the physical structure of the asteroid belt (see above), its small mass of the asteroid belt and its eccentricity and inclination distribution.

## 3. Inventories and isotopic compositions of volatiles in terrestrial planets, meteorites, and comets

Major volatile elements are constantly recycled into Earth's mantle at subduction zones and are degassed into the atmosphere in volcanic regions. We review here the inventory of volatile elements in terrestrial and extraterrestrial reservoirs (Tables 1-3). Earth's core is absent from this discussion because it is not accessible and its volatile element content is unknown. This section describes how volatile inventories are estimated from various measurements; it can be skipped entirely without affecting readability.

3.1. Earth
The Earth is an active planet where volatile elements can exchange between the atmosphere, ocean, crust, and mantle. These exchanges have played a major role in maintaining habitable conditions at the surface of our planet. This is best exemplified by the carbon cycle, where $CO_2$ degassing, rock alteration, carbonate deposition, and recycling into the mantle must have regulated surface temperatures through weathering of silicate rocks (written here as $CaSiO_3$) by $CO_2$ from the atmosphere (dissolved in water as carbonic acid), to make carbonate ($CaCO_3$) and silica ($SiO_2$) that precipitate in the ocean: $CaSiO_3+CO_2 \rightarrow CaCO_3+SiO_2$ (*i.e.*, the Urey reaction). When some forcing acts to increase the temperature (*e.g.*, increase of the solar luminosity with time), the rate of chemical weathering increases, more carbonates are formed and the $CO_2$ is drawn down, which creates a negative greenhouse feedback. It is therefore important to assess where major volatile elements reside in the solid Earth, as the atmosphere is not an isolated reservoir. We review hereafter the terrestrial inventories of H, N, C, and noble gases. For Earth's outer portion ("surface reservoirs" in Table 1), we consider the following reservoirs: the atmosphere, the hydrosphere (oceans, porewater, ice, lakes, rivers, and groundwater), the biosphere (marine and land biota), and crust (oceanic crust, continental crust including soils, sediments, igneous and metamorphic rocks).

Mantle volatile budgets are notoriously difficult to establish. For that purpose, we compare volatile elements to other elements that have similar behaviors during mantle melting and magma differentiation. Thus, H (as $H_2O$) is normalized to Ce (MICHAEL, 1995), N (as $N_2$) is normalized to $^{40}Ar$ (produced by decay of $^{40}K$) (MARTY, 1995), C (as $CO_2$) is normalized to Nb (SAAL et al., 2002). We use two approaches that provide lower and upper limits on the budgets of the major volatile elements:

i) We use H/Ce, N/$^{40}Ar$, and C/Nb atomic ratios of mid-ocean ridge basalts (MORBs) with estimates of the Ce, $^{40}Ar$, and Nb concentrations of the depleted mantle to calculate the volatile content of the mantle by considering that the MORB source extends all the way to the core-mantle boundary. This represents a very conservative lower-limit on the volatile budget of the mantle as it neglects the presence of a volatile-rich reservoir in the mantle source of plume-related magmas.

ii) We calculate the H/Ce, N/$^{40}Ar$, and C/Nb ratios in the crust-hydrosphere-biosphere-atmosphere system. Assuming that H, C, and N were derived from mantle melting and that they were not decoupled from Ce, $^{40}Ar$, and Nb during magma generation and recycling, this approach would give good estimates of bulk silicate Earth (BSE) H/Ce, N/$^{40}Ar$, and C/Nb ratios. However, during subduction, H and C could have been less efficiently recycled than Ce and Nb. Therefore, H/Ce and C/Nb ratios of surface reservoirs represent upper-limits on the BSE ratios. Similarly, N might have been degassed from the mantle before significant decay of $^{40}K$, so the N/$^{40}Ar$ ratio of surface reservoirs represents an upper-limit on the N/$^{40}Ar$ ratio of the BSE. In Table 1, we only compile abundance data derived from this second approach, as we believe that it provides a better estimate of the mantle volatile budget.

The noble gas data for Earth are from a compilation by Ozima & Podosek (2001). While approximately half of $^{40}Ar$ is in the solid Earth (ALLÈGRE et al., 1996), most of the inventory of other noble gases except helium is in the atmosphere. For example, the $^{40}Ar/^{36}Ar$ ratio of the silicate Earth is at least a factor of 10 higher than that of the atmosphere (*i.e.*, >3,000 in the mantle vs. 295.5 in air), so the atmosphere must contain >90 % of the global inventory of $^{36}Ar$ (MARTY et al., 1998; VALBRACHT et al., 1997a). This crude estimate assumes a negligible volume for volatile-rich, low $^{40}Ar/^{36}Ar$ (<3,000) mantle reservoirs.

### 3.1.1 Hydrogen

The water content of Earth's atmosphere is small and is spatially and temporally variable. Lécuyer et al. (1998) and Mottl et al. (2007) reviewed the water inventories and D/H ratios of Earth's surface reservoirs (see Table 3 of MOTTL et al., 2007). The hydrosphere contains $1.80 \times 10^{23}$ mol H, 85 % of which is in the oceans ($M_{ocean}=1.37 \times 10^{21}$ kg). The δD value of the hydrosphere is slightly negative (-6 ‰), due to a small contribution of water from ice that is characterized by δD~-400 ‰. The biosphere contains only $1.51 \times 10^{20}$ mol H (0.001 $M_{ocean}$) with a δD value of -100 ‰. The crust contains $3.43 \times 10^{22}$ mol H (0.22 $M_{ocean}$) with a δD value of -75 ‰. Most of that inventory is in shales. Unsurprisingly, the hydrosphere represents the largest reservoir of water at Earth's surface.

The water content of Earth's mantle has been the focus of considerable work as it affects the viscosity of the mantle (HIRTH and KOHLSTEDT, 1996), which has important implications for the dynamical evolution of our planet. Lack of water in the Venusian mantle may explain the absence of plate tectonics on that planet (*e.g.*, RICHARDS et al., 2001). The Ce concentration of the depleted mantle is $3.93 \times 10^{-9}$ mol/g (WORKMAN and HART, 2005a). The H/Ce atomic ratio of MORBs is $\sim 3.1 \times 10^{3}$ (200 ppm $H_2O$ ppm Ce, MICHAEL, 1995), which translates into a H-concentration of the source of MORBs of $1.2 \times 10^{-5}$ mol/g (~110 ppm $H_2O$). A lower limit on the H content of the mantle is therefore $4.9 \times 10^{22}$ mol ($0.4 \times 10^{21}$ kg $H_2O$; 0.3 $M_{ocean}$).

The masses of the continental and oceanic crusts are $21.3 \times 10^{21}$ kg and $6.4 \times 10^{21}$ kg, respectively (MURAMATSU and WEDEPOHL, 1998). The Ce concentration in the continental crust is $\sim 3.1 \times 10^{-7}$ mol/g (43 ppm Ce, RUDNICK and GAO, 2003) while that of the oceanic crust is $\sim 1.1 \times 10^{-7}$ mol/g (15 ppm Ce, STRACKE et al., 2003). We thus estimate that the bulk crust contains $7.2 \times 10^{18}$ mol of Ce. Using the H data from Table 1, we calculate an H/Ce atomic ratio for surface reservoirs of $3.0 \times 10^{4}$. The Ce concentration of the BSE is $4.37 \times 10^{-9}$ mol/g (0.613 ppm Ce, McDONOUGH and SUN, 1995a), which corresponds to a total Ce content of $1.75 \times 10^{19}$ mol. This translates into a H-content of $5.2 \times 10^{23}$ mol H in the BSE (3.4 $M_{ocean}$). Subtracting the H amount in surface reservoirs, we calculate that the mantle must contain at most $3.05 \times 10^{23}$ mol H (2.0 $M_{ocean}$; $2.7 \times 10^{21}$ kg $H_2O$, corresponding to a concentration of ~690 ppm). The δD value of the mantle is ~-80 ‰ (LECUYER et al., 1998).

### 3.1.2. Carbon

Earth's surface carbon cycle has been the subject of much attention due to the societal importance of that element as a fuel and a greenhouse gas. MacKenzie & Lerman (2006) and Sundquist & Visser (2003) reviewed the carbon budget of Earth's surface. The C isotopic compositions are from Heimann & Maier-Reimer (1996). The atmosphere contains $6.6 \times 10^{16}$ mol C with a $\delta^{13}C$ of ~-8 ‰. The land and ocean biota contain $6 \times 10^{16}$ and $0.025 \times 10^{16}$ mol C, respectively with a $\delta^{13}C$ of ~-25

‰. The hydrosphere contains ~$320\times10^{16}$ mol C (most as dissolved inorganic carbon) with a $\delta^{13}C$ value of ~0 ‰. In the crust, soils contain $24\times10^{16}$ mol C, methane hydrates contain $83\times10^{16}$ mol C, coal, oil and natural gas contain $42\times10^{16}$ mol C, sedimentary organic matter contains $105,000\times10^{16}$ mol C, sedimentary carbonates contain $544,000\times10^{16}$ mol C, the igneous oceanic crust contains $7,660\times10^{16}$ mol C, and the igneous-metamorphic continental crust contains $21,400\times10^{16}$ mol C. The $\delta^{13}C$ values of sedimentary organic matter and carbonates, which dominate the crustal budget, are ~-25 ‰ and ~0 ‰, respectively.

The C/Nb atomic ratio of MORBs is ~1,120 (*i.e.*, $CO_2$/Nb~530 ppm/ppm, Saal et al. 2002; CARTIGNY et al., 2008). The Nb concentration of the MORB source is $1.60\times10^{-9}$ mol/g (0.1485 ppm Nb, WORKMAN and HART, 2005a), which translates into a C concentration of $1.79\times10^{-6}$ mol/g (~79 ppm $CO_2$). This corresponds to a lower limit on the C content of the mantle of $7.2\times10^{21}$ mol ($0.3\times10^{21}$ kg $CO_2$, 75 ppm $CO_2$).

The Nb concentration in the continental crust is ~$9\times10^{-8}$ mol/g (8 ppm Nb, RUDNICK and GAO, 2003), while that of the oceanic crust is ~$2.7\times10^{-8}$ mol/g (2.5 ppm Nb, STRACKE et al., 2003). We thus estimate that the bulk crust contains $2.01\times10^{18}$ mol of Nb. Using the C data from Table 1, we calculate a C/Nb atomic ratio for surface reservoirs of $3.4\times10^{3}$. The Nb concentration of the bulk silicate Earth (BSE) is ~$2.6\times10^{-9}$ mol/g (0.240 ppm Nb, MCDONOUGH and SUN, 1995a), the mass of the BSE (mantle+crust) is $4.03\times10^{24}$ kg, the BSE must contain $1.0\times10^{19}$ mol Nb. Using a C/Nb ratio of $3.4\times10^{3}$, we calculate a C content of the BSE of $33.8\times10^{21}$ mol. Subtracting the amount of C in surface reservoirs, we estimate that the mantle must contain at most $27\times10^{21}$ mol C ($1.2\times10^{21}$ kg $CO_2$, corresponding to a $CO_2$ concentration of 300 ppm). The carbon isotopic compositions of samples from the mantle are variable with an average value of $\delta^{13}C$=-5 ‰ (Deines 1980; Pineau and Javoy 1983; MARTY and ZIMMERMAN, 1999; CARTIGNY et al., 2001).

### 3.1.3. Nitrogen

Galloway (2003) reviewed the global geochemical cycle of nitrogen. Boyd (2001) estimated the N isotopic compositions of the various reservoirs involved in that cycle. Houtlon and Bai (2009) focused on the $\delta^{15}N$ value of the biosphere. The atmosphere contains $2.82\times10^{20}$ mol N with a $\delta^{15}N$ of 0 ‰ (by definition). The oceans contain $1.47\times10^{18}$ mol N ($1.43\times10^{18}$ mol N as dissolved $N_2$ and $4.1\times10^{16}$ mol N as dissolved $NO_3^-$) with a positive $\delta^{15}N$ value of +6 ‰ imparted by denitrification. The biosphere contains $7.50\times10^{14}$ mol N ($7.14\times10^{14}$ and $3.6\times10^{13}$ mol N in land and marine biota, respectively) with a $\delta^{15}N$ of ~0 ‰ for the vegetation. The crust contains $7.14\times10^{19}$ mol N (mostly as sedimentary rocks and $1.4\times10^{16}$ as soil organics). The $\delta^{15}N$ of sediments is ~+6 ‰.

The nitrogen content of the mantle is fairly well known from $N_2$-$^{40}Ar$ systematics. MORBs have an N/$^{40}Ar$ ratio of 248 and the depleted mantle has a $^{40}Ar$ concentration of $2.8\times10^{-11}$ mol/g (MARTY and DAUPHAS, 2003, updated with the $^4He$ degassing rate of Bianchi et al. 2010). We can thus calculate a lower-limit on the N content of the mantle of $0.3\times10^{20}$ mol (0.1 ppm $N_2$).

In magmatic systems, $N_2$ and Ar tend to have similar chemical behaviors. The amount of radiogenic $^{40}Ar$ in the whole Earth can be calculated using the amount of $^{40}K$ in the mantle and crust, the half-life of that nuclide, and the known branching ratio between $^{40}K$ and $^{40}Ca$ (ALLÈGRE et al., 1996). We thus estimate that the Earth contains $3.6 \times 10^{18}$ mol $^{40}Ar$. The $^{40}Ar$ content of the atmosphere-crust is $1.9 \times 10^{18}$ mol ($1.65 \times 10^{18}$ mol in the atmosphere and $0.25 \times 10^{18}$ mol in the crust). The $N/^{40}Ar$ ratio of Earth's surface reservoir is therefore 187. Using the amount of $^{40}Ar$ in the whole Earth, this would translate into a total N content of $6.7 \times 10^{20}$ mol. Subtracting the amount of N in surface reservoirs, we calculate an upper-limit on the mantle N content of $3.15 \times 10^{20}$ mol (1.1 ppm $N_2$). This upper-limit is probably close to the actual N content of the mantle because the $N/^{40}Ar$ ratio is approximately constant between surficial (atmosphere-crust) and mantle reservoirs (MARTY and DAUPHAS, 2003).

The nitrogen isotopic composition of the mantle is uncertain. The source of MORBs has a $\delta^{15}N$ of ~-5 ‰ while the source of plume-related magmas has a $\delta^{15}N$ of ~+3 ‰ (Dauphas et al. 1999; MARTY and DAUPHAS, 2003). Plume-related sources may dominate the nitrogen inventory of the mantle, so we ascribe a $\delta^{15}N$ value of +3 ‰ to the whole mantle.

### 3.2 Solar, Mars, Venus, meteorites, and comets

#### 3.2.1 Solar composition

The H, C, N and noble gas concentrations are from a compilation by Lodders (2010). Early on, the sun experienced deuterium burning and the present-day solar wind D/H ratio is not representative of that of the protosolar nebula. The solar D/H ratio pre-deuterium burning is estimated to be $25 \times 10^{-6}$ (ROBERT et al., 2000). The current best estimate of the solar carbon isotopic composition is from a spectroscopic determination, which gave $^{13}C/^{12}C=0.01152$ (SCOTT et al., 2006). The solar wind nitrogen isotopic was recently measured in collector material recovered from the Genesis mission, yielding a $^{15}N/^{14}N$ ratio of 0.002178 ($\delta^{15}N$=-407 ‰, MARTY et al., 2011). The solar $^{20}Ne/^{22}Ne$, $^{21}Ne/^{22}Ne$, and $^{38}Ar/^{36}Ar$ ratios are from measurements of Genesis collector material (HEBER et al., 2009). The solar $^{40}Ar/^{36}Ar$ ratio is from Lodders (2010). Lunar samples provide the best estimates of the solar Kr and Xe isotopic compositions (PEPIN et al., 1995; WIELER, 2002).

#### 3.2.2 Venus and Mars

The inventories of major volatile elements in Venus and Mars are difficult to establish as a significant fraction of those elements may be residing in crustal and mantle reservoirs (*e.g.*, see Sect. 3.1 for Earth, illustrating how complex such estimates can be).

On Venus, C and N are two major atmospheric constituents and their abundances are well known (VON ZAHN et al., 1983). Hydrogen is present as a trace gas in the venusian atmosphere (~30 ppm $H_2O$) and is characterized by a D/H ratio ~157×SMOW (DE BERGH et al., 2006). Noble gas abundances and isotopic compositions in the venusian atmosphere are from Donahue & Russell (1997).

Mars missions and studies of SNC meteorites have given us a better insight into the cycles of major volatile elements on that planet. Most of the water at the surface of Mars is locked up in the cryosphere. Christensen (2006) estimated that it represents ~$5\times10^{18}$ kg of water. This corresponds to an H concentration of $9\times10^{-7}$ mol/g-planet. The D/H ratio was only measured in the atmosphere and it is unsure whether this is representative of the bulk surface composition, which is dominated by ice. For example, Montmessin et al. (2005) proposed a D/H value of 6.5×SMOW for martian surface water, which is higher than that measured in the atmosphere (5.6×SMOW). SNC meteorites have also revealed the presence of a component enriched in deuterium in the martian crust (Leshin Watson 1994; Leshin et al. 1996). The martian atmosphere contains 95.32 % $CO_2$ (OWEN et al., 1977), corresponding to a C concentration of $8.6\times10^{-10}$ mol/g-planet. Carbonates have been found at the surface of Mars and in martian meteorites (BOYNTON et al., 2009; BRIDGES et al., 2001; EHLMANN et al., 2008; MORRIS et al., 2010). However, the amount of C trapped as carbonate in the martian crust is currently unknown. Pollack et al. (1987) showed that in order to have liquid water stable during the early history of Mars a $PCO_2$ of 0.75 to 5 bar would be needed. For reference, a $PCO_2$ of 1 bar would correspond to a C concentration of $1.4\times10^{-7}$ mol/g-planet. Niles et al. (2010) recently measured the C isotopic composition of atmospheric $CO_2$ with good precision ($\delta^{13}C$=-2.5±4.3 ‰). The martian atmosphere contains 2.7 % $N_2$ (OWEN et al., 1977), which correspond to a N concentration of $4.8\times10^{-11}$ mol/g-planet. Manning et al. (2008) suggested that as much $1.2\times10^{19}$ mol N as nitrate might be present in the martian regolith. However, no direct evidence for the presence of such a large reservoir has been found yet. The N isotopic composition of the martian atmosphere is from a compilation by Bogard et al. (2001). The noble gas composition of the martian atmosphere is well known from measurements of gases trapped SNC meteorites. Bogard et al. (2001) reviewed all constraints on the martian noble gas compositions. The $^{20}Ne/^{22}Ne$ ratio of the martian atmosphere is uncertain and values between 7 and 10 have been proposed. The higher values may reflect contamination with terrestrial air ($^{20}Ne/^{22}Ne$=9.8, GARRISON and BOGARD, 1998). For that reason, we adopt a value of 7 for the martian atmosphere, keeping in mind that the actual value may be different. The martian $^{21}Ne/^{22}Ne$ is unknown as $^{21}Ne$ is affected by cosmogenic effects during exposure of SNC meteorites to cosmic rays in space. The Ar isotopic composition is that recommended by Bogard et al. (2001). The Kr isotopic composition is from Garrison and Bogard (1998). Note that $^{78}Kr$ is not reported in Table 3 as the value given by Garrison and Bogard (1998) is unusually high and these authors concluded that this could be due to an analytical artifact (*e.g.*, an isobaric interference of hydrocarbon). Approximately 9 % of $^{80}Kr$ in the atmosphere was produced by neutron capture on $^{79}Br$ in the martian regolith (RAO et al., 2002). The martian Xe isotopic composition is from Swindle et al. (1986).

3.2.3. CI1 meteorites and comets

The H, C, and N concentrations and isotopic compositions of CI1 chondrites are from Kerridge (1985), noble gas abundances as well as Ne and Ar isotopic ratios are from Mazor et al. (1970), Kr isotopic ratios are from Eugster et al. (1967), and Xe isotopic ratios are from Pepin (2000b). In chondrites, $^{21}$Ne is affected by cosmogenic effects and $^{40}$Ar is affected by $^{40}$K decay. In Table 3, we list the $^{21}$Ne/$^{22}$Ne and $^{40}$Ar/$^{36}$Ar ratios of the dominant trapped components present in primitive chondrites (O$_{TT}$, 2002).

The H, C, and N concentrations of comets are from Marty & Dauphas (2002), based on estimates of the dust and gas compositions reported by Jessberger et al. (1988) and Delsemme (1988). Jehin et al. (2009) compiled and discussed measured D/H, $^{13}$C/$^{12}$C, and $^{15}$N/$^{14}$N isotopic ratios measured in comets. The water D/H ratios of the 6 Oort-cloud comets analyzed thus far are higher than the terrestrial ratio by a factor of ~2. Given that water is the dominant H-bearing species, its D/H ratio is representative of the bulk cometary composition, yielding an average of 0.00034. Note that HCN has a high D/H ratio but its contribution to the bulk composition is minor. The $^{13}$C/$^{12}$C ratios measured in CN, HCN, and C$_2$ are close to the chondritic ratio. A significant fraction of C may be present in refractory dust, which has the same C isotopic isotopic composition as the volatile component so this adds no uncertainty. However, the C isotopic composition of important molecules like CO is presently unknown. The $^{15}$N/$^{14}$N ratio was measured in CN and HCN. The main N-bearing species in comets is NH$_3$ so one should bear in mind that the measurements done so far may not be representative of the bulk cometary composition. Jewitt et al. (1997) had found a $^{15}$N/$^{14}$N ratio in HCN of Hale-Bopp that was similar to the terrestrial ratio. However, subsequent measurements did not confirm that result (B$_{OCKELEE}$-M$_{ORVAN}$ et al., 2008) but found instead that the $^{15}$N/$^{14}$N ratio in both HCN and CN of comets was approximately twice the terrestrial ratio (i.e., 0.0068). Recently, the D/H ratio of a Kuiper-belt comet (103P/Hartley 2) was analyzed and against all expectations, was found to be identical to Earth (Hartogh et al. 2011). Meech et al. (2011) reported a $^{13}$C/$^{12}$C ratio of 0.0105±0.0017 for the CN molecules in that comet, consistent with the terrestrial ratio (~0.11). However, the $^{15}$N/$^{14}$N ratio of CN (0.00645±0.00104) was found to be significantly higher than the terrestrial ratio (0.00367). Further work on Kuiper belt comets is needed to documents their volatile element isotopic compositions. The noble gas concentrations in comets are unknown. In Table 2, we list the noble gas concentration data derived from trapping experiments in amorphous ice (B$_{AR}$-N$_{UN}$ et al., 1988; B$_{AR}$-N$_{UN}$ and O$_{WEN}$, 1998; D$_{AUPHAS}$, 2003). Trapping experiments have shown limited isotopic fractionation for Ar, Kr, and Xe, so their isotopic compositions in comets might be solar (N$_{OTESCO}$ et al., 1999).

## 4. Modeling the origin of noble gases in the terrestrial atmosphere

Noble gases are chemically inert and as such they can provide an unparalleled record of the origin and evolution of the atmosphere. They are relatively easy to extract from rocks, show large abundance and isotopic variations, and were the focus of numerous studies since the dawn of geochemistry. For instance, the first radiometric age was reported by Rutherford in 1905 by measuring α-particles produced by U-decay in some pitchblende. However, there is still no consensus on the origin of noble gases in the atmospheres of Earth and other terrestrial planets. A complete model of the origin of terrestrial noble gases must account for the following observations (Tables 2, 3; Fig. 6):

- When normalized to solar composition, noble gases –with the exception of Xe- show a pattern of elemental depletion of light versus heavy noble gases (*e.g.*, the air Ne/Ar ratio is lower than the solar Ne/Ar ratio).
- This depletion is accompanied by larger isotopic fractionation for the lighter noble gases Ne, Ar, and Kr (*e.g.*, the isotopic fractionation in ‰/amu relative to solar is higher for Ne than for Ar).
- Xenon is an exception to the rules outlined above. Despite being heavier, Xe is more depleted than Kr (*i.e.*, the Xe/Kr in air is lower than solar) and it is more isotopically fractionated than Kr (*i.e.*, 38 ‰/amu for Xe *vs* 8 ‰/amu for Kr). This is known as the missing Xe problem and the peculiarities of all the models proposed thus far are related to solving this issue.

More constraints on the origin and timing of the formation of the atmosphere can be derived from noble gas isotopes of radiogenic origin. These can be examined independently of the stable non-radiogenic isotopes and are discussed in more detail in Sect. 5 of this chapter, as well as Chapter 2 (Zhang, 2012) of the present volume. Most models proposed so far involve an episode of hydrodynamic escape to space, which is presented in more detail below. This is followed by a presentation of the three main models that can explain the characteristics of Earth's atmosphere, underlining their strengths and weaknesses.

### 4.1. Hydrodynamic escape and Earth's missing Xe problem

Noble gases like Xe are too heavy to escape from a planet such as Earth by thermal escape following Jean's theory. Instead, they can escape Earth's atmosphere by hydrodynamic escape (ZAHNLE and KASTING, 1986; HUNTEN et al., 1987; SASAKI and NAKAZAWA, 1988; PEPIN, 1991; DAUPHAS, 2003). Individual $H_2$ molecules moving upwards can drag along heavy noble gas atoms that would otherwise not be able to escape Earth's gravity. For such hydrodynamic escape to proceed, a significant inventory of $H_2$ must be present on the protoplanet. Possible origins for $H_2$ include trapped nebular gases or redox reaction between $H_2O$ and a reductant like metallic iron (*i.e.*, $H_2O+Fe^0 \rightarrow H_2+Fe^{2+}O$). The energy required to sustain an outward flux of $H_2$ could be provided by EUV radiation from the young active T-Tauri sun or by gravitational energy released upon impact with large extraterrestrial bodies. Extrasolar planets that are close to their central star can be subject to such hydrodynamic escape (BALLESTER et al., 2007; TIAN et al., 2005; VIDAL-MADJAR et al., 2003). Following Hunten et al. (1987), the single most important quantity that determines whether an isotope *i* can be lost by hydrodynamic escape is the cross-over mass,

$$m_{ci} = m + \frac{kT\Phi}{b^i g X}, \quad (6)$$

where $k$ is the Boltzmann constant, $g$ is the standard gravity, $T$ is the temperature, $m$, $\Phi$ and $X$ are the mass, the escape flux and mole fractions of the major light constituent (i.e., $H_2$), $b^i$ is the diffusion parameter of *i* (*e.g.*, $b^{Xe} \approx 14.4 \times 10^{18}$, 847 $b^{Kr}\approx 16.3\times 10^{18}$, $b^{Ar}\approx 18.8\times 10^{18}$, and $b^{Ne}\approx 26.1\times 10^{18}$ cm$^{-1}$s$^{-1}$ for diffusion in H$_2$ at 270 K;
848 ZAHNLE and KASTING, 1986; PEPIN, 1991) . If the crossover mass is known for one
849 isotope, it is straightforward to derive it for any another isotope j,

$$m_{cj} = m + (m_{ci} - m)\frac{b^i}{b^j}. \quad (7)$$

When $m_i > m_{ci}$, the escape flux is null and the mixing ratio of the trace constituent follows a diffusive equilibrium profile with the scale height augmented by a term corresponding to the flux of H$_2$. When $m_i < m_{ci}$, the trace constituent can escape to space. It is assumed that X remains constant through time at 1 (*i.e.*, the atmosphere is dominated by H$_2$) and that the escape flux evolved as a function of time following the functional form $\Phi = \Phi_0 \Psi(t)$, where $\Psi$ is a function that starts at $\Psi(0)=1$ and decreases with time. If we write $N$ and $N_i$ the column densities of the major (H$_2$) and minor constituents, we can relate the flux of the minor constituent to that of H$_2$ through,

$$\Phi_i = \frac{N_i}{N}\left(\frac{m_{ci} - m_i}{m_{ci} - m}\right)\Phi. \quad (8)$$

Let us introduce $\mu_i = (m_i - m)/(m_{ci}^0 - m)$. Note that in this expression, the cross-over mass $m_{ci}^0$ is taken at time t=0. Using Eq. 1, we have $(m_{ci}-m)=(m_{ci0}-m)\Psi(t)$. Taking $\Phi_0$ positive for a net escape flux, the previous equation can be rewritten as,

$$\frac{dN_i}{N_i} = -\frac{\Phi_0}{N}\big[\Psi(t) - \mu_i\big]dt. \quad (9)$$

Integration of this equation is only meaningful until the time when the atmosphere becomes retentive for that constituent. This happens at a closure time, $t_{ci} = \Psi^{-1}(\mu_i)$. Therefore, the general equation that should be used to calculate the evolution of a trace constituent during hydrodynamic escape is (DAUPHAS, 2003; HUNTEN et al., 1987; PEPIN, 1991),

$$\ln\frac{N_i}{N_i^0} = -\int_0^{\Psi^{-1}(\mu_i)} \frac{\Phi_0}{N}\big[\Psi(t) - \mu_i\big]dt. \quad (10)$$

To integrate this equation, one has to specify $N$ as well as the functional form of $\Psi$. For example, one can assume that $N$ is constant at a value $N_0$ (H$_2$ is constantly replenished) and that the escape flux decreases with time following an exponential $\Psi(t)=\exp(-t/\tau)$. Under these assumptions, the number of free parameters is reduced to 2 (*i.e.*, $\tau \Phi_0/N_0$ and $m_{c,Xe}^0$ if Xe is chosen as the reference noble gas) and the abundance of an isotope in the atmosphere has a simple analytical expression,

$$\ln\frac{N_i}{N_{i,0}} = \frac{\Phi_0 \tau}{N_0}(\mu_i - 1 - \mu_i \ln \mu_i). \quad (11)$$

The degree of curvature in ln $N$ vs $m$ space in hydrodynamic escape is set by $\Psi$, which directly influences the evolution of the cross-over mass with time. Thus, changing the parameterization of the escape flux offers additional freedom in hydrodynamic escape models to reproduce measured data.

    One may wonder what it means to reproduce elemental and isotopic data. Are the two constraints independent? Expressing isotopic fractionation in ‰/amu, we have,

885 $$F = \left(\frac{N_{i2}/N_{i1}}{N_{i2}^0/N_{i1}^0} - 1\right) \times \frac{1000}{m_{i2} - m_{i1}}. \qquad (12)$$

886 We now introduce an alternative quantity to express isotopic fractionation that is
887 sometimes used in isotope geochemistry (δ' notation; Criss and Farquhar 2008),

888 $$F' = \ln\left(\frac{N_{i2}/N_{i1}}{N_{i2}^0/N_{i1}^0}\right) \times \frac{1000}{m_{i2} - m_{i1}}. \qquad (13)$$

889 The general relationship between these two functions is,

890 $$F = \left[e^{F'(m_{i2}-m_{i1})/1000} - 1\right] \times \frac{1000}{m_{i2} - m_{i1}}. \qquad (14)$$

891 When isotopic fractionation is small and the isotopes cover a narrow mass range, F'
892 is almost identical to F. Because isotopic variations are small and affect isotopes of
893 similar masses, $\ln(N/N^0)$ in Eq. 13 can be expanded in $m$ through a Taylor series
894 truncated at the second order,

895 $$F' \approx 1000 \times \frac{d\ln(N_i/N_i^0)}{dm_i}. \qquad (15)$$

896 We recognize the derivative in mass of Eq. 10. Therefore, reproducing the
897 abundance and isotopic composition of the terrestrial atmosphere is mathematically
898 equivalent to fitting a curve in ln $N$ vs $m$ that has the right position and the right
899 derivatives at each mass corresponding to the four noble gases. The general
900 equation to calculate the isotopic composition of a trace constituent during
901 hydrodynamic escape is (this is a new development of the present work),

902 $$F' \approx 1000 \times \frac{\Phi_0}{m_{ci}^0 - m} \int_0^{\Psi^{-1}(\mu_i)} \frac{1}{N} dt. \qquad (16)$$

903 Note that this formula is strictly valid in the limit that the isotopic fractionation is
904 small, that the relative mass difference between isotopes is small, so that Eq. 15 can
905 approximate Eq. 13. Even for light noble gases with large relative mass difference
906 and large isotopic fractionation such as Ne, we have found that this formula gives a
907 good estimate of the isotopic fractionation produced by hydrodynamic escape. Using
908 the model presented previously where $N = N_0$ and $\Psi(t) = \exp(-t/\tau)$ we have,

909 $$F' \approx -1000 \times \frac{\Phi_0}{(m_{ci}^0 - m)N_0} \ln(\mu_i). \qquad (17)$$

910 Expectedly, this equation has the same free parameters as before, i.e., $\tau \Phi_\circ/N_0$ and
911 $m_{c,Xe}^0$. Using $\tau \Phi_\circ/N_0 = 16.04$ and $m_{c,Xe}^0 = 345$ (Hunten et al. 1987; Dauphas 2003), the
912 exact calculation from Eq. 11 gives $F_{Xe}=48.7$, $F_{Kr}=72.0$, $F_{Ar}=109.1$, and $F_{Ne}=161.4$
913 while the approximate calculation using Eq. 15 and 12c gives $F_{Xe}=48.3$, $F_{Kr}=71.6$,
914 $F_{Ar}=110.5$, and $F_{Ne}=164.4$

915 For hydrodynamic escape, we expect the abundance curve to increase with
916 mass (heavier noble gases are more efficiently retained, Eq. 10). This is what is
917 observed for Ne, which is more depleted than Ar, which is in turn more depleted
918 than Kr. However, Xe is more depleted than Kr, which cannot be explained by
919 hydrodynamic escape alone. We also expect light noble gases to be more isotopically
920 fractionated than heavier ones (Eq. 16). This is true for Ne-Ar-Kr, with Ne more
921 isotopically fractionated than Ar, which is in turn more isotopically fractionated
922 than Kr. However, Xe is more isotopically fractionated than Kr, which cannot be explained by hydrodynamic escape alone. Below, we discuss the three models than can account for these puzzling observations, *i.e.*, the missing Xe problem.

### 4.2. Hydrodynamic escape and preferential Xe retention (Pepin, 1991, 1997).

As will be reviewed in Sect. 5, Earth has lost most of radiogenic $^{129}$Xe produced by decay of $^{129}$I. Such loss was possibly driven by hydrodynamic escape and could have been accompanied by elemental and isotopic fractionation of noble gases. An important idea behind the model proposed by Pepin (1991, 1997) is that during core formation, mantle melting, and degassing, Xe can be preferentially retained in the Earth. This idea is partially backed by laboratory experiments as well as ab initio studies showing that under pressure; Xe could be retained in the mantle (Brock and Schrobilgen, 2010; Jephcoat, 1998; Lee and Steinle-Neumann, 2006; Sanloup et al., 2005). In the original 1991 version, Pepin assumed that hydrodynamic escape was powered primarily by EUV radiation from the active young T-Tauri Sun (Zahnle and Walker, 1982). This model was subsequently revised in 1995 to account for the possibility that the Moon-forming giant impact powered noble gas escape to space. The different steps involved in this later model are outlined below (Fig. 7).

In Pepin's model, the initial atmosphere starts with a noble gas abundance pattern that is close to solar except for a significant depletion in Ne. No reason is provided for this depletion other than the same feature is required to explain the composition of the venusian atmosphere. The assumed isotopic compositions are solar for all noble gases. Collision between the proto-Earth and a planetary embryo leads to an episode of hydrodynamic escape. Noble gases are lost to space according to the prescriptions outlined in Sect. 4.1. The lighter noble gases are more depleted and isotopically fractionated than heavier ones. Subsequently, noble gases are degassed from Earth but Xe is preferentially retained in the mantle (or core). Therefore, all Xe in the atmosphere at that time is from the first escape episode and its isotopic composition is highly fractionated. Krypton on the other hand is derived from mantle degassing of juvenile noble gases and it shows limited isotopic fractionation. By adjusting the proportions of leftover gases from the escape episode and juvenile gases from mantle degassing, Pepin (1997) was able to reproduce the near-solar Xe/Kr ratio of the atmosphere. Argon is mainly derived from the escape episode. In a late escape episode driven by EUV radiation, some Ne is lost and is fractionated isotopically. Heavier noble gases are not affected by this episode.

This model can reproduce most of the features of Earth's atmosphere. However, an important difficulty is that no mantle reservoir has been documented yet that could host the missing Xe, which Pepin (1991; 1997) argued was retained in the Earth. High-pressure experiments have shown that the core or deep mantle could serve as a possible repository for the missing Xe. However, Mars possess the same missing Xe problem as Earth despite the fact that its internal pressure is much lower (*e.g.*, ~23 GPa at the core-mantle boundary on Mars vs ~135 GPa on Earth) and the mineral physics involved would presumably be different. Finally, the model predicts that Kr in Earth's mantle should be less isotopically fractionated compared to the atmosphere, which is inconsistent with the measured Kr isotopic composition of $CO_2$ well gases (Holland et al., 2009).

### 4.3. Hydrodynamic escape and solubility-controlled Xe degassing (Tolstikhin and O'Nions 1994).

Of all noble gases, Xe has the lowest solubility in silicate melts (LUX, 1987); *S(Ar)* = 5 x10$^{-5}$ cm$^3$ STP / (g atm), *S(He)/S(Ar)* = 10, *S(Ne)/S(Ar)* = 4, *S(Kr)/S(Ar)* = 0.55 and *S(Xe)/S(Ar)* = 0.3. During mantle degassing, it is conceivable that Xe would be preferentially degassed relative to Ne, Ar, and Kr. Tolstikhin and O'Nions (1994) followed in the footsteps of Pepin (1991) and proposed a model to solve the missing Xe problem that relies on this idea of solubility controlled degassing during Earth's accretion (Fig. 8). Ideal gas behaviour, equilibrium partitioning between melt and gas phases, and infinitesimal small gas/melt ratio were assumed. In this scenario, the starting composition of Earth's heavy noble gases is chondritic. During mantle degassing, insoluble Xe is preferentially degassed into the atmosphere, where it is lost and isotopically fractionated by hydrodynamic escape. In subsequent stages, remaining mantle noble gases are degassed and lost to space. This model can account for the elemental and isotopic characteristics of noble gases in Earth's atmosphere. Tolstikhin and Kramers (2008) have shown that the conditions of solubility controlled ingassing/degassing and weakening hydrodynamic escape can be achieved in the context of a realistic model of post-giant impact magma ocean evolution on the early Earth. The model cannot explain the heavy Kr isotopic composition of Earth's mantle relative to the atmosphere (HOLLAND et al., 2009).

### 4.4. Hydrodynamic escape and cometary input (Dauphas 2003).

Except for measurements of dust returned from comet 81P/Wild 2 by the Stardust spacecraft (MARTY et al., 2008), the noble gas composition of comets is completely unknown (BOCKELEE-MORVAN et al., 2004). The only indirect information on the composition of cometary ice is from laboratory trapping experiments in amorphous ice (BAR-NUN et al., 1988; Owen and Bar-Nun 1992; BAR-NUN and OWEN, 1998; NOTESCO et al., 2003). These experiments revealed that at a certain temperature, Xe is less efficiently trapped than Kr. No satisfactory microphysical explanation has been provided for this depletion and further experimental work will be required to understand this aspect of noble gas trapping experiments in amorphous ice. Owen and Bar-Nun (1992) showed that this could explain one important aspect of Earth's missing Xe, namely why the Xe/Kr ratio in Earth's atmosphere is lower than the solar ratio. Dauphas (2003) showed that such a cometary input could explain all the abundance and isotopic characteristics of noble gases in Earth's atmosphere (Fig. 9). In this model, atmospheric noble gases start with solar isotopic compositions and relative abundances (actually, some of the noble gases may have been derived from a chondritic source but this was not considered in the model to avoid unnecessary complication). The noble gases are then lost to space by hydrodynamic escape powered by EUV radiation from the young T-Tauri Sun. Following this escape episode; lighter noble gases are more depleted and isotopically fractionated than heavier ones. In a second stage, comets deliver noble gases with near-solar isotopic compositions but with an abundance pattern corresponding to that measured in trapping experiments at ~50 K (or ~25

K based on experiments performed at lower deposition rates; Notesco and Bar-Nun, 2003). As Xe trapped in amorphous ice is possibly deficient relative to Kr, comets could have delivered isotopically solar Kr without disturbing the Xe isotopic signature produced by the hydrodynamic escape episode. This model can also explain the abundances and isotopic compositions of other noble gases. In the model of Dauphas (2003), the isotopic composition of Xe was assumed to be solar but low temperature condensation is prone to creating measurable isotopic fractionation. The Hertz-Knudsen equation derived from the kinetic theory of gases gives the flux of molecules impinging on a surface $F = P/\sqrt{2\pi m k T}$ (molecule m$^{-2}$ s$^{-1}$). Only a fraction γ (sticking coefficient) of the impacting molecules will be able to adhere permanently to the surface. So the fractionation of two isotopes 1 and 2 upon trapping is simply $\alpha_{2/1} = (\gamma_2/\gamma_1)\sqrt{m_1/m_2}$. Assuming a sticking coefficient that does not depend on mass, the isotopic fractionation for trapped vs gas-phase Xe is -3.8 ‰/amu. This is opposite in sign and is much smaller than what is measured in Earth's atmosphere of ~+38 ‰/amu, meaning that comets did not deliver already fractionated Xe or that more complex ice trapping mechanisms must be considered. For example, it remains to be seen if trapping or processing of Xe in ice under UV (in the interstellar or outer solar system) can fractionate Xe isotopes and abundance relative to other noble gases. While lighter noble gases have ionization energies higher than hydrogen (Ne=21.56 eV, Ar=15.76 eV, Kr=14.00 eV *vs.* H=13.60 eV), xenon has an ionization energy (12.13 eV) that is lower than hydrogen, which could have affected its trapping efficiency in conditions relevant to cometary ice formation. Laboratory experiments have shown that ionized Xe could be fractionated isotopically during trapping in refractory solids (Dziczkaniec et al. 1981; Bernatowicz and Hagee 1987; Ponganic et al. 1997; Hohenberg et al. 2002; Marrocchi et al. 2011) but the effects are small in regard of the large Xe isotopic fractionation measured in air. Experiments of noble gas trapping in ice under ionizing radiation remain to be performed. In the different context of hydrodynamic escape from Earth's atmosphere, the low ionization energy of Xe had been identified as a feature that could potentially explain its fractionation relative to lighter noble gases (Zahnle 2000; Pujol et al. 2011). Indeed, this increases the cross-section of Xe for collisions with hydrogen and facilitates its escape to space.

The composition of Jupiter can provide some clues on the noble gas composition of comets. Indeed, the Galileo probe measured noble gas abundances in Jupiter and found that these were enriched by a factor of 2 to 3 relative to hydrogen and solar composition (except for Ne that can be sequestered in the planet's interior). Such enrichments can be explained if these noble gases were delivered to Jupiter in the form of comets (Owen et al. 1999). The enrichment is uniform for Ar, Kr, and Xe (Ar/Kr/Xe are in solar proportions) suggesting that the planetesimals that formed Jupiter trapped volatiles at low temperature, *i.e.* <30 K. To explain the missing Xe in Earth's atmosphere, heavy noble gases have to be fractionated relative to each other, which would require higher trapping temperatures. Although Jupiter's atmosphere can inform us on the composition of icy planetesimals in the giant-planet forming region, these comets may not be relevant to the icy planetesimals that delivered noble gases to Earth's atmosphere.

One difficulty with the model proposed by Dauphas (2003) is that it relies on the composition of hypothetical comets based on trapping experiments that are not fully understood. Like the other two models, the similarity between the terrestrial and martian atmospheres must be taken as a coincidence. However, this model can explain why mantle Kr is enriched in the heavy isotopes relative to the atmosphere (HOLLAND et al., 2009) because it predicts that mantle Kr should have been fractionated in the early Earth while atmospheric Kr was delivered later by the accretion of cometary material with near-solar isotopic composition. Following a similar line of reasoning, Marty and Meibom (2007) argued that accretion of extraterrestrial material by Earth during the late heavy bombardment at ~3.9 Ga could have modified the noble gas composition of Earth's atmosphere.

### 4.5. Standing issues

While several models can explain all the elemental and isotopic characteristics of Earth's atmosphere, there are a number of standing issues that will need to be addressed to make progress in our understanding of the origin of Earth's atmosphere. One of those is to understand why Mars and Earth both present the same missing Xe problem. The similarity can be taken as a coincidence but this is not very satisfactory as Earth and Mars have very different masses and should have followed different evolutionary tracks. In particular, all of the three models discussed above (Sect. 4.2, 4.3, and 4.4) invoke hydrodynamic escape to fractionate Xe isotopes, the efficiency of which should depend on the planet's gravity. One possible solution is that Earth inherited its missing Xe problem from one or several Mars-size embryos that collided to form our planet during the stage of chaotic growth (Dauphas and Pourmand 2011). Indeed, Mars accreted in a few million years, while Earth's accretion was not completed until >30 Myr after solar system birth. Thus, it is conceivable that Earth formed from Mars-like embryos and that some of the features measured in the terrestrial atmosphere reflect evolution on these embryos. The Xe composition of the Venusian atmosphere is completely unknown. This is unfortunate, as this would provide important constraints on the origin of the terrestrial planet atmospheres. If Venus has the same Xe isotopic composition as Earth and Mars, resorting to coincidental conditions to explain this similarity will be even less tenable. The single most important data that is needed at present is an estimate of the noble gas composition of cometary ice. Indeed, it is the only major noble gas planetary reservoir that has not been measured; yet laboratory experiments seem to indicate that they could have played a major role in establishing the noble gas composition of the atmospheres of terrestrial planets. This is an issue that the Rosina instrument on board the Rosetta mission (Balsiger et al. 2007) may be able to settle by measuring the composition of gases emanating from the surface of comet 67P/Churyumov-Gerasimenko.

Pujol et al. (2011) recently reported the discovery of xenon with isotopic composition intermediate between modern air and chondritic/solar in fluid inclusions in ~3.5 Ga quartz and barite. These measurements suggest that the fractionated Xe isotopic composition of the modern atmosphere was established over an extended period of several billions of years. If correct, this may call for a revision of the evolution of the atmosphere. Pujol et al. (2011) suggested that the isotopic fractionation of Xe was related to its low ionization potential compared to other noble gases. Ionized Xe has a much larger size than the neutral form, which could have promoted its atmospheric loss (Zahnle 2000). In this scenario, noble gases Ne, Ar, and Kr would have been derived from mixing between solar and meteoritic components (Marty 2012). A proper mechanism/setting to selectively lose ionized Xe remains to be identified. Given that Mars shows the same missing Xe problem, fractionating Xe on precursors of planets (*e.g.*, embryos or comets) may be more appealing.

## 5. Nature and timing of noble gas degassing and escape

Several noble gases possess parent radioactive isotopes, which can be used to establish the timing of mantle degassing and escape to space. $^{40}$K decays into $^{40}$Ar with a half-life of 1.248 Gy. $^{129}$I is an extinct nuclide that decays into $^{129}$Xe with a half-life of 15.7 My. $^{244}$Pu is another extinct radionuclide that can produce fissiogenic Xe isotopes ($^{131}$Xe-$^{136}$Xe) with a half-life of 80.0 My and a branching ratio of 0.012, 5.6 % of which goes to $^{136}$Xe. Finally, fissiogenic Xe isotopes can also be produced by decay of $^{238}$U ($t_{1/2}$=4.468 Gy) with a spontaneous fission branching of $5.45 \times 10^{-7}$, 6.3 % of which goes to $^{136}$Xe. Both $^{232}$Th and $^{235}$U have very low spontaneous fission probabilities ($1.1 \times 10^{-11}$ and $7.0 \times 10^{-11}$, respectively), so they did not contribute significantly to fissiogenic Xe isotopes on Earth. SHUKOLYUKOV et al. (1994) and MESHIK et al. (2000) argued that some of Xe isotopes might have been contributed by neutron-induced fission of $^{235}$U in the form of CFF-Xe (Chemical Fractionation of Fission Xe) if Oklo-type natural reactor were abundant on early Earth. However, Ballentine et al. (2002) presented mass balance arguments against a significant contribution of CFF-Xe to Earth's atmosphere. At large scales and over long periods of time, it is unlikely that CFF-Xe would be preferentially released into the atmosphere relative to the complementary enriched $^{136}$Xe component.

### 5.1. Mantle noble gases

Several reviews have discussed this topic at length (BALLENTINE and HOLLAND, 2008; FARLEY and NERODA, 1998; LUPTON, 1983; Ozima and Podosek 2001; Graham 2002) and a brief summary is provided below (the depleted upper-mantle composition is compiled in Table 4). This section (5.1) describes how mantle volatile inventories can be inferred from measurements of mantle-derived samples; it can be skipped entirely without affecting readability.

#### 5.1.1. MORBs and $CO_2$ well gases

The elemental and isotopic compositions of the depleted upper-mantle are known from measurements of mid-ocean ridge basalts (MORBs). The noble gas that has been most extensively studied is helium (Fig. 10). MORBs have a near-constant $^3$He/$^4$He ratio (R/R$_A$~8, where R is the $^3$He/$^4$He ratio of a sample and R$_A$ is the $^3$He/$^4$He ratio in air, i.e. $1.38 \times 10^{-6}$), which is more radiogenic than that of most plume-related sources (KURZ et al., 1982). This indicates that MORBs are from a relatively well-homogenized mantle source that has been extensively degassed (*i.e.*, the depleted convective upper-mantle). Because He is constantly lost to space by thermal escape and its concentration is low in the atmosphere (its residence is only ~1 Myr), the $^3$He flux from MORBs can be detected in the form of a plume in the ocean above ridges (CRAIG et al., 1975; LUPTON and CRAIG, 1981). This can be used in turn to infer a $^3$He flux at ridges of 527±102 mol/year (BIANCHI et al. 2010). Using a global ridge production rate of 21 km$^3$/year (with a density of 3 g/cm$^3$) and a degree of partial melting of 10 % for MORBs, it is possible to calculate the $^3$He content of their mantle source of $(8.4±1.6)×10^{-16}$ mol/g. For a $^3$He/$^4$He ratio of 7.3 $R_A$ (Fig. 10), this gives a $^4$He concentration of $(8.3±1.6)×10^{-11}$ mol/g. Using Th and U concentrations of 7.9 and 3.2 ppb respectively for the depleted MORB mantle (WORKMAN and HART, 2005b), this amount of $^4$He corresponds to ~2.5 Gyr of accumulation.

Noble gases heavier than He are abundant in the atmosphere, and MORB samples are prone to contamination by air. In that respect, the gas-rich samples known as popping-rocks have played a tremendous role in establishing the composition of the upper-mantle for noble gases. These rocks are named after the noise that they make on the ship deck when the vesicles burst due to decompression. Equally important are $CO_2$ well gases, which sample a mixture between crustal and mantle-derived gases (BALLENTINE and HOLLAND, 2008). Neon isotope systematics reveals interesting facts about Earth (Fig. 11). The $^{20}$Ne/$^{22}$Ne ratio can only vary due to mass fractionation. Sarda et al. (1988), Marty (1989) and Honda et al. (1991) showed that the mantle $^{20}$Ne/$^{22}$Ne ratio was higher than the air ratio (9.8). In plume-derived samples from the Kola peninsula (Russia), the $^{20}$Ne/$^{22}$Ne ratio is higher ($\geq 13.0±0.2$) and approaches the solar ratio of 13.7 (YOKOCHI and MARTY, 2004). Other parts of the mantle, including the source of MORBs, may be characterized by a lower $^{20}$Ne/$^{22}$Ne ratio of ~12.5 (BALLENTINE et al., 2005; TRIELOFF et al., 2000). Regardless of these complications, the large difference in Ne isotopic composition between the silicate Earth and air is the strongest evidence that atmospheric noble gases could not have been derived solely from degassing of the mantle without further processing. The $^{21}$Ne/$^{22}$Ne ratio can vary due to nucleogenic production in the mantle through the Wetherill reactions, in particular $^{18}O(\alpha,n)^{21}Ne$, where $\alpha$ particles are from decay of U and Th (YATSEVICH and HONDA, 1997). The $^{21}$Ne/$^{22}$Ne therefore reflects the long-term evolution of the U/$^{22}$Ne ratio and like the $^3$He/$^4$He ratio can inform us on the degree of degassing of mantle reservoirs. At a $^{20}$Ne/$^{22}$Ne ratio of 12.5-13.7, the $^{21}$Ne/$^{22}$Ne ratio of MORBs is higher than the solar ratio and is also higher than that seen in some plume-related magmas. This confirms the view that the source of MORBs is a degassed reservoir. The $^{40}$Ar/$^{36}$Ar ratio of the MORB source is much higher than air (295.5) and is probably around 25,000-40,000 (BURNARD et al., 1997; MOREIRA et al., 1998, Fig. 12). The stable isotope ratio $^{38}$Ar/$^{36}$Ar is indistinguishable from the atmospheric ratio (Ballentine and Holland 2008; RAQUIN and MOREIRA, 2009). Krypton isotopes in MORBs show indistinguishable isotopic composition from air. However, more recent work has shown, through high-precision measurements of $CO_2$ well gases, that the Kr isotopic composition of Earth's mantle was probably different compared to that of the atmosphere (HOLLAND et al., 2009, Fig. 13). The atmosphere is slightly enriched in the heavy isotopes of Kr by ~8

‰/amu relative to solar. Holland et al. (2009) found that mantle Kr was even more fractionated isotopically, which is consistent with a meteoritic or fractionated solar origin and addition of solar-type Kr by late cometary accretion, as was suggested by Dauphas (2003). Xenon isotopes have been the focus of much work, as they comprise several decay systems that can be used as chronometers of mantle degassing and atmospheric loss. Xenon non-radiogenic isotopes in MORBs are indistinguishable from air. However, higher precision measurements of $CO_2$ well-gases have revealed that mantle Xe has a different isotopic composition relative to the atmosphere (Caffee et al., 1999; Holland and Ballentine, 2006, Fig. 13). Because solar and meteoritic reservoirs have similar Xe isotopic compositions, it is not possible to tell one end-member from the other using non-radiogenic Xe isotopes alone. MORBs show excess $^{129}Xe$ from decay of $^{129}I$ as well as excess fissiogenic isotopes (*e.g.*, $^{136}Xe$) from decay of $^{238}U$ and $^{244}Pu$ (Kunz et al., 1998; Moreira et al., 1998; Staudacher and Allègre, 1982, Fig. 14). The $^{129}Xe/^{130}Xe$ ratio of the upper-mantle is estimated to be 7.9 based on both MORB and $CO_2$-well gas measurements (Holland and Ballentine, 2006; Moreira et al., 1998). An important parameter is the fraction of $^{136}Xe$ derived from $^{244}Pu$ *vs.* $^{238}U$, which is difficult to estimate because the fission spectra of these two actinides are close. Kunz et al. (1998) estimated that 32±10 % of fissiogenic $^{136}Xe$ (*i.e.*, excess relative to air) was from $^{244}Pu$ and that the rest was from $^{238}U$, similar within uncertainties to the decomposition obtained for plume-related magmas by Yokochi and Marty (2005). Assuming non-radiogenic mantle Xe to be U-Xe (a hypothetical component described in Sect. 5.2; Pepin, 2000a) or solar wind-Xe, Pepin and Porcelli (2006) obtained a higher fraction of Pu-derived Xe of 63-75 %.

    The elemental compositions of the MORB source is compiled in Table 4 (Moreira et al., 1998). The inferred composition of the source of $CO_2$ well gases is identical to that of MORBs (Holland and Ballentine, 2006). The $^3He/^{22}Ne$ is not very far from the solar ratio; i.e., 4.9 (Moreira et al., 1998) or 10.2 (Honda and McDougall 1998) vs 2.65 for the solar composition. Yokochi and Marty (2004) showed that the source $^3He/^{22}Ne$ ratio varied systematically between mantle reservoirs and that this ratio correlated with the time-integrated excess in nucleogenic $^{21}Ne^*$, presumably reflecting mantle differentiation processes. Heavier noble gases $^{36}Ar$, $^{84}Kr$, and $^{130}Xe$ present an abundance pattern that is more akin to a meteoritic or a fractionated solar component. Yet, the non-radiogenic isotopic composition is dominantly atmospheric and distinct from meteorite values. Holland and Ballentine (2006) suggested that this pattern best matched the noble gas composition of seawater, implying that the heavy noble gas inventory of the mantle could have been derived from recycling. This conclusion was corroborated by subsequent studies that found seawater-derived noble gases in mantle wedge peridotites (Sumino et al. 2010) and subducted serpentinites (Kendrick et al. 2011). Thus, trapped noble gases may have had two sources. Light noble gases (*i.e.*, He and Ne) could have been derived from solar composition by ion implantation or dissolution in the magma ocean of nebular gases, while heavier noble gases could have been derived from recycling of air.

    In Table 4, we summarize our knowledge of the upper-mantle volatile composition based on MORB and $CO_2$ well gas measurements.

        5.1.2. Plumes

Plume-derived magmas can have very high $^3$He/$^4$He ratios (Fig. 10). The current record holder is a sample from the Baffin Island with a R/R$_A$ ratio of 49.5 (STUART et al., 2003). These elevated $^3$He/$^4$He ratios indicate that the samples originate from a source region that is less degassed than the source of MORBs (see Anderson 1998 for an alternative view point). OIB samples from Loihi (Hawaii) as well as intrusive samples from the Kola Peninsula (Russia) have provided a wealth of data on the composition of plume-related sources. The $^{20}$Ne/$^{22}$Ne ratio in samples from Kola is close to solar (YOKOCHI and MARTY, 2004). The $^{21}$Ne/$^{22}$Ne ratios of Loihi (HIYAGON et al., 1992; HONDA et al., 1993; VALBRACHT et al., 1997b) and Kola (MARTY et al., 1998; YOKOCHI and MARTY, 2004) are lower than MORBs, indicating that their mantle sources are less degassed than the convective upper-mantle (Fig. 11). The $^{40}$Ar/$^{36}$Ar ratio of the source of plume samples is around 4,500-8,000 (Farley and Craig 1994; MARTY et al., 1998; VALBRACHT et al., 1997a; Trieloff et al. 2000, 2002), confirming that their source mantles are not as efficiently degassed as the MORB mantle (Fig. 12). Krypton isotope ratios are indistinguishable from those in the atmosphere. Stable Xe isotopes are again indistinguishable from the atmosphere. However, excess $^{129}$Xe and fissiogenic isotopes are found (Poreda and Farley 1992; TRIELOFF et al., 2000; YOKOCHI and MARTY, 2004). The precision of Xe isotope measurements on plume samples is insufficient to disentangle what results from decay of $^{244}$Pu from decay of $^{238}$U. This decomposition is critical to establish the timescale of mantle degassing as the two parent nuclides have considerably different half-lives. Yokochi and Marty (2004) proposed an elegant method to estimate the relative contributions of these two nuclides to Xe fission isotopes based on the $^{136}$Xe*/$^4$He* ratio (Fig. 14). The rationale is that $^{244}$Pu-decay produces a $^{136}$Xe/$^4$He ratio that is very different from that produced by decay of $^{238}$U. They concluded that 33-60 % of fission $^{136}$Xe was from decay of $^{244}$Pu while the rest was from decay of $^{238}$U. This is similar (within error bars) to the estimate of ~32±10 % of fission $^{136}$Xe from $^{244}$Pu for MORBs (KUNZ et al., 1998).

### 5.2. Model age of atmosphere retention

Wetherill (1975) calculated a model age of the atmosphere that has influenced all discussions on Earth's early evolution. The atmosphere contains excess $^{129}$Xe* that must come from the decay of $^{129}$I. In the simplest case of a two-stage model, radiogenic Xe is entirely lost to space until some retention time $t_r$, after which time it is completely retained. Stated otherwise, the amount of $^{129}$Xe* found in the atmosphere (6.8±0.5 % of atmospheric $^{129}$Xe, PEPIN, 2000a) should correspond to the amount of $^{129}$I that was in the crust when the atmosphere became retentive for Xe. Accordingly, we have,

$$\frac{^{129}\text{Xe}^*_{atm}}{^{127}\text{I}_{crust}} = \left(\frac{^{129}\text{I}}{^{127}\text{I}}\right)_{t_r} = e^{-\lambda_{129} t_r}. \qquad (18)$$

The amount of radiogenic $^{129}$Xe in Earth's atmosphere is 2.8×10$^{11}$ mol (PEPIN, 2000a). The $^{127}$I content of the crust is 2.6×10$^{17}$ mol (DÈRUELLE et al., 1992). These surface inventories correspond to a retention time of ~100 Myr. Several studies have refined this approach and have found similar closure times (*e.g.,* ZHANG, 1998).

In the first 100 My, 1-exp(-$\lambda_{129}\times 100$)=99 % of $^{129}$I would have decayed and the corresponding inventory of $^{129}$Xe is missing from the present Earth. This means that at least the same amount and possibly more of the initial inventory of Xe non-radiogenic isotopes on Earth was lost to space. This result is consistent with the large isotopic fractionation of Xe isotopes seen in the atmosphere, which could have been created by hydrodynamic escape of an H$_2$-rich protoatmosphere (HUNTEN et al., 1987). A similar dating method could potentially be used based on fissiogenic Xe isotopes. This approach however faces an important difficulty. If one corrects atmospheric Xe for mass fractionation using light Xe isotopes, one obtains an isotopic composition that has less of some of the heavy isotopes of Xe than the solar composition (PEPIN, 2000a). This is an odd result as it is difficult to envision how a nebular component could be deficient in some isotopes compared to solar composition. Pepin and Phinney (1978, "Components of Xenon in the Solar System", unpublished preprint) speculated that such a component existed and named it U-Xe (here U does not mean uranium; it stands for "Ur", a German word for indigenous). This component is not found on Mars (martian Xe seems to be derived from solar Xe, SWINDLE and JONES, 1997) and was also not detected in any meteoritic component (BUSEMANN and EUGSTER, 2002). Cometary ice is one of the few noble gas reservoirs in the solar system that has not been documented yet and it might be the carrier of the elusive U-Xe. The amounts of fissiogenic Xe in Earth's atmosphere depends heavily on the assumed starting composition, which is not well defined. Therefore, the contributions of plutonogenic and uranogenic Xe in Earth's atmosphere are uncertain.

5.3. Model age of mantle degassing

The radiogenic nuclides that could potentially be used to constrain the timescale of atmosphere formation are $^4$He, $^{21}$Ne, $^{40}$Ar, $^{129}$Xe, and fissiogenic Xe isotopes (e.g., $^{136}$Xe). $^4$He is of limited use as it is continuously lost to space. $^{21}$Ne is also of limited use as the atmospheric composition is near the mass fractionation line that runs through the solar composition, so the amount of nucleogenic $^{21}$Ne in the atmosphere is highly uncertain. This leaves us with $^{40}$Ar, $^{129}$Xe, and $^{136}$Xe. Because the different parent radionuclides to these isotopes have a range of half-lives from 15.7 My ($^{129}$Xe) to 4.47 Gy ($^{238}$U), noble gas isotopes can be used to investigate the degassing history of the Earth over different timescales. Models of mantle degassing often assume that this follows a first order rate, where the rate of degassing is proportional to the amount in the mantle. This can be understood if the volume of mantle tapped by melting is approximately constant and the mantle is well mixed.

Early degassing is best investigated using Xe as it possess several parent radionuclides with short half-lives (KUNZ et al., 1998; YOKOCHI and MARTY, 2005). Mantle degassing timescales are plagued by the fact that the initial Xe isotopic composition is uncertain (*i.e.*, "U"-Xe component, Sect. 5.2). However, Earth's mantle has excess fissiogenic Xe compared to air, which makes the decomposition into plutonogenic and uranogenic components more robust (*e.g.*, one can assume conservatively that the initial Xe isotopic composition of the mantle lies between U-Xe and air-Xe). The most refined model at the present time is that proposed by Yokochi and Marty (2005), who used their decomposition of $^{136}$Xe into $^{238}$U and $^{244}$Pu-derived components to constrain the degassing history of the Earth (also see COLTICE et al., 2009). If no degassing had taken place after solar system formation, then the ratio of $^{136}$Xe$_{Pu}$/$^{136}$Xe$_{U}$ should be identical to the chondritic ratio, *i.e.* 27. On the other hand, if most degassing had taken place after $^{244}$Pu decay, then mantle Xe would have been dominated by $^{238}$U-derived fission and the $^{136}$Xe$_{Pu}$/$^{136}$Xe$_{U}$ would be close to zero. A similar reasoning can be applied to the ratio of $^{129}$Xe$^*$/$^{136}$Xe$_{Pu}$. The fact that a significant fraction of mantle $^{136}$Xe is derived from $^{244}$Pu therefore indicates that most Xe degassing took place while $^{244}$Pu was live. A concordant age for Pu-U and I-Pu cannot be obtained by using a single stage degassing history (KUNZ et al., 1998; TOLSTIKHIN and HOFMANN, 2005; YOKOCHI and MARTY, 2005). A solution to this problem is to consider a model where the degassing occurs over a period $\Delta t$ after solar system birth and to assume that the rate of degassing during that time was proportional to the amount in the mantle,

$$\frac{d^iXe^*}{dt} = P(t)\lambda Y - \alpha^i Xe^*, \quad t < \Delta t$$

$$\frac{d^iXe^*}{dt} = P(t)\lambda Y, \quad t > \Delta t \tag{19}$$

where $P(t)$ is the abundance of the parent nuclide, $\lambda$ is its decay constant, $Y$ is the decay yield, and $\alpha$ is a degassing constant. Using the two ratios $^{136}$Xe$_{Pu}$/$^{136}$Xe$_{U}$ and $^{129}$Xe$^*$/$^{136}$Xe$_{Pu}$, Yokochi and Marty (2005) were able to write down two equations and solve the system for the two unknowns $\alpha$ and $\Delta t$. They obtained $\alpha \sim 1\times10^{-8}$ to $2\times10^{-8}$. According to this analysis, mantle Xe was degassed early with an e-folding timescale of ~50-100 My. By 100 My, 60-90 % of mantle Xe would have been degassed. The degassing history of the Earth recorded by $^{40}$Ar$^*$ (OZIMA, 1975; Turner 1989; ZHANG and ZINDLER, 1989) also supports very early degassing of the mantle. A more extensive discussion of the degassing history of the Earth is provided in Chapter 2 of this volume (Zhang 2012). It is worth noting that the history recorded by radiogenic isotopes could be largely decoupled from that of non-radiogenic isotopes in the atmosphere. For examples, most noble gases in air could be derived from a proto-atmosphere of trapped nebular gases with no relationship with the degassing history of the mantle.

## 6. The origin of major volatile elements in Earth

Laboratory analyses show a clear correlation between the water content in meteorites and the heliocentric distance of the parent bodies from which the meteorites are thought to come from (Fig. 15). Although there is the theoretical possibility that water-rich planetesimals formed in the hot regions of the disk by water-vapor absorption on silicate grains (Muralidharan et al., 2008; King et al., 2010), empirical evidence suggests that the planetesimals in the terrestrial planet region were extremely dry. In contrast, the Earth has a water content that, although small, is non-negligible and definitely larger than what the above-mentioned correlation would suggest for material condensed at 1 AU. In fact, the terrestrial water content is ~$7 \times 10^{-4}$ of Earth's mass (Table 1, Lecuyer et al., 1998). A larger quantity of water might have resided in the primitive Earth and might have been subsequently lost during core formation and impacts. Thus, the current Earth contains more water than enstatite chondrites and possibly the primitive Earth had water amounts comparable to, or larger than, ordinary chondrites. Where did this water come from, if the local planetesimals were dry? There are basically three possibilities for the origin of water on Earth, if water was not present in sufficient quantity in the local planetesimals.

One model is that of the *nebular origin*. Ikoma and Genda (2006) assumed that, at the end of the Earth formation process, there was still some nebular hydrogen in the proto-planetary disk. In this condition, the Earth attracted nebular gas by gravity, forming an hydrogen-rich atmosphere, up to a mass of $10^{21}$ kg. Then, atmospheric hydrogen was oxidized by some oxides such as FeO in the magma ocean to produce water. However, in this model, the initial D/H ratio of the water would be solar. Increasing the D/H ratio by a factor of 6 by the process of hydrodynamic escape discussed in Genda and Ikoma (2008) would require unrealistically long timescales, *i.e.* the hydrodynamic escape of the primitive atmosphere should have been protracted over billions of years, while evidence from $^{129}$I-$^{129}$Xe indicates that this must have occurred in the first ~100-150 Myr of the solar system history (Sect. 5.2). Such a model would also not explain the relatively oxidized nature of Earth's mantle (oxygen fugacity at the FMQ buffer, 3.5 log unit above the IW buffer where metallic Fe is in equilibrium with FeO).

A second possibility is that the water was brought to Earth by the bombardment of comets (Delsemme, 1992, 1999). A first problem with this model is that the D/H ratio in the water vapor released by long period comets from the Oort cloud is about twice that on Earth (Balsiger et al., 1995; Eberhardt et al., 1995; Meier et al., 1998; Bockelee-Morvan et al., 1998), and there are no known terrestrial processes that could *decrease* the D/H ratio of the original water on Earth. A similar constraint exists for the $^{15}$N/$^{14}$N ratio (Jehin et al. 2009). Recently, the D/H ratio of a Jupiter-family comet originating from the Kuiper belt (103P/Hartley 2) was measured and the value ($1.61 \pm 0.24 \times 10^{-4}$) was found to be indistinguishable from terrestrial composition (Hartogh et al. 2011). This result came as a surprise as Kuiper belt comets are thought to have formed in a more external region of the solar system than Oort-cloud comets and a gradient in D/H ratio with heliocentric distance is expected (Drouart et al. 1999). Oort cloud comets are presently farther away from the Sun than Kuiper belt comets but they are thought to have formed in more internal regions (*i.e.*, between Uranus and Saturn) and have been subsequently ejected to their present location (Fernandez and Ip, 1981). The discovery of a comet with a terrestrial D/H ratio reopens the possibility that such objects delivered water to the Earth. However, the $^{15}$N/$^{14}$N ratio of CN in comet 103P/Hartley 2 was also measured and the value is twice that of Earth's atmosphere (Meech et al. 2011; Marty 2012) and is similar to $^{15}$N/$^{14}$N ratios measured in other comets (Jehin et al. 2009). Furthermore, comets are probably rich in noble gases compared to Earth and delivery of major volatile elements by cold icy bodies would have completely obliterated the noble gas composition of the atmosphere (Dauphas 2003). Therefore, geochemical evidence suggests that comets did not deliver Earth's oceans but further work is needed to characterize the elemental and isotopic compositions of H, N, and noble gases in comets from diverse regions. A second problem with a cometary origin of Earth's oceans is that the collision probability of comets with the Earth is very small. Of the planetesimals scattered by the giant planets from the proto-planetary disk, only 1 in a million would strike our planet (Morbidelli et al., 2000), which is far from enough fort he following reasons. From studies on the range of radial migration that the giant planets should have suffered, due to their interaction with planetesimals after the disappearance of the disk of gas (Malhotra, 1995; Hahn and Malhotra, 1999; Gomes et al., 2004, 2005), it is expected that the total mass of the cometary disk was 35-50 Earth masses; moreover, from measurements of the ice/dust ratio in comets (Kuppers et al., 2005), it is now believed that less than half of the mass of a comet is in water-ice. Putting all these elements together, the mass of water delivered by comets to the Earth should have been $\sim 2.5 \times 10^{-5}$ Earth masses (neglecting impact losses), i.e. 10% of the crustal water. This corresponds to the upper-limit allowed by constraints on the D/H ratio on Oort-cloud comets (Dauphas et al. 2000).

The third possibility and the one that we favor is that the Earth accreted water from primitive planetesimals and/or planetary embryos originally from the outer asteroid belt (Morbidelli et al., 2000; Raymond et al., 2004, 2005, 2006, 2007; O'brien et al., 2006; Dauphas, 2003). The abundance pattern of major volatile elements as well as their isotopic compositions are very close to those observed in carbonaceous chondrites (Fig. 16). Major elements could have been delivered during the main stage of Earth's accretion by one or several partially hydrated embryos. An alternative possibility, however, is that H, C and N were delivered as part of a late veneer of volatile-rich asteroids, *i.e.* after formation of Earth's core. Evidence for a late veneer of the Earth-Moon system by extraterrestrial materials comes from measurements of highly siderophile elements in Earth's mantle, which abundances in Earth's mantle are too high to be explained by core-mantle equilibration and must have been supplied after Earth's core had formed (CHOU, 1978; JAGOUTZ et al., 1979; MORGAN, 1986). It is thus estimated that $0.7 \times 10^{22}$ to $2.7 \times 10^{22}$ kg of matter impacted the Earth after the completion of core formation. Using a CI chondrite composition of ~6 wt% $H_2O$, such a late veneer would have delivered $0.4 \times 10^{21}$ to $1.6 \times 10^{21}$ kg of water, which is comparable to the amount in the oceans of $\sim 1.4 \times 10^{21}$ kg. However, the highly siderophile abundance pattern of the mantle, is different from that of carbonaceous chondrites and all other chondrite groups (Fig. 17; BECKER et al., 2006; FISCHER-GÖDDE et al. 2010; WALKER, 2009), questioning the idea that the late veneer was chondritic or that the nature of the late veneer can be inferred from the pattern of highly siderophile elements in Earth's mantle. Furthermore, the fact that Earth's mantle plots on the correlation defined by chondrites for Mo and Ru suggests that the nature of the material accreted by the Earth did not change drastically before and after core formation (DAUPHAS et al., 2004a), so the late veneer was relatively dry (Fig. 17).

A strong argument that water was delivered to Earth as part of its main accretion stage rather than as part of a later veneer comes from examination of the $H_2O$-Xe budgets (DAUPHAS, 2003). Accretion of the late veneer probably proceeded by impacts that had little erosive capability on the atmosphere. Indeed, even the giant moon forming impact that is thought to have punctuated the main stage of accretion of the Earth could only remove 10-30 % of the atmosphere (Genda and Abe, 2002, 2005). Subsequent smaller impacts would have induced lower ground motions resulting in little atmospheric loss. The $H_2O$/Xe of Earth is much higher than that of chondrites, including water-rich types such as CI (Fig. 18). Thus, if one were to deliver all water on Earth by a late veneer, too much Xe would have been delivered to the atmosphere compared to the present inventory. In addition, meteoritic Xe does not have the appropriate stable isotopic composition (Earth's Xe is fractionated isotopically by ~40 ‰/amu relative to solar and chondritic compositions). A solution to this conundrum is that water was delivered to the Earth at a time when impacts could deposit enough energy to remove some of the gases accumulated in the protoatmosphere. Water, which reacted with rocks and formed oceans was preferentially retained (Genda and Abe 2005). Xenon, which is chemically inert, accumulated in the atmosphere and was lost by impacts. Some C and N present in Earth's protoatmosphere might have been lost at that time. Late accretion of cometary volatiles delivered noble gases depleted in Xe, thus explaining the missing Xe problem. To summarize, from a geochemical point of view, it is likely that Earth's major volatile inventory was acquired during the main stage of planetary growth by accretion of water-rich embryos and planetesimals, otherwise too much Xe with inappropriate isotopic composition would have been delivered.

Water was recently found in appreciable quantities in lunar rocks (Saal et al. 2008; Boyce et al. 2010; McCubbin et al. 2010; Greenwood et al. 2011; Hauri et al. 2011). This suggests that Earth accreted water before the Moon-forming impact. However, Greenwood et al. (2011) measured high water D/H ratios in lunar apatite crystals and argued that this was indicative of a cometary origin unrelated to Earth's oceans. Water-rich water glasses have D/H ratios that partially overlap with terrestrial values and the higher D/H ratios measured in some samples could reflect isotopic fractionation upon water loss (Hauri et al. 2010; Saal et al. 2011). The water content of the Moon can potentially provide important constraints on the timing and nature of water delivery to Earth but uncertainties in the D/H ratio of juvenile lunar water prevent a definitive conclusion to be reached.

Albarède (2009) recently argued based on volatile element isotopic ratios (*e.g.*, $^{66}Zn/^{64}Zn$) that the Earth must have accreted dry and that volatile elements were delivered late in Earth's accretion history. However, Zn isotopes are poor tracers of the accretion history of water on our planet. The silicate Earth contains $2.2 \times 10^{20}$ kg of Zn (55 ppm Zn, MCDONOUGH and SUN, 1995b) with a $\delta^{66}Zn$ value of ~+0.25 ‰. To deliver Earth's oceans, $2.3 \times 10^{22}$ kg of CI material would be needed, which would deliver $7.3 \times 10^{18}$ kg of Zn (CI contain ~312 ppm Zn) with a $\delta^{66}Zn$ value of ~+0.45 ‰. Adding this amount to the Earth would deliver only ~3 % of the Zn inventory and would change its isotopic composition by only ~0.006 ‰, which is unresolvable. On the contrary, adding this amount would deliver ~3 times the Xe inventory of the atmosphere and would completely obliterate its isotopic composition. This shows that Zn isotopes are insensitive to the presence of water-bearing planetary bodies in the mix of material that made the Earth. Xenon isotopes and $H_2O$/Xe ratios provide much tighter constraints on the origin of Earth's oceans. Mann et al. (2009) and Wood et al. (2010) found evidence that the accretion of moderately volatile elements to the Earth occurred mostly before core formation was complete. Wood et al. (2010) noted that the relative abundance of elements in our planet is correlated not only with condensation temperature but also with chemical affinity. Highly siderophile volatile elements are more depleted in the mantle than moderately siderophile elements or lithophile elements with the *same* condensation temperature. This implies that these volatile elements saw the formation of the Earth's core.

To address the question of the delivery of water by asteroids from a modeling point of view, we need to distinguish between the classical scenario, in which the outer belt is originally inhabited by primitive objects that are removed by mutual scattering and interactions with resonances with Jupiter, and the Grand Tack scenario. In the first case, as we have seen above, the amount of material accreted by the terrestrial planets from the asteroid belt depends on the eccentricity of the orbit of Jupiter. If Jupiter's orbit was almost circular, the terrestrial planets should have accreted 10-20 % of their mass from beyond 2.5 AU (O'brien et al., 2006), most of which should have been of carbonaceous chondritic nature. Thus, the terrestrial planets should have been originally very water-rich, possibly even as much as envisioned by Abe et al. (2000), and should have lost most of their water by impacts. However, the amount of material accreted from the outer asteroid belt drops with increasing eccentricity of Jupiter. If Jupiter had an original orbit with an eccentricity comparable to, or larger than the current one, no material would have been accreted from the outer asteroid belt, and the terrestrial planets would have been dry (O'brien et al., 2006; Raymond et al., 2009). In the Grand Tack scenario, the primitive asteroids have been implanted into the asteroid belt from in between the orbits of the giant planets and beyond. In the simulations carried out by Walsh et al. (2011), for every primitive planetesimal implanted in the outer asteroid belt, 10–30 planetesimals end up on orbits that cross the terrestrial planet forming region, for a total of $3–11 \times 10^{-2}$ Earth masses. O'Brien et al. (2010) showed that, in this situation, the Earth could accrete about 0.5-2 % of its mass from these objects, enough to supply a few times the current amount of water on our planet (assuming that the primitive planetesimals were 5-10 % water by mass). Walsh et al. and O'Brien et al. did not consider primitive planetary embryos in their simulations, so in principle the total amount of primitive material supplied to the Earth could be somewhat larger than the reported estimate.

A common feature of the classical and Grand Tack scenarios for the asteroidal delivery of water to the terrestrial planets is that the water is accreted *during* the formation of the planets and not in a *late veneer* fashion (*i.e.*, after core formation is complete), which is consistent with geochemical evidence (*i.e.*, Mo-Ru isotope anomalies and $H_2O$/Xe budgets). The accretion of water, though, is not uniform throughout the planet accretion history; instead it accelerates towards the end, as shown in Fig. 19. Although many issues have yet to be resolved, we are approaching a coherent and global view of the terretrial planet formation history indicating that water was acquired by our planet very early in its history.

## 7. The late heavy bombardment

The end of the terrestrial formation process was presumably characterized by a gradual decline of the bombardment rate, due to the decrease of the number of planetesimals still on planet-crossing orbits. Numerical simulations show that the number of planet-crossing planetesimals decays as exp[-t/T(t)], with T(t) initially equal to ~10 My and then increasing up to ~100 My (Morbidelli et al., 2000b; Bottke et al., 2007). The abundance of higly siderophile elements in the Earth's mantle constrains that at most 1 % of an Earth mass was accreted during this declining bombardment, after the formation of the Earth's core. Presumably, the last episode of core-mantle (partial) re-equilibration on the Earth occurred during the last giant impact suffered by our planet, most likely associated with the formation of the Moon. This event is dated to have occurred about 4.5 Gy ago (TOUBOUL et al., 2007). The Moon, however, shows evidence of a surge in the bombardment rate, approximately 4 Gy ago, *i.e.* ~500 My after its formation (TERA et al., 1974). In fact, about 10-15 basins, *i.e.* impact structures larger than 300 km in diameter, formed in an interval of time ranging at most from 4.1 to 3.8 Gy ago, Nectaris basin being the oldest and Imbrium and Orientale the most recent ones (Wilhelms et al., 1987). The formation of so many basins in such a short time-range cannot be due to a bombardment declining since the onset of terrestrial planet formation (even assuming T=100 Myr; Bottke et al., 2007). The surge of the bombardment rate 4 Gyr ago is called the "Late Heavy Bombardment" (LHB). Several lines of evidence suggest that the LHB concerned all the objects in the inner solar system (including asteroid Vesta) and possibly even the satelllites of the giant planets.

From a modeling point of view, the origin of the LHB is most likely related to a sudden change in the orbits of the giant planets. We have seen above that, at the disappearance of the gas-disk, the giant planets should have had orbits different from the current ones: the orbital separation among the planets, the eccentricities and inclinations were significantly smaller; the planets were probably in resonance with each other. If this is true, the orbits of the giant planets must have changed at some point of the history of the solar system, so to achieve the current configuration. The so-called "Nice model" is probably the one that gives the most comprehensive description of how this happened (see Morbidelli, 2011 for an extensive review). According to this model, at the disappearance of the gas-disk, the giant planets were surrounded by a massive disk of planetesimals. This trans-planetary disk was the surviving part of the original icy planetesimal disk that did not participate in the construction of the giant planet themselves and was not depleted by the migration of the giant planets during the gas-disk phase (Gomes et al., 2005). The gravitational interaction between the planets and this disk slowly modified the orbits of the former until eventually the planets became unstable (Tsiganis et al., 2005; Morbidelli et al., 2007; Batygin and Brown, 2010). The planets started to have close encounters with each other. Their orbits became more eccentric and inclined. In particular, Uranus and Neptune acquired very eccentric orbits that started to penetrate through the planetesimal disk. The planetesimals got scattered away by these planets. In response, the dynamical dispersal of the disk damped the eccentricities and inclinations of the orbits of Uranus and Neptune, bringing these planets to orbits very similar to the current ones (Fig. 20). By this process, the current orbits of the giant planets can be remarkably well reproduced (Tsiganis et al., 2005; Batygin and Brown, 2010). Moreover, if the inner edge of the trans-planetary disk was located a few AU beyond the orbit of the last planet, the trigger of the giant planet instability generically occurs after several hundreds of millions of years of quiet evolution, so that the dispersal of the disk could occur at the LHB time (Gomes et al., 2005; Levison et al., 2011). The Nice model explains not only the dynamical transition of the giant planets from their primordial orbits to the current orbits, but also the origin and orbital distribution of three populations of small objects in the outer solar system: Jupiter's Trojans (Morbidelli et al., 2005), the irregular satellites of the giant planets (Nesvorny et al., 2007; Bottke et al., 2009) and the Kuiper belt (Levison et al., 2008; Batygin et al., 2011).

In the framework of the Nice model, the LHB is caused by two populations of impactors. The first population is that of the icy planetesimals from the trans-planetary disk, that we can identify with "comets" given that this same disk generated also the Kuiper belt. When the disk was dispersed, most of the population was ultimately ejected onto hyperbolic orbit. About 1/3 of the population, though, passed temporarily through the inner solar system and about $10^{-6}$ of the objects entered in collisions with the Earth. Thus, given that the trans-planetary disk is estimated to have contained ~35 Earth masses in planetesimals, about $2 \times 10^{23}$ g of cometary material should have hit the Earth at the LHB time. This, however, should be regarded as an upper estimate. In fact, it is well known that comets can disrupt on their way into the inner solar system. For instance, for the current population of Jupiter family comets, it is estimated that the physical lifetime as active comets is about 1/10 of the dynamical lifetime (Levison and Duncan, 1997). Notice, though, that many "dead comets" just survive as inactive objects that can still have physical collisions with the planets. Such comets would have delivered little of Earth's water inventory but could have modified its noble gas, in particular supplying Ar and Kr as explained in sect. 4.4 (Dauphas 2003, Marty and Meibom 2007).

The second population of impactors is that of asteroids. Asteroids from the current main belt boundaries should have delivered to the Earth only $\sim 10^{22}$ g of material (Morbidelli et al., 2010). However, it is possible that before the LHB the asteroid belt extended further inwards, towards the orbit of Mars. The population situated between 1.8 AU and the current inner border of the asteroid belt (2.2. AU) would have been heavily decimated during the radial displacement of the orbits of Jupiter and Saturn and could have caused about 10 basins on the Moon and ~200 on the Earth (Bottke et al., 2011), delivering to our planet a total mass of about $2 \times 10^{23}$ g. The left-over of this population are the asteroids in the Hungaria region. As said previously, the Hungaria asteroids are mostly E-type, which are probably linked to enstatite chondrites. Thus, the dominant asteroidal contribution would have been of enstatite nature and would not have contributed to the water inventory of the Earth.

## 8. Conclusion: a not so rare Earth?

Geochemical and dynamical approaches offer complementary perspectives on the origin of volatile elements on Earth. Hydrogen, carbon, and nitrogen are present in proportions that correspond approximately to those found in volatile-rich meteorites. In addition, they have similar D/H, $^{13}C/^{12}C$, and $^{15}N/^{14}N$ ratios. Both observations strongly support the view that they were derived from the accretion of volatile-rich bodies akin to carbonaceous chondrites. Little is known about the composition of comets and further work is warranted before a definitive conclusion can be reached on their contribution to Earth's volatile elements. Still, available evidence suggests that comets are unlikely water sources for the Earth. A remarkable feature of the Earth is that its $H_2O$/Xe ratio is higher than all possible extraterrestrial sources. This is consistent with the idea that water was delivered during the main stage of planetary growth when erosive processes were still significant and inert Xe could be decoupled from reactive water. Dynamical modeling also supports this view that Earth's water was delivered by accretion of partially hydrated planetesimals and embryos during the main building stage of the Earth.

Overall, the circumstances that prevailed to the delivery of water to the Earth are probably shared in many other planet-forming stellar systems (Raymond et al., 2007). However, numerical simulations show that the delivery of water from the outer planetesimal disk to the forming terrestrial planets can be inhibited in some giant planet orbital configurations and evolutions, particularly those involving large orbital eccentricities (Raymond et al., 2004). With improvements in observations, detection of oceans on extrasolar planets might be feasible by looking for specular reflection of starlight on the planet's ocean or detection of water vapor in their atmospheres. Such, observations of remote planetary systems may inform us on the evolution of the young Earth, for which we have limited geological record. Closer to us, significant progress in our understanding of water delivery to terrestrial planets could be achieved by studying in more detail solar system objects like comets and Venus.

**Acknowledgements.** We thank R. Yokochi, I.N. Tolstikhin, F.J. Ciesla, B. Marty, K.J. Zahnle, B. Schmitt, E. Quirico, P. Beck, and U. Marboeuf for discussions. Insightful and thorough reviews by Chris Ballentine and John Chambers were greatly appreciated. I.N. Tolstikhin provided Fig. 8 and part of the caption that accompanies it. Part of this review was written during a sabbatical stay of N.D. at Institut de Planétologie et d'Astrophysique de Grenoble. Eric Quirico and Florence Lelong are thanked for their hospitality. This work was supported by a Packard fellowship and NASA through grant NNX09AG59G to N.D.

**Figure captions**

**Fig. 1.** Patterns of isotopic anomalies in bulk planetary materials (enstatite, carbonaceous, and ordinary chondrites). Note that all isotopic anomalies are normalized relative to the composition of the silicate Earth, which has a composition of 0 in this diagram (by convention). $\Delta^{17}O$ data are from Clayton et al. (1991), Clayton and Mayeda (1999, 1984); $\varepsilon^{50}Ti$ from Trinquier et al. (2009); $\varepsilon^{54}Cr$ from Trinquier et al. (2007), Qin et al. (2010); $\varepsilon^{92}Mo$ from Dauphas et al. (2002a, c), Burkhardt et al. (2011). Other elements show isotopic anomalies at a bulk scale (e.g., Ba and Ru) but the systematics of these elements is less developed. Only volatile-poor enstatite chondrites are a match to the terrestrial composition for all elements.

**Fig. 2.** An illustration of the process of runaway growth. Each panel represents a snapshot of the system at a different time. The coordinates represent the semi major axis a and the eccentricity e of orbits of the objects in a portion of the disk centered at 1 AU. The size of the dots is proportional to the physical radius of the objects. Initially, the system is made of a planetesimal population, in which two objects are 2 times more massive than the others. These objects accrete planetesimals very fast, increasing exponentially their mass ratio relative to the individual planetesimals. They become planetary embryos. Notice how the eccentricity of the planetary embryos remain low, while the eccentricities of the planetesimals are excited with time. The embryos also separate from each other as they grow. At the end, the embryos have grown by a factor 200, whereas the mean mass of the planetesimals has grown only by a factor of 2. From Kokubo and Ida (1998).

**Fig. 3.** The growth of terrestrial planets from a disk of planetary embryos. Each panel shows the semi-major axis and eccentricity of the bodies in the system at a given time, reported on top of the panel. Embryos and protoplanets are represented with red dots, whose size is proportional to the cubic root of their mass. Planetesimals are represented in green. The big blue dot represents Jupiter. A system of 3 terrestrial planets, the most massive of which has approximately an Earth mass, is eventually formed inside of 2 AU, whereas only a small fraction of the original planetesimal population survives within the asteroid belt boundaries (sketched with dashed curves). From O'Brien et al. (2006).

**Fig. 4.** The orbits of embryos (green full dots) and planetesimals (red dots) at the end of the inward-then-outward migration of Jupiter, when the gas is fully removed. The dash curve in the right bottom corner marks the inner boundary of the asteroid belt. From this state, the system evolves naturally in a timescale of a few $10^7$ y into two Earth-mass planets at ~0.7 and 1 AU and a small Mars at 1.5 AU. (see Fig. 4).

**Fig. 5.** The mass distribution of the synthetic terrestrial planets produced in the Walsh et al. (2010) simulations. The open symbols represent the planets produced in different runs starting from different initial conditions. The horizontal lines denote the perihelion-aphelion excursion of the planets on their eccentric final orbits. The black squares show the real planets of the solar system. The large mass ratio between the Earth and Mars is statistically reproduced.

**Fig. 6.** Noble gas concentrations (Table 2) and isotopic compositions (Table 3) in Mars, Earth, CI-chondrites, Comets, and Solar composition. The cometary concentrations were derived from noble gas trapping experiments in amorphous ice (calculated at 55 K). The composition of Venus' atmosphere is not plotted as it is very uncertain. The concentrations (left panel) are normalized to solar composition $Log(C/C_\odot)$. The isotopic compositions are given using the fractionation factor (‰/amu); $F_{Ne}= [(^{22}Ce/^{20}Ne)_{reservoir}/(^{22}Ce/^{20}Ne)_{Solar}-1]\times 1000/(22-20)$, $F_{Ar}= [(^{38}Ar/^{36}Ar)_{reservoir}/(^{38}Ar/^{36}Ar)_{Solar}-1]\times 1000/(38-36)$, $F_{Kr}= [(^{83}Kr/^{84}Kr)_{reservoir}/(^{83}Kr/^{84}Kr)_{Solar}-1]\times 1000/(83-84)$, $F_{Xe}= [(^{128}Xe/^{130}Xe)_{reservoir}/(^{128}Xe/^{130}Xe)_{Solar}-1]\times 1000/(128-130)$. Mars and Earth have different noble gas abundance patterns and isotopic compositions relative to possible progenitors.

**Fig. 7.** Hydrodynamic escape-preferential Xe retention model of the origin of Earth's noble gases (Pepin 1991, 1995). The isotopic compositions are given next to the data points in ‰/amu relative to solar. The initial atmosphere (empty circles) has solar isotopic composition (F=0 ‰/amu) and near solar elemental abundances except for a severe depletion in Ne. Hydrodynamic escape driven by one or several giant impact leads to noble gas removal with preferential depletion and isotopic fractionation of the lighter gases (red circles). The large isotopic fractionation is established at this stage. Mantle noble gases with solar isotopic compositions (dashed line) are degassed into the atmosphere where they are mixed with the fractionated noble gases remaining from the escape episode (mixture shown with green circles). During this stage, Xe is assumed to be retained in the mantle/core. Most Kr in Earth's atmosphere is from mantle degassing. In a final stage, Ne is partially lost to space and isotopically fractionated by hydrodynamic escape driven by EUV-radiation from the young Sun. Heavier noble gases are unaffected. The final modeled composition is shown in blue circles. A difficulty with this model is that no mantle reservoir has been documented yet that could host the missing Xe. Furthermore, this scenario cannot explain the measured Kr isotopic composition of $CO_2$ well gases (HOLLAND et al., 2009).

**Fig. 8.** Hydrodynamic escape-solubility controlled degassing model of the origin of Earth's noble gases (Tolstikhin & O'Nions, 1994). The isotopic compositions are given next to the data points in ‰/amu relative to solar. Proto-terrestrial material (thick black line) has Xe, Kr and Ar as in the South Oman E-chondrite (Pepin, 1991) and has solar-wind-like $^3He/^{22}Ne$ ratio. All these gases in these materials are isotopically non-fractionated (F=0 ‰/amu). Stage 1 of the degassing-dissipation process: Xe has been lost from the atmosphere, mass fractionated and set in this reservoir; its amount is slightly below the present day value as a small portion is still retained in the mantle and is degassed later. All lighter noble gases have been lost from the atmosphere quantitatively, but their amounts in the mantle exceed the atmospheric values. Stage 2: Kr has been degassed, partially lost from the atmosphere, slightly fractionated and set in this reservoir; only a small amount of Kr still resides in the mantle, whereas amounts of the lighter gases still exceed those in the atmosphere. Stage 3: Ar has been degassed, partially lost from the atmosphere, fractionated and set in this reservoir. Ne abundance in the mantle still exceeds the air abundance and Ne degasses, releases from the atmosphere and fractionates later on. Helium isotopes dissipate from the atmosphere continuously, with a mean residence time of ~1 Myr, so that He concentration is negligibly small in the present day atmosphere. Solar He abundance in non-dissipated atmosphere is shown in this Figure. This model cannot explain the heavy Kr isotopic composition of Earth's mantle relative to the atmosphere (HOLLAND et al., 2009).

**Fig. 9.** Hydrodynamic escape-cometary input model of the origin of Earth's noble gases (Dauphas 2003). The isotopic compositions are given next to the data points in ‰/amu relative to solar. The initial atmosphere (empty circles) has solar isotopic composition (F=0 ‰/amu) and solar elemental abundances. Following an episode of hydrodynamic escape possibly driven by EUV radiation from the T-Tauri sun, noble gases are depleted and isotopically fractionated (red circles). The isotopic composition of Xe is established at that stage. Comets deliver noble gases trapped in amorphous ice that display a depletion in Xe relative to Kr (dashed line; Bar-Nun and Owen). This late cometary bombardment can contribute most of the inventories of Ar and Kr without affecting much the isotopic composition of Ne and Xe (blue circles). One difficulty with this model is that it relies on the composition of hypothetical comets based on trapping experiments that are not fully understood.

**Fig. 10.** Histograms of $^3$He/$^4$He ratio (R/R$_A$) of arcs, MORBs, and intraplate magmatism. Most data are from a compilation published in 2006 (http://pubs.usgs.gov/ds/2006/202/) updated with data published since then. MORBs have a narrow distribution of R/R$_A$ values that peaks at a mode of 7.3. The mean of 8.9 is higher, which reflects the fact that the distribution has a tail towards high R/R$_A$ values corresponding to an enriched source. Intraplate magmatism includes plume sources that are characterized by elevated $^3$He/$^4$He ratios (i.e., less degassed than the MORB source).

**Fig. 11.** Neon isotope systematics of Earth's mantle. **A**. Mantle Ne is prone to contamination at Earth's surface by air. This produces linear mixing arrays in $^{20}$Ne/$^{22}$Ne vs $^{21}$Ne/$^{22}$Ne space between air and mantle end-members. The different mixing lines point to different end-members with elevated $^{20}$Ne/$^{22}$Ne (i.e., towards solar composition) but characterized by various contributions of nucleogenic $^{21}$Ne*. The more degassed samples (MORBs) have higher U/Ne ratios and therefore higher $^{21}$Ne/$^{22}$Ne ratios compared to plume-related samples such as those from Loihi,

Hawaii. The highest $^{20}$Ne/$^{22}$Ne ratios in plumes are found in Kola peninsula and approach the solar ratio. Other samples have lower $^{20}$Ne/$^{22}$Ne ratios and it is uncertain at present whether this reflects the presence of a non-solar end-member in the mantle ($^{20}$Ne/$^{22}$Ne ~12.5) or if this limit reflects residual air contamination. Data from http://pubs.usgs.gov/ds/2006/202/ for most MORBs (light green circles), Moreira et al. (1998) for popping-rocks (large dark green triangles in the MORB field), Yokochi and Marty (2004) for Kola (large orange circles), Trieloff et al. (2000) for Hawaii and Iceland (large blue squares). MFL stands for mass fractionation line. **B**. Attempt to derive the Ne isotopic composition of the mantle (see Ballentine et al. 2005; Holland and Ballentine 2006 for details). $CO_2$ well gas samples define a mixing line between a mantle end-member and a mixture between crust and air. The depleted mantle end-member can be inferred by taking the intersection between the MORB line and that derived by CO2 well gases. Using a more refined scheme, Holland and Ballentine (2006) obtained $^{20}$Ne/$^{22}$Ne and $^{21}$Ne/$^{22}$Ne ratios of 12.49±0.04 and 0.0578±0.0003, respectively, for the depleted mantle.

**Fig. 12.** Radiogenic $^{40}$Ar in the mantle. The $^{20}$Ne/$^{22}$Ne ratio is a proxy of air contamination as there is a contrast between the Ne isotopic composition of air (~9.8) and that of the atmosphere (12.5-13.7). In this diagram, popping rocks and plume-related magmas (i.e., Loihi seamount, Kola peninsula, and Iceland) define correlations that point to different $^{40}$Ar/$^{36}$Ar ratios end-members between the depleted upper-mantle (25,000-44,000) and the plume-source (~4,500), corresponding to different extents of degassing. Data sources: Popping rocks (Moreira et al. 1998), Loihi, Hawaii (Valbracht et al. 1997; Trieloff et al. 2000); Iceland (Trieloff et al. 2000); Kola (Marty et al. 1998; Yokochi and Marty 2004, 2005).

**Fig. 13.** Kr-Xe non-radiogenic isotopic compositions of $CO_2$ well-gases (data from Holland et al. 2009). The samples define a correlation corresponding to mixing between air and the mantle source. The black arrow shows the direction expected for mass fractionation. As shown, air cannot be derived from mantle gases by simple mass fractionation. Instead, the data can be explained if the atmosphere received a late addition of cometary materials that modified its heavy noble gas isotopic composition (Dauphas, 2003). Holland et al. (2009) argued that the mantle member of the mixing derives from a trapped meteoritic component. This end-member could also be nebular gases of a transient atmosphere fractionated by hydrodynamic escape while the Earth had not yet reached its full size.

**Fig. 14.** Radiogenic Xe isotope systematics in the mantle. **A, B**. Determination of the $^{129}$Xe/$^{130}$Xe ratio of the source of MORBs (Moreira et al. 1998; Holland and Ballentine 2006). Popping rocks define a correlation corresponding to mixing between air and the mantle (A, Moreira et al. 1998). At a $^{20}$Ne/$^{22}$Ne ratio of 12.5-13.7, the $^{129}$Xe/$^{130}$Xe is estimated to be 7.6-8.2. $CO_2$-well gases define a correlation corresponding to mixing between mantle and air+crustal gases (B, Holland and Ballentine 2006). It intersects the MORB correlation at a $^{129}$Xe/$^{130}$Xe ratio of 7.90±0.14. **C, D.** Decomposition of $^{136}$Xe into $^{238}$U and $^{244}$Pu-fission derived components (Kunz et al. 1998; Yokochi and Marty 2005). $^{238}$U-decay and $^{244}$Pu decay produce $^{136}$Xe and $^4$He in different ratios. Yokochi and Marty (2005) used the $^{21}$Ne*/$^4$He* ratio to correct for $^{136}$Xe/$^4$He during degassing and concluded that 33-60 % of total fission $^{136}$Xe was from $^{244}$Pu-decay (C). $^{244}$Pu and $^{238}$U-fission produce Xe isotopes with different spectra, which Kunz et al. (1998) used to conclude that 32±10 % of total fission $^{136}$Xe was from $^{244}$Pu-decay (D).

**Fig. 15.** CI and CM meteorites are the most rich in water; water amounts to about 5 to 10% of their total mass (Robert and Epstein, 1982; Kerridge, 1985). They are expected to come from C-type asteroids, predominantly in the asteroid belt and possibly accreted even further out (Walsh et al., 2010). Water in ordinary chondrites amounts for only 0.1% of the total weight (Robert, 1977; Robert et al., 1979; McNaughton et al. 1981), or a few times as much (Jaresewich 1966); they are spectroscopically linked to S-type asteroids, predominant between 2 and 2.5 AU. Finally, enstatite chondrites are very dry, with only 0.01% of their total mass in water (ref.); they are expected to come from E-type asteroids, which dominate the Hungaria region in the very inner asteroid belt at 1.8 AU.

**Fig. 16.** Major volatile element concentrations (Table 2) and isotopic compositions (Table 3) in Earth, CI-chondrites, Comets, and Solar composition. The concentrations (left panel) are normalized to solar composition $\text{Log}(C/C_\odot)$. The isotopic compositions are given in δ-notation (‰) ; $\delta D=[(D/H)_{reservoir}/(D/H)_{VSMOW}-1]\times 1000$, $\delta^{13}C=[(^{13}C/^{12}C)_{reservoir}/(^{13}C/^{12}C)_{VPDB}-1]\times 1000$, and $\delta^{15}N=[(^{15}N/^{14}N)_{reservoir}/(^{15}N/^{14}N)_{VPDB}-1]\times 1000$, where VSMOW (Vienne Standard Mean Ocean Water) and VPDB (Vienna Pee Dee belemnite) are two terrestrial reference materials stored in Vienna. Although at a lower concentration, Earth has the same abundance pattern as CI-chondrites. The isotopic compositions of the major volatile elements on Earth are also very close to CI-chondrites.

**Fig. 17.** Constraints from moderately and highly siderophile elements on the nature of the late veneer. **A.** Highly siderophile element pattern of the terrestrial primitive mantle, EH, EL, O, and CV chondrites (Becker et al. 2006, Fischer-Gödde et al. 2011). The primitive mantle has approximately chondritic composition except for large excess in Ru. This excess could be due to accretion of a late veneer with non-chondritic composition (Puchtel et al. 2008) or mixing of a chondritic late-veneer with fractionated highly siderophile elements remaining in the mantle after core formation (Dauphas et al. 2002b). Walker (2009) recently provided a very detailed review of that question. **B.** Correlation between Mo and Ru isotopic anomalies in meteorites (Dauphas et al. 2004; Chen et al. 2010; Burkhardt et al. 2011). The dashed line corresponds to mixing between terrestrial and s-process end-members. Molybdenum is a moderately siderophile element and most of its inventory in Earth's mantle was delivered during the main stage of Earth's accretion. Ruthenium is a highly siderophile element and most of its inventory win Earth's mantle was delivered after completion of core formation, as part of the late veneer. The fact that Earth's mantle plots on the correlation defined by meteorites suggests that the nature of the material accreted by the Earth did not change drastically before and after core formation.

**Fig. 18.** $H_2O$/Xe constraints on the origin of Earth's atmosphere (data from Tables 1 and 2). Earth and presumably Mars have $H_2O$/Xe ratios that are much higher than those of chondrites and comets. If such objects had delivered Earth's oceans as part of a late veneer, they would have delivered too much Xe with the wrong isotopic composition compared to what is in the atmosphere. Most likely, water was delivered during the main stage of terrestrial growth when Xe could be lost from the protoatmosphere by impact erosion (and hydrodynamic escape) while reactive water was more efficiently retained.

**Fig. 19.** The fraction of the total mass of a terrestrial planet as a function of the said planet 's total mass, from a simulation of O'Brien et al. (2010). In this case, 50 % of the water is accreted when the planet is at least 90 % of it's final mass. Assuming that the "wet material" is 5 % water by mass, consistent with CM carbonaceous chondrites, this planet would have at the end 4 times the amount of crustal water on Earth.

**Fig. 20.** The evolution of the giant planets in the Nice model. Here, each planet is represented by a pair of curves of distinctive color (Jupiter in blue, Saturn in red, Uranus in green and Neptune in magenta). The inner curve marks the perihelion distance and the outercuve the aphelion distance, as a function of time. Thus, when the two curves are very close, the planet's orbit is almost circular. When the two curves separate from each other, the eccentricity is large. In this simulaiton, the instability of the giant planets occurs at 762 My. From Levison et al. (2011)

**References**

Abe, Y., Ohtani, E., Okuchi, T., Righter, K., Drake, M. (2000). Water in the Early Earth. Origin of the Earth and Moon 413-433.

Adachi, I., Hayashi, C., and Nakazawa, K., 1976. Gas Drag Effect on Elliptic Motion of a Solid Body in Primordial Solar Nebula. *Progress of Theoretical Physics* **56**, 1756-1771.

Agnor, C.B., Canup, R.M., Levison, H.F. (1999). On the Character and Consequences of Large Impacts in the Late Stage of Terrestrial Planet Formation. Icarus 142, 219-237.

Albarede, F., 2009. Volatile accretion history of the terrestrial planets and dynamic implications. *Nature* **461**, 1227-1233.

Alexander , R. (2008) From discs to planetesimals: evolution of gas and dust discs. New. Astron. Rev. 52, 60-77.

Allegre, C.J., Manhes, G., Gopel, C. (1995). The age of the Earth. Geochimica et Cosmochimica Acta 59, 2445-1456.

Allegre, C. J., Hofmann, A., and O'Nions, K., 1996. The argon constraints on mantle structure. *Geophys. Res. Lett.* **23**, 3555-3557.

Anderson, D.L. (1998) The helium paradoxes. Proceedings of the National Academy of Sciences 95, 4822-4827.

Ballentine, C.J. and Burnard P.G. 2002. Production, release and transport of noble gases in the continental crust. Reviews in Mineralogy and Geochemistry 47, 481-538.

Ballentine, C. J. and Holland, G., 2008. What CO2 well gases tell us about the origin of noble gases in the mantle and their relationship to the atmosphere. *Philosophical Transactions of the Royal Society A: Mathematical, Physical and Engineering Sciences* **366**, 4183-4203.

Ballentine, C. J., Marty, B., Sherwood Lollar, B., and Cassidy, M., 2005. Neon isotopes constrain convection and volatile origin in the Earth's mantle. *Nature* **433**, 33-38.

Ballester, G. E., Sing, D. K., and Herbert, F., 2007. The signature of hot hydrogen in the atmosphere of the extrasolar planet HD 209458b. *Nature* **445**, 511-514.

Balsiger H., Altwegg K. and Geiss J. (1995). D/H and $^{18}O/^{16}O$ ratio in hydronium ion and in neutral water from in situ ion measurements in Comet P/Halley. J. Geophys. Res., 100, 5834-5840.

Balsiger, H., Altwegg, K., Bochsler, P., Eberhardt, P., Fischer, J., Graf, S., Jäckel, A., Kopp, E., Langer, U., Mildner, M., Müller, J., Riesen, T., Rubin, M., Scherer, S., Wurz, P., Wüthrich, S., Arijs, E., Delanoye, S., Keyser, J., Neefs, E., Nevejans, D., Rème, H., Aoustin, C., Mazelle, C., Médale, J. L., Sauvaud, J., Berthelier, J. J., Bertaux, J. L., Duvet, L., Illiano, J. M., Fuselier, S., Ghielmetti, A., Magoncelli, T., Shelley, E., Korth, A., Heerlein, K., Lauche, H., Livi, S., Loose, A., Mall, U., Wilken, B., Gliem, F., Fiethe, B., Gombosi, T., Block, B., Carignan, G., Fisk, L., Waite, J., Young, D., and Wollnik, H., 2007. Rosina – Rosetta Orbiter Spectrometer for Ion and Neutral Analysis. *Space Science Reviews* **128**, 745-801.

Bar-Nun, A., Kleinfeld, I., and Kochavi, E., 1988. Trapping of gas mixtures by amorphous water ice. *Physical Review B* **38**, 7749.

Bar-Nun, A. and Owen, T., 1998. Trapping of gases in water ice and consequences to comets and the atmospheres of the inner planets. In: Schmitt, B. B., De Bergh, C., and Festou, M. Eds.), *Solar System Ices*. Kluwer, Dordrecht.

Batygin, K., Brown, M.E. (2010). Early Dynamical Evolution of the Solar System: Pinning Down the Initial Conditions of the Nice Model. The Astrophysical Journal 716, 1323-1331.

Batygin, K., Brown, M.E., Fraser, W.C. (2011). In-situ Formation of the Cold Classical Kuiper Belt. ApJ, submitted.

Becker, H., Horan, M. F., Walker, R. J., Gao, S., Lorand, J. P., and Rudnick, R. L., 2006. Highly siderophile element composition of the Earth's primitive upper mantle: Constraints from new data on peridotite massifs and xenoliths. *Geochimica et Cosmochimica Acta* **70**, 4528-4550.

Bianchi, D., Sarmiento, J. L., Gnanadesikan, A., Key, R. M., Schlosser, P., and Newton, R., 2010. Low helium flux from the mantle inferred from simulations of oceanic helium isotope data. *Earth and Planetary Science Letters* **297**, 379-386.

Binzel, R.P., Bus, S.J., Burbine, T.H., Sunshine, J.M. (1996). Spectral Properties of Near-Earth Asteroids: Evidence for Sources of Ordinary Chondrite Meteorites. Science 273, 946-948.

Bizzarro, M. and et al., 2005. Rapid Timescales for Accretion and Melting of Differentiated Planetesimals Inferred from 26Al-26Mg Chronometry. *The Astrophysical Journal Letters* **632**, L41.

Blum, J. and Wurm, G., 2008. The growth mechanisms of macroscopic bodies in protoplanetary disks. *Annual Review of Astronomy and Astrophysics* **46**, 21-56.

Bockelee-Morvan D., Gautier D., Lis D.C., Young K., Keene J., Phillips T., Owen T., Crovisier J., Goldsmith P.F., Bergin E.A., Despois D. and Wooten A. (1988). Deuterated Water in Comet C/1996 B2 (Hyakutake) and its Implications for the Origin of Comets. Icarus, 193, 147-162.

Bockelee-Morvan, D., Biver, N., Jehin, E., Cochran, A. L., Wiesemeyer, H., Manfroid, J., Hutsemekers, D., Arpigny, C., Boissier, J., Cochran, W., Colom, P., Crovisier, J., Milutinovic, N., Moreno, R., Prochaska, J. X., Ramirez, I., Schulz, R., and Zucconi, J. M., 2008. Large excess of heavy nitrogen in both hydrogen cyanide and cyanogen from comet 17P/Holmes. *Astrophysical Journal Letters* **679**, L49-L52.

Bockelee-Morvan, D., Crovisier, J., Mumma, M. J., and Weaver, H. A., 2004. The composition of cometary volatiles. In: Festou, M. C., Keller, H. U., and Weaver, H. A. Eds.), *Comets II*. University of Arizona Press, Tucson.

Bogard, D. D., Clayton, R. N., Marti, K., Owen, T., and Turner, G., 2001. Martian volatiles: isotopic composition, origin, and evolution. *Space Science Reviews* **96**, 425-458.

Bottke, W.F., Levison, H.F., Nesvorny, D., Dones, L. (2007). Can planetesimals left over from terrestrial planet formation produce the lunar Late Heavy Bombardment?. Icarus 190, 203-223.

Bottke, W.F., Nesvorny, D., Vokrouhlicky, D., Morbidelli, A. (2010). The Irregular Satellites: The Most Collisionally Evolved Populations in the Solar System. The Astronomical Journal 139, 994-1014.

Bottke, W.F., Vokrouhlicky, D., Minton, D., Nesvorny, D., Morbidelli, A., Brasser, R., Simonson, B. (2011). The Great Archean Bombardment, or the Late Heavy Bombardment. *42nd Lunar and Planetary Science Conference, Abstract #2591*

Boyce J.W., Liu Y., Rossman G.R., Guan Y., Eiler J.M. Stolper E.M., Taylor L.A. (2010) Lunar apatite with terrestrial volatile abundances. Nature 466, 466-469.

Boyd, S. R., 2001. Nitrogen in future biosphere studies. *Chemical Geology* **176**, 1-30.

Boynton, W. V., Ming, D. W., Kounaves, S. P., Young, S. M. M., Arvidson, R. E., Hecht, M. H., Hoffman, J., Niles, P. B., Hamara, D. K., Quinn, R. C., Smith, P. H., Sutter, B., Catling, D. C., and Morris, R. V., 2009. Evidence for Calcium Carbonate at the Mars Phoenix Landing Site. *Science* **325**, 61-64.

Bridges, J. C., Catling, D. C., Saxton, J. M., Swindle, T. D., Lyon, I. C., and Grady, M. M., 2001. Alteration assemblages in martian meteorites: Implications for near-surface processes. *Space Science Reviews* **96**, 365-392.

Brock, D. S. and Schrobilgen, G. J., 2010. Synthesis of the missing oxide of xenon, XeO2, and its implications for Earth's missing xenon. *Journal of the American Chemical Society* **dx.doi.org/10.1021/ja110618g**.

Bromley, B. C. and Kenyon, S. J., 2006. A hybrid N-body-coagulation code for planet formation. *Astronomical Journal* **131**, 2737-2748.

Burbine, T.H., Binzel, R.P., Bus, S.J., Buchanan, P.C., Hinrichs, J.L., Hiroi, T., Meibom, A., Sunshine, J.M. (2000). Forging Asteroid-Meteorite Relationships Through Reflectance Spectroscopy. Lunar and Planetary Institute Science Conference Abstracts 31, 1844.

Burkhardt, C., Kleine, T., Oberli, F., Pack, A., Bourdon, B., Wieler, R., 2011. Nucleosynthetic Mo isotopic anomalies in planetary materials as tracers of circumstellar disk processes. Lunar and Planetary Science Conference 42, #2554.

Burnard, P., Graham, D., and Turner, G., 1997. Vesicle-Specific Noble Gas Analyses of "Popping Rock": Implications for Primordial Noble Gases in Earth. *Science* **276**, 568-571.

Busemann, H. and Eugster, O., 2002. The trapped noble gas component in achondrites. *Meteoritics & Planetary Science* **37**, 1865-1891.

Caffee, M. W., Hudson, G. B., Velsko, C., Huss, G. R., Alexander, E. C., and Chivas, A. R., 1999. Primordial Noble Gases from Earth's Mantle: Identification of a Primitive Volatile Component. *Science* **285**, 2115-2118.

Campins, H., Hargrove, K., Pinilla-Alonso, N., Howell, E.S., Kelley, M.S., Licandro, J., Mothe-Diniz, T., Fernandez, Y., Ziffer, J. (2010). Water ice and organics on the surface of the asteroid 24 Themis. Nature 464, 1320-1321.

Canup, R.M., Esposito, L.W. (1996). Accretion of the Moon from an Impact-Generated Disk. Icarus 119, 427-446.

Canup, R.M., Asphaug, E. (2001). Origin of the Moon in a giant impact near the end of the Earth's formation. Nature 412, 708-712.

Canup, R.M., Ward, W.R. (2002). Formation of the Galilean Satellites: Conditions of Accretion. The Astronomical Journal 124, 3404-3423.

Cartigny, P., Boyd, S. R., Harris, J. W., and Javoy, M., 1997. Nitrogen isotopes in peridotitic diamonds from Fuxian, China: the mantle signature. *Terra Nova* **9**, 175-179.

Cartigny P., Harris J.W., Javoy M. (2001) Diamond genesis, mantle fractionations and mantle nitrogen content: a study of d13C-N concentrations in diamonds. Earth and Planetary Science Letters 185, 85-98.

Cartigny, P., Pineau, F., Aubaud, C., and Javoy, M., 2008. Towards a consistent mantle carbon flux estimate: Insignts from volatile systematics (H2O/Ce, dD, CO2/Nb) in the North Atlantic mantle (14° N and 34° N) *Earth and Planetary Science Letters* **265**, 672-685.

Cassen, P., 2001. Nebular thermal evolution and thermal properties of primitive planetary materials. *Meteoritics & Planetary Science* **36**, 671-700.

Cates, N. L. and Mojzsis, S. J., 2007. Pre-3750†Ma supracrustal rocks from the Nuvvuagittuq supracrustal belt, northern QuÈbec. *Earth and Planetary Science Letters* **255**, 9-21.

Chambers, J.E., Wetherill, G.W. (1998). Making the Terrestrial Planets: N-Body Integrations of Planetary Embryos in Three Dimensions. Icarus 136, 304-327.

Chambers, J.E., Wetherill, G.W. (2001). Planets in the asteroid belt. Meteoritics and Planetary Science 36, 381-399.

Chambers, J.E. (2010). Planetesimal formation by turbulent concentration. Icarus 208, 505-517.

Chen, J.H., Papanastassiou, D.A., Wasserburg, G.J. (2010) Ruthenium endemic isotope effects in chondrites and differentiated meteorites. Geochimica et Cosmochimica Acta 74, 3851-3862.

Chou, C.-L., 1978. Fractionation of siderophile elements in the Earth's upper mantle. *Proceedings of the Lunar and Planetary Science Conference* **9**, 219-230.

Christensen, P. R., 2006. Water at the poles and in permafrost regions of Mars. *Elements* **2**, 151-155.

Chyba, C. F., 1991. Terrestrial mantle siderophiles and the lunar impact record. *Icarus* **92**, 217-233.

Clayton, R.N., Mayeda, T.K. (1984) Oxygen isotopic compositions of enstatite chondrites and aubrites. Journal of Geophysical Research (Supplement), C245-C249.

Clayton, R.N., Mayeda, T.K., Goswami, J.N., Olsen, E.J. (1991) Oxygen isotope studies of ordinary chondrites. Geochimica et Cosmochimica Acta 55, 2317-2337.

Clayton, R.N., Mayeda, T.K. (1999) Oxygen isotope studies of carbonaceous chondrites. Geochimica et Cosmochimica Acta 63, 2089-2104.

Clayton, R. N., 1993. Oxygen isotopes in meteorites. *Annual Review of Earth and Planetary Sciences* **21**, 115-149.

Coltice, N., Marty, B., and Yokochi, R., 2009. Xenon isotope constraints on the thermal evolution of the early Earth. *Chemical Geology* **266**, 4-9.

Craig, H., Clarke, W. B., and Beg, M. A., 1975. Excess 3He in deep water on the East Pacific Rise. *Earth and Planetary Science Letters* **26**, 125-132.

Craig, H. and Lupton, J. E., 1976. Primordial neon, helium, and hydrogen in oceanic basalts. *Earth and Planetary Science Letters* **31**, 369-385.

Criss R.E., Farquhar J. 2008. Abundance, notation, and fractionation of light stable isotopes. Reviews in Mineralogy & Geochemistry 68, 15-30.

Cuzzi, J.N., Hogan, R.C., Shariff, K. (2008). Toward Planetesimals: Dense Chondrule Clumps in the Protoplanetary Nebula. The Astrophysical Journal 687, 1432-1447.

Cuzzi, J.N., Hogan, R.C., Bottke, W.F. (2010). Towards initial mass functions for asteroids and Kuiper Belt Objects. Icarus 208, 518-538.

Dauphas, N., Marty, B. 1999. Heavy nitrogen in carbonatites of the Kola Peninsula: A possible signature of the deep mantle. Science 286, 2488-2490.

Dauphas, N., 2003. The dual origin of the terrestrial atmosphere. *Icarus* **165**, 326-339.

Dauphas, N., Cates, N. L., Mojzsis, S. J., and Busigny, V., 2007. Identification of chemical sedimentary protoliths using iron isotopes in the > 3750 Ma Nuvvuagittuq supracrustal belt, Canada. *Earth and Planetary Science Letters* **254**, 358-376.

Dauphas, N. and Chaussidon, M., 2011. A perspective from extinct radionuclides on a young stellar object: The Sun and its accretion disk. *Annual Review of Earth and Planetary Sciences* **39**.

Dauphas, N., Davis, A. M., Marty, B., and Reisberg, L., 2004a. The cosmic molybdenum-ruthenium isotope correlation. *Earth and Planetary Science Letters* **226**, 465-475.

Dauphas, N. and Marty, B., 2002. Inference on the nature and the mass of Earth's late veneer from noble metals and gases. *J. Geophys. Res.* **107**, 5129.

Dauphas, N., Marty, B., and Reisberg, L., 2002a. Inference on terrestrial genesis from molybdenum isotope systematics. *Geophysical Research Letters* **29**, -.

Dauphas N., Reisberg L., Marty B. (2002b) An alternative explanation for the distribution of highly siderophile elements in the Earth. Geochemical Journal 36, 409-419.

Dauphas N., Marty B., Reisberg L., 2002c. Molybdenum evidence for inherited planetary scale isotope heterogeneity of the protosolar nebula. The Astrophysical Journal 565, 640-644.

Dauphas, N. and Pourmand, A., 2011. Hf-W-Th evidence for rapid growth of Mars and clues on the birth of an oligarch. *Nature* **in press**.

Dauphas, N., Robert, F., and Marty, B., 2000. The Late Asteroidal and Cometary Bombardment of Earth as Recorded in Water Deuterium to Protium Ratio. *Icarus* **148**, 508-512.

Dauphas, N., van Zuilen, M., Wadhwa, M., Davis, A. M., Marty, B., and Janney, P. E., 2004b. Clues from Fe isotope variations on the origin of early Archean BIFs from Greenland. *Science* **306**, 2077-2080.

de Bergh, C., Moroz, V. I., Taylor, F. W., Crisp, D., BÈzard, B., and Zasova, L. V., 2006. The composition of the atmosphere of Venus below 100†km altitude: An overview. *Planetary and Space Science* **54**, 1389-1397.

Deines P. (1980) The carbon isotopic composition of diamonds: relationship to diamond shape, color, occurrence and vapor composition. Geochimica et Cosmochimica Acta 44, 943-961.

Delsemme, A. H., 1988. The chemistry of comets. *Philosophical Transactions of the Royal Society, Series A* **325**, 509-523.

Delsemme, A.H. (1992). Cometary origin of carbon and water on the terrestrial planets. Advances in Space Research 12, 5-12.

Delsemme, A.H. (1999). The deuterium enrichment observed in recent comets is consistent with the cometary origin of seawater. Planetary and Space Science 47, 125-131.

DÈruelle, B., Dreibus, G., and Jambon, A., 1992. Iodine abundances in oceanic basalts: implications for Earth dynamics. *Earth and Planetary Science Letters* **108**, 217-227.

Donahue, T. M. and Russell, C. T., 1997. The Venus atmosphere and ionosphere and their interaction with the solar wind: an overview. In: Bouger, S., Hunter, D. M., and Phillips, R. J. Eds.), *Venus II*. University of Arizona Press.

Graham, D.W. (2002) Noble gas isotope geochemistry of Mid-Ocean Ridge and Ocean Island Basalts: characterization of mantle source reservoirs. Reviews in Mineralogy and Geochemistry 47, 247-317.

Eberhardt P., Reber M., Krankowski D. and Hodges R.R. (1995). The D/H and $^{18}O/^{16}O$ ratios in water from Comet P/Halley. Astron. Astrophys., 302, 301—316.

Ehlmann, B. L., Mustard, J. F., Murchie, S. L., Poulet, F., Bishop, J. L., Brown, A. J., Calvin, W. M., Clark, R. N., Marais, D. J. D., Milliken, R. E., Roach, L. H., Roush, T. L., Swayze, G. A., and Wray, J. J., 2008. Orbital Identification of Carbonate-Bearing Rocks on Mars. *Science* **322**, 1828-1832.

Eugster, O., Eberhardt, P., and Geiss, J., 1967. The isotopic composition of krypton in unequilibrated and gas rich chondrites. *Earth and Planetary Science Letters* **2**, 385-393.

Farley K.A., Craig H. 1994. Atmospheric argon contamination of ocean island basalt olivine phenocrysts. Geochimica et Cosmochimica Acta 58. 2509-2517.

Farley, K. A. and Neroda, E., 1998. NOBLE GASES IN THE EARTH'S MANTLE. *Annual Review of Earth and Planetary Sciences* **26**, 189-218.

Fernandez J.A., Ip W.-H. (1981) Dynamical evolution of a cometary swarm in the outer planetary region. Icarus 47, 470-479.

Fischer-Gˆdde, M., Becker, H., and Wombacher, F., Rhodium, gold and other highly siderophile elements in orogenic peridotites and peridotite xenoliths. *Chemical Geology* **280**, 365-383.

Fornasier, S., Migliorini, A., Dotto, E., Barucci, M.A. (2008). Visible and near infrared spectroscopic investigation of E-type asteroids, including 2867 Steins, a target of the Rosetta mission. Icarus 196, 119-134.

Garrison, D. H. and Bogard, D. D., 1998. Isotopic composition of trapped and cosmogenic noble gases in several martian meteorites. *Meteoritics & Planetary Science* **33**, 721-736.

Genda H., Abe Y. (2003) Survival of a proto-atmosphere through the stage of giant impacts: the mechanical aspects. Icarus 164, 149-162.

Genda H., Abe Y. (2005) Enhanced atmospheric loss on protoplanets at the giant impact phase in the presence of oceans. Nature 433, 842-844.

Genda, H., Ikoma, M. (2008). Origin of the ocean on the Earth: Early evolution of water D/H in a hydrogen-rich atmosphere. Icarus 194, 42-52.

Gomes, R.S., Morbidelli, A., Levison, H.F. (2004). Planetary migration in a planetesimal disk: why did Neptune stop at 30 AU?. Icarus 170, 492-507.

Gomes, R., Levison, H.F., Tsiganis, K., Morbidelli, A. (2005). Origin of the cataclysmic Late Heavy Bombardment period of the terrestrial planets. Nature 435, 466-469.

Gounelle, M., Morbidelli, A., Bland, P.A., Spurny, P., Young, E.D., Sephton, M. 2008. Meteorites from the Outer Solar System?. The Solar System Beyond Neptune 525-541.

Gradie, J., Tedesco, E. (1982). Compositional structure of the asteroid belt. Science 216, 1405-1407.

Greenberg, R., Hartmann, W.K., Chapman, C.R., Wacker, J.F. (1978). Planetesimals to planets - Numerical simulation of collisional evolution. Icarus 35, 1-26.

Greenwood J.P., Itoh S., Sakamoto N., Warren P., Taylor L., Yurimoto H. (2011) Hydrogen isotope ratios in lunar rocks indicate delivery of cometary water to the moon. Nature Geoscience 4, 79-82.

Greenzweig, Y., Lissauer, J.J. (1992). Accretion rates of protoplanets. II - Gaussian distributions of planetesimal velocities. Icarus 100, 440-463.

Grossman, L., 1972. Condensation in Primitive Solar Nebula. *Geochimica et Cosmochimica Acta* **36**, 597-&.

Grossman, L. and Larimer, J. W., 1974. Early Chemical History of Solar-System. *Reviews of Geophysics* **12**, 71-101.

Hahn, J.M., Malhotra, R. (1999). Orbital Evolution of Planets Embedded in a Planetesimal Disk. The Astronomical Journal 117, 3041-3053.

Halliday, A.N., Wanke, H., Birck, J.L., Clayton, R.N. (2001). The Accretion, Composition and Early Differentiation of Mars. Space Science Reviews 96, 197-230.

Hansen, B.M.S. (2009). Formation of the Terrestrial Planets from a Narrow Annulus. The Astrophysical Journal 703, 1131-1140.

Harrison, T. M., 2009. The hadean crust: evidence from >4 Ga zircons. *Annual Review of Earth and Planetary Sciences* **37**, 479-505.

Hauri E., Saal, A.E., Rutherford M.C., van Orman J.A. (2010) Hydrogen isotope similarity of the Earth and Moon revealed by water in lunar volcanic glasses. Bulletin of the American Astronomical Society 42, 987.

Hauri E.H., Weinreich T., Saal, A.E., Rutherfore M.C., van Orman J.A. (2011) High pre-eruptive water contents preserved in lunar melt inclusions. Science 333, 213-215.

Heber, V. S., Wieler, R., Baur, H., Olinger, C., Friedmann, T. A., and Burnett, D. S., 2009. Noble gas composition of the solar wind as collected by the Genesis mission. *Geochimica et Cosmochimica Acta* **73**, 7414-7432.

Heimann, M. and Maier-Reimer, E., 1996. On the relations between the oceanic uptake of $CO_2$ and its carbon isotopes. *Global Biogeochemical Cycles* **10**, 89-110.

Hirth, G. and Kohlstedt, D. L., 1996. Water in the oceanic upper mantle: implications for rheology, melt extraction and the evolution of the lithosphere. *Earth and Planetary Science Letters* **144**, 93-108.

Hiyagon, H., Ozima, M., Marty, B., Zashu, S., and Sakai, H., 1992. Noble gases in submarine glasses from mid-oceanic ridges and Loihi seamount: Constraints on the early history of the Earth. *Geochimica et Cosmochimica Acta* **56**, 1301-1316.

Holland, G. and Ballentine, C. J., 2006. Seawater subduction controls the heavy noble gas composition of the mantle. *Nature* **441**, 186-191.

Holland, G., Cassidy, M., and Ballentine, C. J., 2009. Meteorite Kr in Earth‚Äôs Mantle Suggests a Late Accretionary Source for the Atmosphere. *Science* **326**, 1522-1525.

Hollenbach D.J., Yorke H.W., Johnstone D. 2000. Disk dispersal around young stars. In Protostars and Planets IV. ed. Mannings W., Boss A.P., Russell S.S., pp. 401-428. Tucson, AZ, Univ. Arizona Press.

Honda, M., McDougall, I,, Patterson, D.B., Doulgeris, A., Clague, D.A. (1991) Possible solar noble-gas component in Hawaiian basalts. Nature 349, 149-151.

Honda, M., McDougall, I., Patterson, D. B., Doulgeris, A., and Clague, D. A., 1993. Noble gases in submarine pillow basalt glasses from Loihi and Kilauea, Hawaii: A solar component in the Earth. *Geochimica et Cosmochimica Acta* **57**, 859-874.

Honda M., McDougall I. (1998) Primordial helium and neon in the Earth- a speculation on early degassing. Geophysical Research Letters 25, 1951-1954.

Houlton, B. Z. and Bai, E., 2009. Imprint of denitrifying bacteria on the global terrestrial biosphere. *Proceedings of the National Academy of Sciences* **106**, 21713-21716.

Hsieh, H.H., Jewitt, D. (2006). A Population of Comets in the Main Asteroid Belt. Science 312, 561-563.

Hunten, D. M., Pepin, R. O., and Walker, J. C. G., 1987. Mass fractionation in hydrodynamic escape. *Icarus* **69**, 532-549.

HutsemÈkers, D., Manfroid, J., Jehin, E., and Arpigny, C., 2009. New constraints on the delivery of cometary water and nitrogen to Earth from the 15N/14N†isotopic ratio. *Icarus* **204**, 346-348.

Ida, S., Makino, J. (1993). Scattering of planetesimals by a protoplanet - Slowing down of runaway growth. Icarus 106, 210.

Ikoma, M., Genda, H. (2006). Constraints on the Mass of a Habitable Planet with Water of Nebular Origin. The Astrophysical Journal 648, 696-706.

Jagoutz, E., Palme, H., Baddenhausen, H., Blum, K., Cendales, M., Dreibus, G., Spettel, B., Lorenz, V., and Wanke, H., 1979. The abundances of major, minor and trace elements in the earth's mantle as derived from primitive ultramafic nodules. *Proc. Lunar Planet. Sci. Conf.* **10**, 2031-2050.

Jaresewich, E. (1966). Chemical Analysis of Ten Stony Meteorites. Geochimca et Cosmochimica Acta, 30, 1261-1265

Javoy, M., Kaminski, E., Guyot, F., Andrault, D., Sanloup, C., Moreira, M., Labrosse, S., Jambon, A., Agrinier, P., Davaille, A., and Jaupart, C., 2010. The chemical composition of the Earth: Enstatite chondrite models. *Earth and Planetary Science Letters* **293**, 259-268.

Jehin, E., Manfroid, J., Hutsemekers, D., Arpigny, C., and Zucconi, J.-M., 2009. Isotopic ratios in comets: status and perspectives. *Earth Moon and Planets*, 167-180.

Jephcoat, A. P., 1998. Rare-gas solids in the Earth's deep interior. *Nature* **393**, 355-358.

Jessberger, E. K., Christoforidis, A., and Kissel, J., 1988. Aspects of the major element composition of Halley's dust. *Nature* **332**, 691-695.

Jewitt, D. C., Matthews, H. E., Owen, T., and Meier, R., 1997. Measurements of 12C/13C, 14N/15N, and 32S/34S Ratios in Comet Hale-Bopp (C/1995 O1). *Science* **278**, 90-93.

Johansen, A., Oishi, J.S., Mac Low, M.M., Klahr, H., Henning, T., Youdin, A. (2007). Rapid planetesimal formation in turbulent circumstellar disks. Nature 448, 1022-1025.

Johansen, A., Youdin, A., Mac Low, M.M. (2009). Particle Clumping and Planetesimal Formation Depend Strongly on Metallicity. The Astrophysical Journal 704, L75-L79.

Kendrick M.A., Scambelluri M., Honda M., Phillips D. (2011) High abundances of noble gas and chlorine delivered to the mantle by serpentinite subduction. Nature Geoscience 4, 807-812.

Kerridge, J. F., 1985. Carbon, hydrogen and nitrogen in carbonaceous chondrites: abundances and isotopic compositions in bulk samples. *Geochimica et Cosmochimica Acta* **49**, 1707-1714.

King, H.E., Stimpfl, M., Deymier, P., Drake, M.J., Catlow, C.R.A., Putnis, A., de Leeuw, N.H. (2010). Computer simulations of water interactions with low-coordinated forsterite surface sites: Implications for the origin of water in the inner solar system. Earth and Planetary Science Letters 300, 11-18.

Kita, N. T., Huss, G. R., Tachibana, S., Amelin, Y., Nyquist, L. E., and Hutcheon, I. D., 2005. Constraints on the origin of chondrules and CAIs from short-lived and long-lived radionuclides. In: Krot, A. N., Scott, E. R. D., and Reipurth, B. Eds.), *Chondrites and the Protoplanetary Disk*. Astronomical Society of the Pacific, San Francisco.

Kleine, T., Munker, C., Mezger, K., and Palme, H., 2002. Rapid accretion and early core formation on asteroids and the terrestrial planets from Hf-W chronometry. *Nature* **418**, 952-955.

Kleine, T., Touboul, M., Bourdon, B., Nimmo, F., Mezger, K., Palme, H., Jacobsen, S.B., Yin, Q.Z., Halliday, A.N. (2009). Hf-W chronology of the accretion and early evolution of asteroids and terrestrial planets.\ Geochimica et Cosmochimica Acta 73, 5150-5188.

Kokubo, E., Ida, S. (1998). Oligarchic Growth of Protoplanets. Icarus 131, 171-178.

Krasinsky, G.A., Pitjeva, E.V., Vasilyev, M.V., Yagudina, E.I. (2002). Hidden Mass in the Asteroid Belt. Icarus 158, 98-105.

Kunz, J., Staudacher, T., and All√®gre, C. J., 1998. Plutonium-Fission Xenon Found in Earth's Mantle. *Science* **280**, 877-880.

Kuppers, M., and 40 colleagues 2005. A large dust/ice ratio in the nucleus of comet 9P/Tempel 1. Nature 437, 987-990.

Kurz, M. D., Jenkins, W. J., and Hart, S. R., 1982. Helium isotopic systematics of oceanic islands and mantle heterogeneity. *Nature* **297**, 43-47.

Lecuyer, C., Gillet, P., and Robert, F., 1998. The hydrogen isotope composition of seawater and the global water cycle. *Chemical Geology* **145**, 249-261.

Lee, K. K. M. and Steinle-Neumann, G., 2006. High-pressure alloying of iron and xenon: "Missing" Xe in the Earth's core? *J. Geophys. Res.* **111**, B02202.

Leshin Watson L., Hutcheon, I.D., Epstein S., Stolper E.M. (1994) Water on Mars: clues from deuterium/hydrogen and water contents of hydrous phases in SNC meteorites. Science 265, 86-90.

Leshin L.A., Epstein S., Stolper E.M. (1996) Hydrogen isotope geochemistry of SNC meteorites. Geochimica et Cosmochimica Acta 60, 2635-2650.

Levison, H.F., Duncan, M.J. (1997). From the Kuiper Belt to Jupiter-Family Comets: The Spatial Distribution of Ecliptic Comets. Icarus 127, 13-32.

Levison, H.F., Morbidelli, A., Vanlaerhoven, C., Gomes, R., Tsiganis, K. (2008). Origin of the structure of the Kuiper belt during a dynamical instability in the orbits of Uranus and Neptune. Icarus 196, 258-273.

Levison, H.F., Bottke, W.F., Gounelle, M., Morbidelli, A., Nesvorny, D., Tsiganis, K. (2009). Contamination of the asteroid belt by primordial trans-Neptunian objects. Nature 460, 364-366.

Levison, H.F., Morbidelli, A., Tsiganis, K., Nesvorny, D., Gomes, R. (2011). Reevaluating the Early Dynamical Evolution of the Outer Planets. in preparation

Lin, D.N.C., Papaloizou, J. (1986). On the tidal interaction between protoplanets and the protoplanetary disk. III - Orbital migration of protoplanets. The Astrophysical Journal 309, 846-857.

Lissauer, J.J. (1987). Timescales for planetary accretion and the structure of the protoplanetary disk. Icarus 69, 249-265.

Lodders, K., 2010. Solar system abundances of the elements, *Principles and Perspectives in Cosmochemistry*. Springer Berlin Heidelberg.

Lupton, J. E., 1983. Terrestrial Inert Gases: Isotope Tracer Studies and Clues to Primordial Components in the Mantle. *Annual Review of Earth and Planetary Sciences* **11**, 371-414.

Lupton, J. E. and Craig, H., 1981. A Major Helium-3 Source at 15¬∞S on the East Pacific Rise. *Science* **214**, 13-18.

Lux, G., 1987. The behavior of noble gases in silicate liquids: Solution, diffusion, bubbles and surface effects, with applications to natural samples. *Geochimica et Cosmochimica Acta* **51**, 1549-1560.

Lyra, W., Paardekooper, S.J., Mac Low, M.M. (2010). Orbital Migration of Low-mass Planets in Evolutionary Radiative Models: Avoiding Catastrophic Infall. The Astrophysical Journal 715, L68-L73.

Mackenzie, F. T. and Lerman, A., 2006. *Carbon in the geobiosphere*. Springer, Dordrecht.

Malhotra, R. (1995). The Origin of Pluto's Orbit: Implications for the Solar System Beyond Neptune. The Astronomical Journal 110, 420.

Mann U., Frost D.J., Rubie D.C. (2009) Evidence for high-pressure core-mantle differentiation from the metal-silicate partitioning of lithophile and weakly-siderophile elements. Geochimica et Cosmochimica Acta 73, 7360-7386.

Manning, C. V., McKay, C. P., and Zahnle, K. J., 2008. The nitrogen cycle on Mars: Impact decomposition of near-surface nitrates as a source for a nitrogen steady state. *Icarus* **197**, 60-64.

Markowski, A., Leya, I., QuittÈ, G., Ammon, K., Halliday, A. N., and Wieler, R., 2006. Correlated helium-3 and tungsten isotopes in iron meteorites: Quantitative cosmogenic corrections and planetesimal formation times. *Earth and Planetary Science Letters* **250**, 104-115.

Marty, B., 1989. Neon and xenon isotopes in MORB: implications for the earth-atmosphere evolution. *Earth and Planetary Science Letters* **94**, 45-56.

Marty, B., 1995. Nitrogen content of the mantle inferred from N2-Ar correlation in oceanic basalts. *Nature* **377**, 326-329.

Marty, B. 2012. The origins and concentrations of water, carbon, nitrogen and noble gases on Earth. Earth and Planetary Science Letters, in press.

Marty, B., Chaussidon, M., Wiens, R. C., Jurewicz, A. J. G., and Burnett, D. S., 2011. The lowest 15N/14N end-member of the solar system is the Sun. *Lunar and Planetary Science Conference* **42**, #1870.

Marty, B. and Dauphas, N., 2002. Formation and early evolution of the atmosphere. In: Fowler, C. M. R., Ebinger, C. J., and Hawkesworth, C. J. Eds.), *The Early Earth: Physical, Chemical and Biological Development*.

Marty, B. and Dauphas, N., 2003. The nitrogen record of crust-mantle interaction and mantle convection from Archean to present. *Earth and Planetary Science Letters* **206**, 397-410.

Marty, B. and Meibom, A., 2007. Noble gas signature of the late heavy bombardment in the Earth's atmosphere. *eEarth* **2**, 99-113.

Marty, B., Palma, R. L., Pepin, R. O., Zimmermann, L., Schlutter, D. J., Burnard, P. G., Westphal, A. J., Snead, C. J., Bajt, S. a., Becker, R. H., and Simones, J. E., 2008. Helium and Neon Abundances and Compositions in Cometary Matter. *Science* **319**, 75-78.

Marty, B., Tolstikhin, I., Kamensky, I. L., Nivin, E., Balaganskaya, E., and Zimmermann, J. L., 1998. Plume-derived rare gases in 380 Ma carbonatites from the Kola region (Russian) and the argon isotopic composition in the deep mantle. *Earth and Planetary Science Letters* **164**, 179-192.

Marty, B. and Yokochi, R., 2006. Water in the Early Earth. *Reviews in Mineralogy and Geochemistry* **62**, 421-450.

Marty, B. and Zimmerman, L., 1999. Volatiles (He, C, N, Ar) in mid0ocean ridge basalts: assesment of shallow-level fractionation and characterization of source composition. *Geochimica et Cosmochimica Acta* **63**, 3619-3633.

Masset, F., Snellgrove, M. (2001). Reversing type II migration: resonance trapping of a lighter giant protoplanet. Monthly Notices of the Royal Astronomical Society 320, L55-L59.

Mazor, E., Heymann, D., and Anders, E., 1970. Noble gases in carbonaceous chondrites. *Geochimica et Cosmochimica Acta* **34**, 781-824.

McCubbin F.M., Steele A., Hauri E.H., Nekvasil H., Yamashita S., Hemley R,J, 2010, Nominally hydrous magmatism on the Moon. PNAS 107, 11223-11228.

McDonough, W. F. and Sun, S. s., 1995b. The composition of the Earth. *Chemical Geology* **120**, 223-253.

McNaughton, N.J., Borthwick, J., Fallick, A.E., Pillinger, C.T. (1981). Deuterium/hydrogen ratios in unequilibrated ordinary chondrites. Nature 294, 639-641.

Meech K.J. et al. (2011) EPOXI: Comet 103P/Hartley 2 observations from a wporldwide campaign. The Astrophysical Journal Letters 734, doi:1-.1088/2041-8205/734/1/L1

Meier R., Owen T.C., Matthews H.E., Jewitt D.C., Bockelee-Morvan D., Biver N., Crovisier J. and Gautier D. (1998) A determination of the DHO/H2O ratio in Comet C/1995 O1 (Hale—Bopp). Science, 279, 842—844.

Meshik, A. P., Kehm, K., and Hohenberg, C. M., 2000. Anomalous xenon in zone 13 Okelobondo. *Geochimica et Cosmochimica Acta* **64**, 1651-1661.

Michael, P., 1995. Regionally distinctive sources of depleted MORB: evidence from trace elements and H2O. *Earth and Planetary Science Letters* **131**, 301-320.

Mojzsis, S. J., Harrison, T. M., and Pidgeon, R. T., 2001. Oxygen-isotope evidence from ancient zircons for liquid water at the Earth's surface 4,300[thinsp]Myr ago. *Nature* **409**, 178-181.

Montmessin, F., Fouchet, T., and Forget, F., 2005. Modeling the annual cycle of HDO in the Martian atmosphere. *Journal of Geophysical Research* **110**, doi:/10.1029/2004JE002357.

Moorbath, S., O'Nions, R. K., and Pankhurst, R. J., 1973. Early Archaean Age for the Isua Iron Formation, West Greenland. *Nature* **245**, 138-139.

Morbidelli, A., Chambers, J., Lunine, J.I., Petit, J.M., Robert, F., Valsecchi, G.B., Cyr, K.E. (2000). Source regions and time scales for the delivery of water to Earth. Meteoritics and Planetary Science 35, 1309-1320.

Morbidelli, A., Petit, J.M., Gladman, B., Chambers, J. (2001). A plausible cause of the late heavy bombardment. Meteoritics and Planetary Science 36, 371-380.

Morbidelli, A., Levison, H.F., Tsiganis, K., Gomes, R. (2005). Chaotic capture of Jupiter's Trojan asteroids in the early Solar System. Nature 435, 462-465.

Morbidelli, A., Crida, A., Masset, F., Nelson, R.P. (2008). Building giant-planet cores at a planet trap. Astronomy and Astrophysics 478, 929-937.

Morbidelli, A., Bottke, W.F., Nesvorny, D., Levison, H.F. (2009). Asteroids were born big. Icarus 204, 558-573.

Morbidelli, A., Crida, A. (2007). The dynamics of Jupiter and Saturn in the gaseous protoplanetary disk. Icarus 191, 158-171.

Morbidelli, A., Tsiganis, K., Crida, A., Levison, H.F., Gomes, R. (2007). Dynamics of the Giant Planets of the Solar System in the Gaseous Protoplanetary Disk and Their Relationship to the Current Orbital Architecture. The Astronomical Journal 134, 1790-1798.

Morbidelli, A., Brasser, R., Gomes, R., Levison, H.F., Tsiganis, K. (2010). Evidence from the Asteroid Belt for a Violent Past Evolution of Jupiter's Orbit. The Astronomical Journal 140, 1391-1401.

Morbidelli, A. (2010b). A coherent and comprehensive model of the evolution of the outer Solar System. Comptes Rendus Physique 11, 651-659.

Moreira, M., Kunz, J., and Allv√®gre, C., 1998. Rare Gas Systematics in Popping Rock: Isotopic and Elemental Compositions in the Upper Mantle. *Science* **279**, 1178-1181.

Morgan, J. W., 1986. Ultramafic Xenoliths: Clues to Earth's Late Accretionary History. *J. Geophys. Res.* **91**, 12375-12387.

Morishima, R., Stadel, J., Moore, B. (2010). From planetesimals to terrestrial planets: N-body simulations including the effects of nebular gas and giant planets. Icarus 207, 517-535.

Morris, R. V., Ruff, S. W., Gellert, R., Ming, D. W., Arvidson, R. E., Clark, B. C., Golden, D. C., Siebach, K., Klingelh√∂fer, G. s., Schr√∂der, C., Fleischer, I., Yen, A. S., and Squyres, S. W., 2010. Identification of Carbonate-Rich Outcrops on Mars by the Spirit Rover. *Science* **329**, 421-424.

Mottl, M., Glazer, B. T., Kaiser, R. I., and Meech, K. J., 2007. Water and astrobiology. *Chemie Der Erde-Geochemistry* **67**, 253-282.

Muralidharan, K., Deymier, P., Stimpfl, M., de Leeuw, N.H., Drake, M.J. (2008). Origin of water in the inner Solar System: A kinetic Monte Carlo study of water adsorption on forsterite. Icarus 198, 400-407.

Muramatsu, Y. and Wedepohl, K. H., 1998. The distribution of iodine in the earth's crust. *Chemical Geology* **147**, 201-216.

Nakamura, T., Noguchi, T., Tanaka, M., and al., 2011. Mineralogy and major element abundance of the dust particles recovered from Muses-C region on the asteroid Itokawa. *Lunar and Planetary Science Conference* **42**, 1766.

Nesvorny, D., Vokrouhlicky, D., Morbidelli, A. (2007). Capture of Irregular Satellites during Planetary Encounters. The Astronomical Journal 133, 1962-1976.

Nesvorny, D., Youdin, A.N., Richardson, D.C. (2010). Formation of Kuiper Belt Binaries by Gravitational Collapse. The Astronomical Journal 140, 785-793.

Nesvorny, D., Jenniskens, P., Levison, H.F., Bottke, W.F., Vokrouhlicky, D., Gounelle, M. (2010b). Cometary Origin of the Zodiacal Cloud and Carbonaceous Micrometeorites. Implications for Hot Debris Disks. The Astrophysical Journal 713, 816-836.

Niles, P. B., Boynton, W. V., Hoffman, J. H., Ming, D. W., and Hamara, D., 2010. Stable isotope measurements of martian atmospheric CO2 at the Phoenix landing site. *Science* **329**, 1334-1337.

Notesco, G., Bar-Nun, A., and Owen, T., 2003. Gas trapping in water ice at very low deposition rates and implications for comets. *Icarus* **162**, 183-189.

Notesco, G., Laufer, D., Bar-Nun, A., and Owen, T., 1999. An Experimental Study of the Isotopic Enrichment in Ar, Kr, and Xe When Trapped in Water Ice. *Icarus* **142**, 298-300.

Nutman, A. P., Mojzsis, S. J., and Friend, C. R. L., 1997. Recognition of >=3850 Ma water-lain sediments in West Greenland and their significance for the early Archaean Earth. *Geochimica et Cosmochimica Acta* **61**, 2475-2484.

O'Brien, D.P., Morbidelli, A., Levison, H.F. (2006). Terrestrial planet formation with strong dynamical friction. Icarus 184, 39-58.

O'Brien, D.P., Morbidelli, A., Bottke, W.F. (2007). The primordial excitation and clearing of the asteroid belt- Revisited. Icarus 191, 434-452.

O'Brien, D.P., Walsh, K.J., Morbidelli, A., Raymond, S.N., Mandell, A.M., Bond, J.C. (2010). Early Giant Planet Migration in the Solar System: Geochemical and Cosmochemical Implications for Terrestrial Planet Formation. Bulletin of the American Astronomical Society 42, 948.

O'Neil, J., Maurice, C., Stevenson, R. K., Larocque, J., Cloquet, C., David, J., Francis, D., Martin J. van Kranendonk, R. H. S., and Vickie, C. B., 2007. Chapter 3.4 The Geology of the 3.8 Ga Nuvvuagittuq (Porpoise Cove) Greenstone Belt, Northeastern Superior Province, Canada, *Developments in Precambrian Geology*. Elsevier.

Ott, U., 2002. Noble gases in meteorites - Trapped components. *Reviews in Mineralogy and Geochemistry* **47**, 71-100.

Owen, T., Bar-Nun, A., and Kleinfeld, I., 1992. Possible cometary origin of heavy noble gases in the atmospheres of Venus, Earth and Mars. *Nature* **358**, 43-46.

Owen, T., Biemann, K., Rushneck, D. R., Biller, J. E., Howarth, D. W., and Lafleur, A. L., 1977. The Composition of the Atmosphere at the Surface of Mars. *J. Geophys. Res.* **82**, 4635-4639.

Ozima, M., 1975. Ar isotopes and Earth-atmosphere evolution models. *Geochimica et Cosmochimica Acta* **39**, 1127-1134.

Ozima, M. and Podosek, F., 2001. *Noble Gas Geochemistry*. Cambridge University Press, Cambridge.

Pepin, R., 2000a. On the Isotopic Composition of Primordial Xenon in Terrestrial Planet Atmospheres. *Space Science Reviews* **92**, 371-395.

Pepin, R. O., 1991. On the origin and early evolution of terrestrial planet atmospheres and meteoritic volatiles. *Icarus* **92**, 2-79.

Pepin, R. O., 1997. Evolution of Earth's Noble Gases: Consequences of Assuming Hydrodynamic Loss Driven by Giant Impact. *Icarus* **126**, 148-156.

Pepin, R. O., 2000b. On the isotopic composition of primordial xenon in terrestrial planet atmospheres. *Space Science Reviews* **92**, 371-395.

Pepin, R. O., 2006. Atmospheres on the terrestrial planets: Clues to origin and evolution. *Earth and Planetary Science Letters* **252**, 1-14.

Pepin, R. O., Becker, R. H., and Rider, P. E., 1995. Xenon and krypton isotopes in extraterrestrial regolith solis and in the solar wind. *Geochimica et Cosmochimica Acta* **59**, 4997-5022.

Petit, J.M., Morbidelli, A., Chambers, J. (2001). The Primordial Excitation and Clearing of the Asteroid Belt. Icarus 153, 338-347.

Pierens, A., Nelson, R.P. 2008. Constraints on resonant-trapping for two planets embedded in a protoplanetary disc. Astronomy and Astrophysics 482, 333-340.

Pineau F., Javoy M. (1983) Carbon isotopes and concentrations in mid-oceanic ridge basalts. Earth and Planetary Science Letters 62, 239-257.

Pollack, J. B., Kasting, J. F., Richardson, S. M., and Poliakoff, K., 1987. The case for a wet, warm climate on early Earth. *Icarus* **71**, 203-224.

Poreda R.J., Farley K.A. (1992) Rare gases in Samoan xenoliths. Earth and Planetary Science Letters 113, 129-144,

Puchtel I.S., Walker, R.J., James, O.B., Kring, D.A., 2008. Osmium isotope and highly siderophile element systematics of lunar impact melt breccias: implications for the late accretion history of the Moon and Earth. Goechimica et Cosmochimica Acta 72, 3022-3042.

Pujol M., Marty B., Burgess R. (2011) Chondritic-like xenon trapped in Archean rocks: a possible signature of the ancient atmosphere. Earth and Planetary Science Letters 308, 298-306.

Qin, L. P., Dauphas, N., Wadhwa, M., Masarik, J., and Janney, P. E., 2008. Rapid accretion and differentiation of iron meteorite parent bodies inferred from Hf-182-W-182 chronometry and thermal modeling. *Earth and Planetary Science Letters* **273**, 94-104. Qin, L., Alexander, C.M.O.'D., Carlson, R.W., Horan, M.F., Yokoyama, T. (2010) Contributors to chromium isotope variation of meteorites. Geochimica et Cosmochimica Acta 74, 1122-1145.

Rao, M. N., Bogard, D. D., Nyquist, L. E., McKay, D. S., and Masarik, J., 2002. Neutron capture isotopes in the martian regolith and implication for martian atmospheric noble gases. *Icarus* **156**, 352-372.

Raquin, A. and Moreira, M., 2009. Atmospheric 38Ar/36Ar in the mantle: Implications for the nature of the terrestrial parent bodies. *Earth and Planetary Science Letters* **287**, 551-558.

Raymond, S.N., Quinn, T., Lunine, J.I. (2004). Making other earths: dynamical simulations of terrestrial planet formation and water delivery. Icarus 168, 1-17.

Raymond, S.N., Quinn, T., Lunine, J.I. (2005). Terrestrial Planet Formation in Disks with Varying Surface Density Profiles. The Astrophysical Journal 632, 670-676.

Raymond, S.N., Quinn, T., Lunine, J.I. (2006). High-resolution simulations of the final assembly of Earth-like planets I. Terrestrial accretion and dynamics. Icarus 183, 265-282.

Raymond, S.N., Quinn, T., Lunine, J.I. (2007). High-Resolution Simulations of The Final Assembly of Earth-Like Planets. 2. Water Delivery And Planetary Habitability. Astrobiology 7, 66-84.

Raymond, S.N., O'Brien, D.P., Morbidelli, A., Kaib, N.A. (2009). Building the terrestrial planets: Constrained accretion in the inner Solar System. Icarus 203, 644-662.

Raymond, S. N., 2010. Formation of terrestrial planets. In: Barnes, R. (Ed.), *Formation and Evolution of Exoplanets*. Wiley, Weinheim, Germany.

Richards, M. A., Yang, W.-S., Baumgardner, J. R., and Bunge, H.-P., 2001. Role of a low-viscosity zone in stabilizing plate tectonics: implications for comparative terrestrial planetology. *Geochemistry Geophysics Geosystems* **2**, 2000GC000115.

Rivkin, A.S., Emery, J.P. (2010). Detection of ice and organics on an asteroidal surface. Nature 464, 1322-1323.

Robert, F., Merlivat, L., Javoy, M. (1977). Water and Deuterium Content in Eight Condrites. Meteoritics 12, 349-356.

Robert, F., Merlivat, L., Javoy, M. (1979). Deuterium concentration in the early solar system - Hydrogen and oxygen isotope study. Nature 282, 785-789.

Robert F. and Epstein S. (1982). The concentration of isotopic compositions of hydrogen carbon and nitrogen in carbonaceous chondrites. Geochim. Cosmochim. Acta, 16, 81-95

Robert, F. (2003). The D/H Ratio in Chondrites. Space Science Reviews 106, 87-101.

Robert, F., Gautier, D., and Dubrulle, B., 2000. The solar system D/H ratio: observations and theories. *Space Science Reviews* **92**, 201-224.

Robinson, T. D., Meadows, V. S., and Crisp, D., 2010. Detecting oceans on extrasolar planets using the glint effect. *The Astrophysical Journal Letters* **721**, L67-L71.

Rudnick, R. L. and Gao, S., 2003. Composition of the continental crust. In: Holland, H. D. and Turekian, K. K. Eds.), *Treatise on Geochemistry*. Elsevier.

Saal, A. E., Hauri, E. H., Langmuir, C. H., and Perfit, M. R., 2002. Vapour undersaturation in primitive mid-ocean-ridge basalt and the volatile content of Earth's upper mantle. *Nature* **419**, 451-455.

Saal, A.E., Hauri E.H., Cascio M.L., Van Orman J.A. (2008) Volatile content of lunar volcanic glasses and the presence of water in the Moon's interior. Nature 454, 192-195.

Saal A.E., Hauri E.H., Rutherford M.J., van Orman J. (2011) The volatile content and D/H ratios of the lunar picrtitic glasses, Wet vs Dry Moon conference, abstract #6034.

Sandor, Z., Lyra, W., Dullemond, C.P. 2011. Formation of Planetary Cores at Type I Migration Traps. The Astrophysical Journal 728, L9.

Sanloup, C. l., Schmidt, B. C., Perez, E. M. C., Jambon, A., Gregoryanz, E., and Mezouar, M., 2005. Retention of Xenon in Quartz and Earth's Missing Xenon. *Science* **310**, 1174-1177.

Sarda, P., Staudacher, T., and AllÈgre, C. J., 1988. Neon isotopes in submarine basalts. *Earth and Planetary Science Letters* **91**, 73-88.

Sasaki, S. and Nakazawa, K., 1988. Origin of isotopic fractionation of terrestrial Xe: hydrodynamic fractionation during escape of the primordial H2He atmosphere. *Earth and Planetary Science Letters* **89**, 323-334.

Schoenberg, R., Kamber, B. S., Collerson, K. D., and Eugster, O., 2002. New W-isotope evidence for rapid terrestrial accretion and very early core formation. *Geochimica et Cosmochimica Acta* **66**, 3151-3160.

Scott, E.R.D. (2006). Meteoritical and dynamical constraints on the growth mechanisms and formation times of asteroids and Jupiter. Icarus 185, 72-82.

Scott, P. C., Asplund, M., Grevesse, N., and Sauval, A. J., 2006. Line formation in solar granulation. VII. CO lines and the solar C and O isotopic abundances. *Astronomy & Astrophysics* **456**, 675-688.

Shukolyukov, Y. A., Jessberger, E. K., Meshik, A. P., Vu Minh, D., and Jordan, J. L., 1994. Chemically fractionated fission-xenon in meteorites and on the earth. *Geochimica et Cosmochimica Acta* **58**, 3075-3092.

Simon, S. B., Joswiak, D. J., Ishii, H. A., Bradley, J. P., Chi, M., Grossman, L., AlÉOn, J., Brownlee, D. E., Fallon, S., Hutcheon, I. D., Matrajt, G., and McKeegan, K. D., 2008. A refractory inclusion returned by Stardust from comet 81P/Wild 2. *Meteoritics & Planetary Science* **43**, 1861-1877.

Sotin, C., Mitri, G., Rappaport, N., Schubert, G., Stevenson, D. (2010). Titan's Interior Structure. Titan from Cassini-Huygens 61-73.

Staudacher, T. and AllÈgre, C. J., 1982. Terrestrial xenology. *Earth and Planetary Science Letters* **60**, 389-406.

Stracke, A., Bizimis, M., and Salters, J. M., 2003. Recycling oceanic crust: Quantitative constraints. *Geochemistry Geophysics Geosystems* **4**, doi:10.1029/2001GC000223.

Stuart, F. M., Lass-Evans, S., Godfrey Fitton, J., and Ellam, R. M., 2003. High 3He/4He ratios in picritic basalts from Baffin Island and the role of a mixed reservoir in mantle plumes. *Nature* **424**, 57-59.

Sumino H., Burgess R., Mizukami T., Wallis S.R., Holland G., Ballentine C.J. (2010) Seawater-derived noble gases and halogens in exhumed mantle wedge peridotite. Earth and Planetary Science Letters 294, 163-172.

Sundquist, E. T. and Visser, K., 2003. The geologic history of the carbon cycle. In: Holland, H. D. and Turekian, K. K. Eds.), *Treatise on Geochemistry*. Elsevier.

Sunshine, J.M., and 22 colleagues (2006). Exposed Water Ice Deposits on the Surface of Comet 9P/Tempel 1. Science 311, 1453-1455.

Swindle, T. D., Caffee, M. W., and Hohenberg, C. M., 1986. Xenon and other noble gases in Shergottites. *Geochimica et Cosmochimica Acta* **50**, 1001-1015.

Swindle, T. D. and Jones, J. H., 1997. The xenon isotopic composition of the primordial Martian atmosphere: Contributions from solar and fission components. *J. Geophys. Res.* **102**, 1671-1678.

Tera, F., Papanastassiou, D. A., and Wasserburg, G. J., 1974. Isotopic evidence for a terminal lunar cataclysm. *Earth and Planetary Science Letters* **22**, 1-21.

Tian, F., Toon, O. B., Pavlov, A. A., and De Sterck, H., 2005. Transonic hydrodynamic escape of hydrogen from extrasolar planetary atmospheres. *The Astrophysical Journal* **621**, 1049-1060.

Tolstikhin, I. and Hofmann, A. W., 2005. Early crust on top of the Earth's core. *Physics of The Earth and Planetary Interiors* **148**, 109-130.

Tolstikhin, I. and Kramers, J., 2008. *The evolution of matter: from the big bang to the present day*. Cambridge University Press, Cambridge.

Tolstikhin, I. N. and O'Nions, R. K., 1994. The Earth's missing xenon: A combination of early degassing and of rare gas loss from the atmosphere. *Chemical Geology* **115**, 1-6.

Touboul, M., Kleine, T., Bourdon, B., Palme, H., and Wieler, R., 2007. Late formation and prolonged differentiation of the Moon inferred from W isotopes in lunar metals. *Nature* **450**, 1206-1209.

Trieloff, M., Kunz, J., Clague, D. A., Harrison, D., and All√®gre, C. J., 2000. The Nature of Pristine Noble Gases in Mantle Plumes. *Science* **288**, 1036-1038.

Trieloff M., Kunz J., Allegre C.J. (2002) Noble gas systematics of the Reunion mantle plume source and the origin of primordial noble gases in Earth's mantle. Earth and Planetary Science Letters 200, 297-313.

Trinquier, A., Birck, J.-L., and Allegre, C. J., 2007. Widespread 54Cr heterogeneity in the inner solar system. *The Astrophysical Journal* **655**, 1179-1185.

Trinquier, A., Elliott, T., Ulfbeck, D., Coath, C., Krot, A. N., and Bizzarro, M., 2009. Origin of Nucleosynthetic Isotope Heterogeneity in the Solar Protoplanetary Disk. *Science* **324**, 374-376.

Tsiganis, K., Gomes, R., Morbidelli, A., Levison, H.F. (2005). Origin of the orbital architecture of the giant planets of the Solar System. Nature 435, 459-461.

Turner, G. (1989) The outgassing history of the Earth's atmosphere. Journal of the Geological Society 146, 147-154.

Valbracht, P. J., Staudacher, A., Malahoff, A., and Allegre, C. J., 1997a. Noble gas systematics of deep rift zone glasses from Loihi Seamount, Hawaii. *Earth and Planetary Science Letters* **150**, 399-411.

Valbracht, P. J., Staudacher, T., Malahoff, A., and AllËgre, C. J., 1997b. Noble gas systematics of deep rift zone glasses from Loihi Seamount, Hawaii. *Earth and Planetary Science Letters* **150**, 399-411.

Vidal-Madjar, A., des Etangs, A. L., Desert, J. M., Ballester, G. E., Ferlet, R., Hebrard, G., and Mayor, M., 2003. An extended upper atmosphere around the extrasolar planet HD209458b. *Nature* **422**, 143-146.

Villeneuve, J., Chaussidon, M., and Libourel, G., 2009. Homogeneous Distribution of 26Al in the Solar System from the Mg Isotopic Composition of Chondrules. *Science* **325**, 985-988.

Von Zahn, U., Kumar, S., Niemann, H., and Prinn, R., 1983. Composition of the Venus atmosphere. In: Hunten, D. M., Colin, L., Donahue, T. M., and Moroz, V. I. Eds.), *Venus*. University of Arizona Press, Tucson, AZ.

Walker, R. J., 2009. Highly siderophile elements in the Earth, Moon and Mars: Update and implications for planetary accretion and differentiation. *Chemie der Erde - Geochemistry* **69**, 101-125.

Walsh, K.J., Morbidelli, A., Raymond, S.N., O'brien, D.P., Mandell, A.M. (2011) Sculpting of the inner Solar System by gas-driven orbital migration of Jupiter. Nature, in press.

Ward, W.R. (1997). Protoplanet Migration by Nebula Tides. Icarus 126, 261-281.

Weidenschilling, S.J. (1977). Aerodynamics of solid bodies in the solar nebula. Monthly Notices of the Royal Astronomical Society 180, 57-70.

Wetherill, G. W., 1975. Radiometric Chronology of the Early Solar System. *Annual Review of Nuclear Science* **25**, 283-328.

Wetherill, G.W. (1989). Origin of the asteroid belt. Asteroids II 661-680.

Wieler, R., 2002. Noble gases in the solar system. *Reviews in Mineralogy and Geochemistry* **47**, 2-70.

Wilde, S. A., Valley, J. W., Peck, W. H., and Graham, C. M., 2001. Evidence from detrital zircons for the existence of continental crust and oceans on the Earth 4.4[thinsp]Gyr ago. *Nature* **409**, 175-178.

Wilhelms D.L., McCanley J.F. and Trask N.J. (1987) The geologic history of the Moon. U.S. Geological survey professional paper, 1348.

Williams, D. M. and Gaidos, E., 2008. Detecting the glint of starlight on the oceans of distant planets. *Icarus* **195**, 927-937.

Wood B.J., Halliday A.N., Rehkamper M. (2010) Volatile accretion history of the Earth. Nature 467, E6-E7.

Workman, R. K. and Hart, S. R., 2005a. Major and trace element composition of the depleted MORB mantle (DMM). *Earth and Planetary Science Letters* **231**, 53-72.

Workman, R. K. and Hart, S. R., 2005b. Major and trace element composition of the depleted MORB mantle (DMM). *Earth and Planetary Science Letters* **231**, 53-72.

Yatsevich, I. and Honda, M., 1997. Production of nucleogenic neon in the Earth from natural radioactive decay. *J. Geophys. Res.* **102**, 10291-10298.

Yin, Q., Jacobsen, S. B., Yamashita, K., Blichert-Toft, J., Telouk, P., and Albarede, F., 2002. A short timescale for terrestrial planet formation from Hf-W chronometry of meteorites. *Nature* **418**, 949-952.

Yokochi, R. and Marty, B., 2004. A determination of the neon isotopic composition of the deep mantle. *Earth and Planetary Science Letters* **225**, 77-88.

Yokochi, R. and Marty, B., 2005. Geochemical constraints on mantle dynamics in the Hadean. *Earth and Planetary Science Letters* **238**, 17-30.

Zahnle, K. J. and Kasting, J. F., 1986. Mass fractionation during transonic escape and implications for loss of water from Mars and Venus. *Icarus* **68**, 462-480.

Zahnle, K. J. and Walker, J. C. G., 1982. The evolution of solar ultraviolet luminosity. *Rev. Geophys.* **20**, 280-292.

Zahnle, K.J. (2000) Hydrodynamic escape of ionized xenon from ancient atmospheres. American Astronomical Society DPS Meeting #32, Buletin of the American Astronomical Society 32, 1044.

Zhang, Y., 1998. The young age of Earth. *Geochimica et Cosmochimica Acta* **62**, 3185-3189.

Zhang, Y. and Zindler, A., 1989. Noble Gas Constraints on the Evolution of the Earth's Atmosphere. *J. Geophys. Res.* **94**, 13719-13737.

Zolensky, M.E., and 74 colleagues (2006). Mineralogy and Petrology of Comet 81P/Wild 2 Nucleus Samples. Science 314, 1735-1739.

**Table 1. Major volatile element abundances and isotopic ratios in Earth (excluding the core)**

|  | H (mol) | $\delta D_{VSMOW}$ (‰) | D/H | C (mol) | $\delta^{13}C_{VPDB}$ (‰) | $^{13}C/^{12}C$ | N (mol) | $\delta^{15}N_{air}$ (‰) | $^{15}N/^{14}N$ |
|---|---|---|---|---|---|---|---|---|---|
| Atmosphere ($5.1\times10^{18}$ kg) |  |  |  | 6.60 (16) | -8 | 0.01115 | 2.82 (20) | 0 | 0.003677 |
| Hydrosphere ($1.6\times10^{21}$ kg) | 1.80 (23) | -6 | 0.000155 | 3.20 (18) | 0 | 0.01124 | 1.47 (18) | 6 | 0.003699 |
| Biosphere ($1.6\times10^{18}$ kg) | 1.51 (20) | -100 | 0.000140 | 6.06 (16) | -25 | 0.01096 | 7.50 (14) | 0 | 0.003677 |
| Crust ($2.8\times10^{22}$ kg) | 3.43 (22) | -75 | 0.000144 | 6.78 (21) | -4 | 0.01119 | 7.14 (19) | 6 | 0.003699 |
| Surface reservoirs (A+H+B+C) | 2.15 (23) | -17 | 0.000153 | 6.78 (21) | -4 | 0.01119 | 3.55 (20) | 1 | 0.003681 |
| Mantle ($4.0\times10^{24}$ kg) | 3.05 (23) | -80 | 0.000143 | 2.70 (22) | -5 | 0.01118 | 3.15 (20) | 3 | 0.003688 |
| **Total Earth ($5.97\times10^{24}$ kg)** | 5.20 (23) | -54 | 0.000147 | 3.38 (22) | -5 | 0.01118 | 6.70 (20) | 2 | 0.003684 |

*Notes.* Powers of ten multipliers in parentheses. See text for details and references.

$\delta D_{VSMOW}=[(D/H)/(D/H)_{VSMOW}-1]\times1000$; $\delta^{13}C_{VPDB}=[(^{13}C/^{12}C)/(^{13}C/^{12}C)_{VPDB}-1]\times1000$; $\delta^{15}N_{air}=[(^{15}N/^{14}N)/(^{15}N/^{14}N)_{air}-1]\times1000$.

**Table 2. Abundances of H, C, N, and noble gases in solar and selected planetary reservoirs (in mol/g).**

| | $^1$H | $^{12}$C | $^{14}$N | | | $^{20}$Ne | $^{36}$Ar | $^{84}$Kr | $^{130}$Xe |
|---|---|---|---|---|---|---|---|---|---|
| Solar | 7.112 (−01) | 1.95 (−04) | 5.80 (−05) | | | 8.40 (−05) | 2.15 (−06) | 8.72 (−10) | 6.42 (−12) |
| Venus (surface) | 1.4 (−10) | 2.2 (−06) | 1.6 (−07) | | | 1.6 (−11) | 6.8 (−11) | 6.10 (−14) | <3.7 (−15) |
| Earth | 8.70 (−05) | 5.60 (−06) | 1.12 (−07) | | | 4.87 (−13) | 9.29 (−13) | 1.92 (−14) | 1.05 (−16) |
| Mars (surface) | 9 (−07) | 8.6 (−10) | 4.8 (−11) | | | 7.9 (−16) | 8.10 (−15) | 1.80 (−16) | 1.30 (−18) |
| CI chondrites | 6.7 (−03) | 3.1 (−03) | 1.1 (−04) | | | 1.64 (−11) | 4.33 (−11) | 4.78 (−13) | 6.47 (−14) |
| Comets | 6.7 (−02) | 1.4 (−02) | 2.9 (−03) | From trapping experiments | 30 K | 1.68 (−05) | 1.30 (−04) | 5.25 (−08) | 3.87 (−10) |
| | | | | | 35 | <1.68 (−05) | 4.60 (−05) | 1.07 (−07) | 4.76 (−10) |
| | | | | | 40 | <1.68 (−05) | 1.71 (−05) | 1.35 (−07) | 4.14 (−10) |
| | | | | | 45 | <1.68 (−05) | 6.21 (−06) | 8.33 (−08) | 2.33 (−10) |
| | | | | | 50 | <3.34 (−08) | 2.25 (−06) | 5.13 (−08) | 1.28 (−10) |
| | | | | | 55 | <3.34 (−08) | 8.19 (−07) | 3.89 (−08) | 7.54 (−11) |
| | | | | | 60 | <3.34 (−08) | 2.97 (−07) | 4.47 (−08) | 7.37 (−11) |
| | | | | | 65 | <3.34 (−08) | 1.10 (−07) | 2.40 (−08) | 8.27 (−11) |
| | | | | | 70 | <3.34 (−08) | 4.01 (−08) | 1.29 (−08) | 9.28 (−11) |

*Notes.* Powers of ten multipliers in parentheses. See text for details and references.
Solar in mol/g-Solar composition; Venus, Earth, and Mars in mol/g-Planet; CI chondrites in mol/g-Meteorite; Comets in mol/g-Comet (ice+dust).

**Table 3. Isotopic compositions of major volatile elements and nobles gases in planetary reservoirs.**

| | Major volatile elements, neon, and argon | | | | | | | | | | | |
|---|---|---|---|---|---|---|---|---|---|---|---|---|
| | D/H | $\delta D$ (‰) | $^{13}C/^{12}C$ | $\delta^{13}C$ (‰) | $^{15}N/^{14}N$ | $\delta^{15}N$ (‰) | $^{20}Ne/^{22}Ne$ | $^{21}Ne/^{22}Ne$ | $F_{Ne}$ (‰/amu) | $^{38}Ar/^{36}Ar$ | $^{40}Ar/^{36}Ar$ | $F_{Ar}$ (‰/amu) |
| Solar (4.56 Ga) | 0.000025 | -839 | 0.0115 | 25 | 0.00218 | -408 | 13.78 | 0.0329 | 0 | 0.1828 | 0.000284 | 0 |
| Venus (surface) | 0.024454 | 156000 | 0.0113 | 8 | 0.00366 | -4 | 11.8 | | 84 | 0.180 | 1.1 | -8 |
| Earth | 0.000147 | -54 | 0.01118 | -5 | 0.003684 | 2 | 9.80 | 0.0290 | 203 | 0.1880 | 295.5 | 14 |
| Mars (surface) | 0.001012 | 5500 | 0.01121 | -2.5 | 0.00596 | 620 | 7 | | 484 | 0.26 | 1800 | 201 |
| CI chondrites | 0.000181 | 161 | 0.01112 | -10 | 0.003831 | 42 | 9.0 | 0.03 | 266 | 0.185 | <0.12 | 6 |
| Comets Oort-cloud | 0.000308 | 977 | 0.0107 | -48 | 0.0068 | 850 | solar? | | | solar? | | |
| Comets Kuiper-belt | 0.000161 | 34 | 0.0105 | -63 | 0.00645 | 755 | solar? | | | solar? | | |
| | **Krypton** | | | | | | | | | | | |
| | $^{78}Kr$ | $^{80}Kr$ | $^{82}Kr$ | $^{83}Kr$ | $^{84}Kr$ | $^{86}Kr$ | $F_{Kr}$ (‰/amu) | | | | | |
| Solar | 0.6365 | 4.088 | 20.482 | 20.291 | 100 | 30.24 | 0 | | | | | |
| Earth | 0.6087 | 3.9599 | 20.217 | 20.136 | 100 | 30.524 | 8 | | | | | |
| Mars | | 4.32 | 20.99 | 20.58 | 100 | 29.75 | -14 | | | | | |
| CI chondrites | 0.5962 | 3.919 | 20.149 | 20.141 | 100 | 30.95 | 7 | | | | | |
| Comets | solar? | | | | | | | | | | | |
| | **Xenon** | | | | | | | | | | | |
| | $^{124}Xe$ | $^{126}Xe$ | $^{128}Xe$ | $^{129}Xe$ | $^{130}Xe$ | $^{131}Xe$ | $^{132}Xe$ | $^{134}Xe$ | $^{136}Xe$ | $F_{Xe}$ (‰/amu) | | |
| Solar | 2.939 | 2.549 | 51.02 | 627.3 | 100 | 498 | 602.0 | 220.68 | 179.71 | 0 | | |
| Earth | 2.337 | 2.180 | 47.15 | 649.6 | 100 | 521.3 | 660.7 | 256.3 | 217.6 | 38 | | |
| Mars | 2.5 | 2.2 | 47.6 | 1555 | 100 | 513.9 | 648.1 | 259.7 | 227.7 | 33 | | |
| CI chondrites | 2.851 | 2.512 | 50.73 | 654.2 | 100 | 504.3 | 615.0 | 235.9 | 198.8 | 3 | | |
| Comets | solar/fractionated? | | | | | | | | | | | |

*Notes.* Powers of ten multipliers in parentheses. See text for details and references.

$\delta D_{VSMOW}=[(D/H)/(D/H)_{VSMOW}-1]\times 1000$; $\delta^{13}C_{VPDB}=[(^{13}C/^{12}C)/(^{13}C/^{12}C)_{VPDB}-1]\times 1000$; $\delta^{15}N_{air}=[(^{15}N/^{14}N)/(^{15}N/^{14}N)_{air}-1]\times 1000$.

$F_E=[(^iE/^jE)_{reservoir}/(^iE/^jE)_{solar}-1]\times 1000$, where (i,j)=(22,20) for Ne; (38,36) for Ar; (83,84) for Kr; (128,130) for Xe.

**Table 4. Proposed volatile element composition of the depleted mantle**

| $^1$H | $^{12}$C | $^{14}$N | $^3$He | $^{20}$Ne | $^{36}$Ar | $^{84}$Kr | $^{130}$Xe |
|---|---|---|---|---|---|---|---|
| 1.2 (-5) | 1.8 (-6) | 6.9 (-9) | 8.4 (-16) | 2.1 (-15) | 2.1 (-15) | 6.0 (-17) | 1.1 (-18) |

| D/H | $^{13}$C/$^{12}$C | $^{15}$N/$^{14}$N | $^4$He/$^3$He | $^{20}$Ne/$^{22}$Ne | $^{21}$Ne/$^{22}$Ne | $^{38}$Ar/$^{36}$Ar | $^{40}$Ar/$^{36}$Ar |
|---|---|---|---|---|---|---|---|
| 0.000143 | 0.01118 | 0.003659 | 99,000 | 12.49 | 0.0578 | 0.188 | 25,000 |

| $^{78}$Kr/$^{84}$Kr | $^{80}$Kr/$^{84}$Kr | $^{82}$Kr/$^{84}$Kr | $^{83}$Kr/$^{84}$Kr | $^{86}$Kr/$^{84}$Kr |
|---|---|---|---|---|
| 0.00595 | 0.03901 | 0.2007 | 0.2006 | 0.3075 |

| $^{124}$Xe/$^{130}$Xe | $^{126}$Xe/$^{130}$Xe | $^{128}$Xe/$^{130}$Xe | $^{129}$Xe/$^{130}$Xe | $^{131}$Xe/$^{130}$Xe | $^{132}$Xe/$^{130}$Xe | $^{134}$Xe/$^{130}$Xe | $^{136}$Xe/$^{130}$Xe |
|---|---|---|---|---|---|---|---|
| 0.02452 | 0.0225 | 0.479 | 7.6 | 5.37 | 6.99 | 2.79 | 2.55 |

The concentrations are in mol/g of rock. The $^3$He concentration was calculated by using the $^3$He flux at ridges (Bianchi et al. 2010), 21 km$^3$/year MORB production and 10 % partial melting. The $^4$He/$^3$He ratio corresponds to the mode in the distribution of MORB values (7.3 $R_A$). The Ne isotopic composition is from Holland and Ballentine (2006). $^{20}$Ne, $^{36}$Ar, $^{84}$Kr, $^{130}$Xe concentrations as well as $^{40}$Ar/$^{36}$Ar and $^{129}$Xe/$^{130}$Xe ratios are from Moreira et al. (1998) assuming a $^{20}$Ne/$^{22}$Ne ratio of ~12.5. Major volatile element concentrations and isotopic compositions are from Sect. 3.1 (i.e., 110 ppm H$_2$O, 79 ppm CO$_2$, 0.1 ppm N$_2$). The $^{38}$Ar/$^{36}$Ar ratio is the air value. The fissiogenic Xe isotopes $^{131}$Xe, $^{132}$Xe, $^{134}$Xe, and $^{136}$Xe are obtained from correlations with $^{129}$Xe/$^{130}$Xe (the later ratio is taken to be 7.6), see Kunz et al. 1998. The $^{128}$Xe/$^{130}$Xe ratio is obtained by regressing the CO$_2$ well gas data (Caffee et al. 1999, Holland and Ballentine 2006) to a $^{129}$Xe/$^{130}$Xe ratio of 7.6. The $^{124}$Xe/$^{130}$Xe and $^{126}$Xe/$^{130}$Xe ratios are calculated assuming similar isotopic fractionation as $^{128}$Xe/$^{130}$Xe relative to air. The $^{82}$Kr/$^{84}$Kr ratio is obtained by regressing the data to a $^{128}$Xe/$^{130}$Xe of 0.479 (Holland et al. 2009) and the inferred isotopic fractionation is applied to other Kr isotopes using an exponential law.

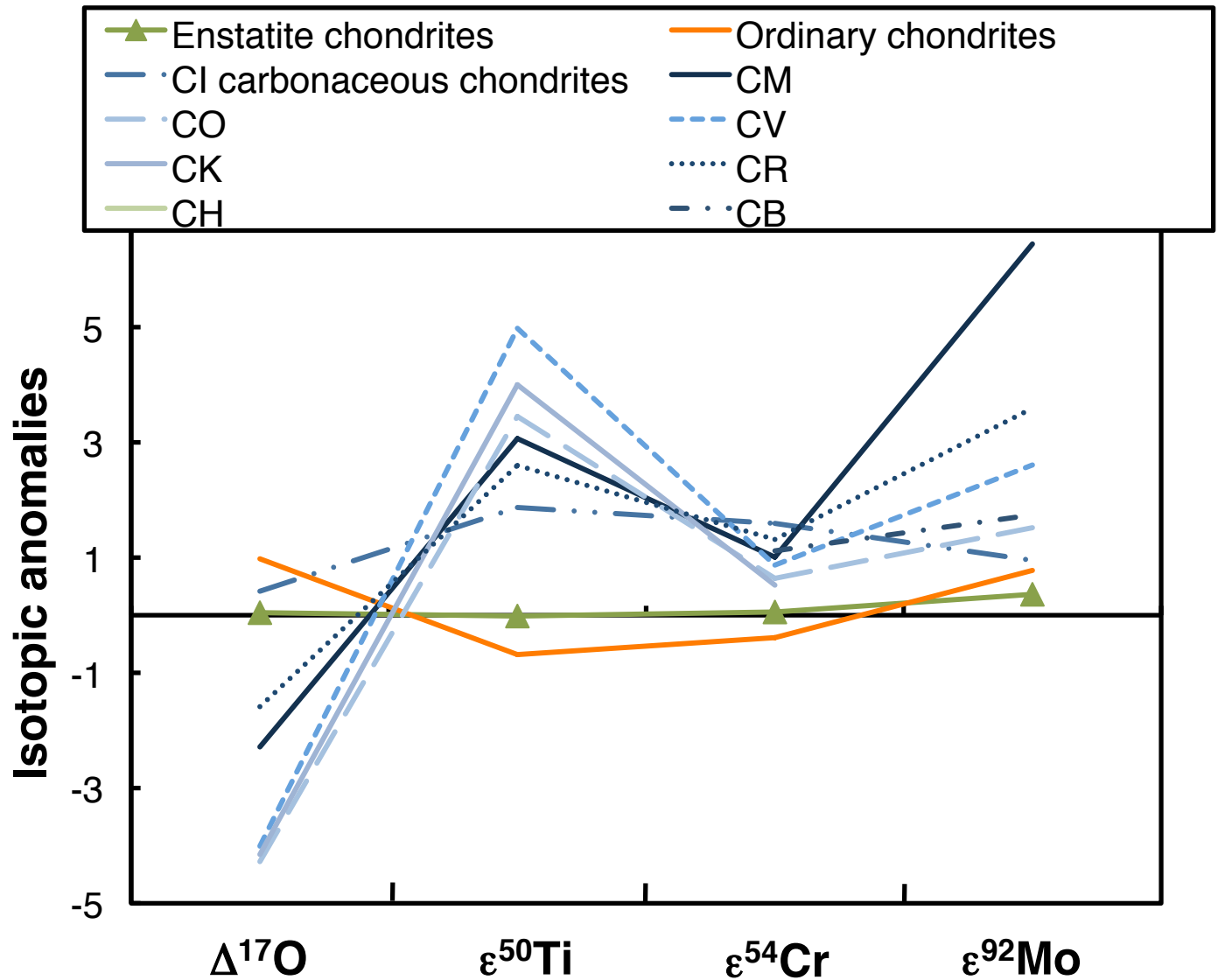

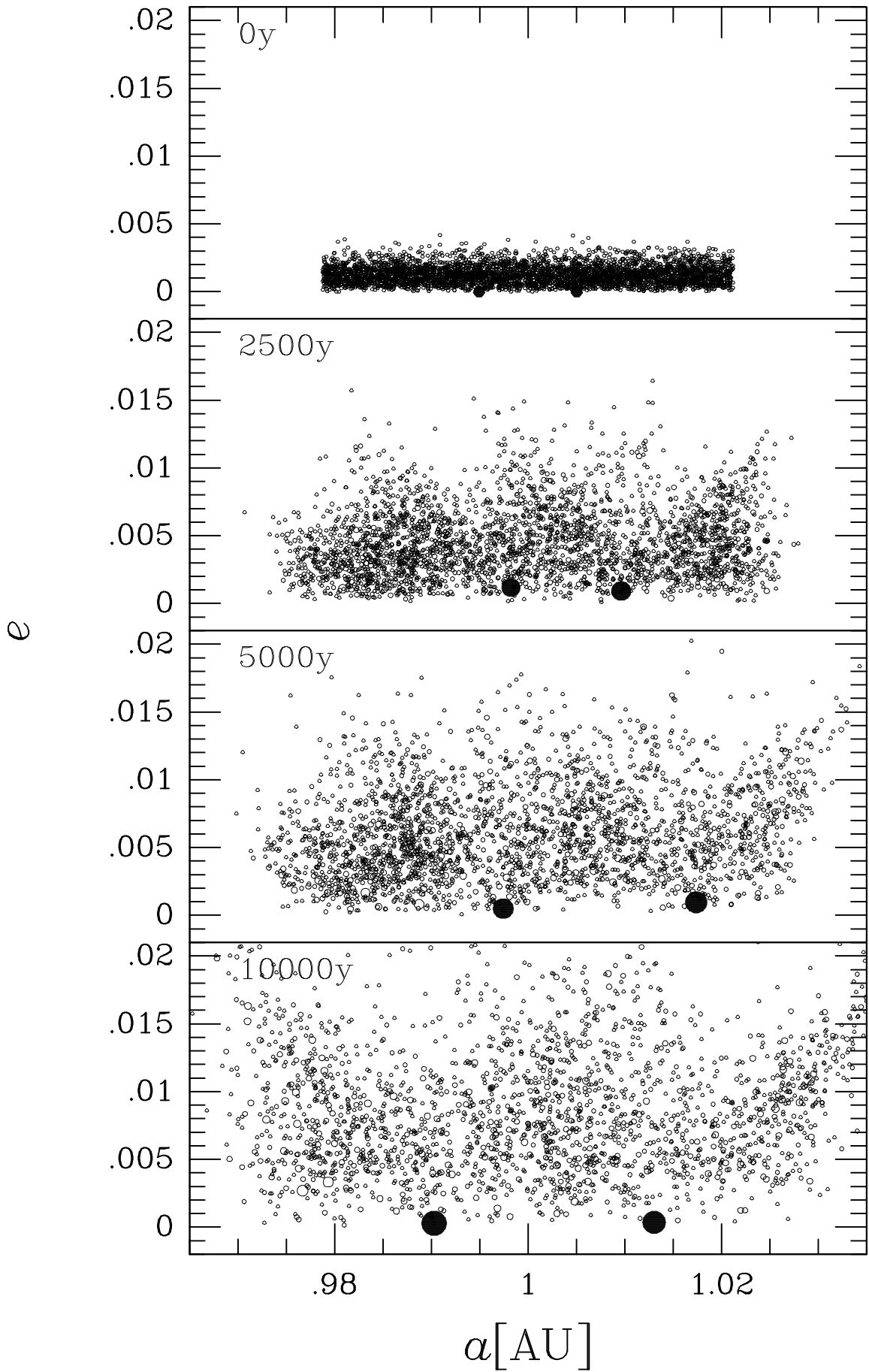

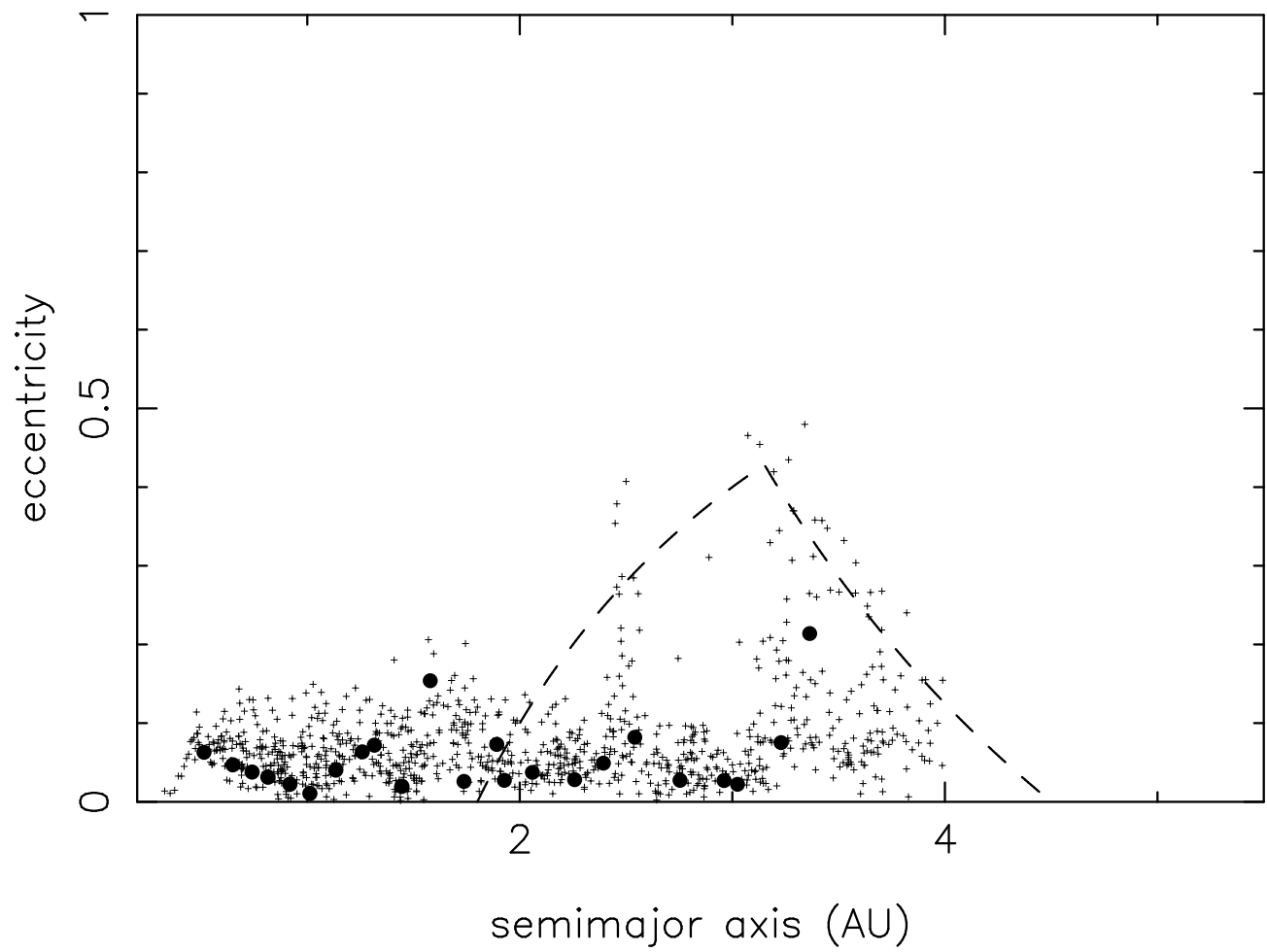
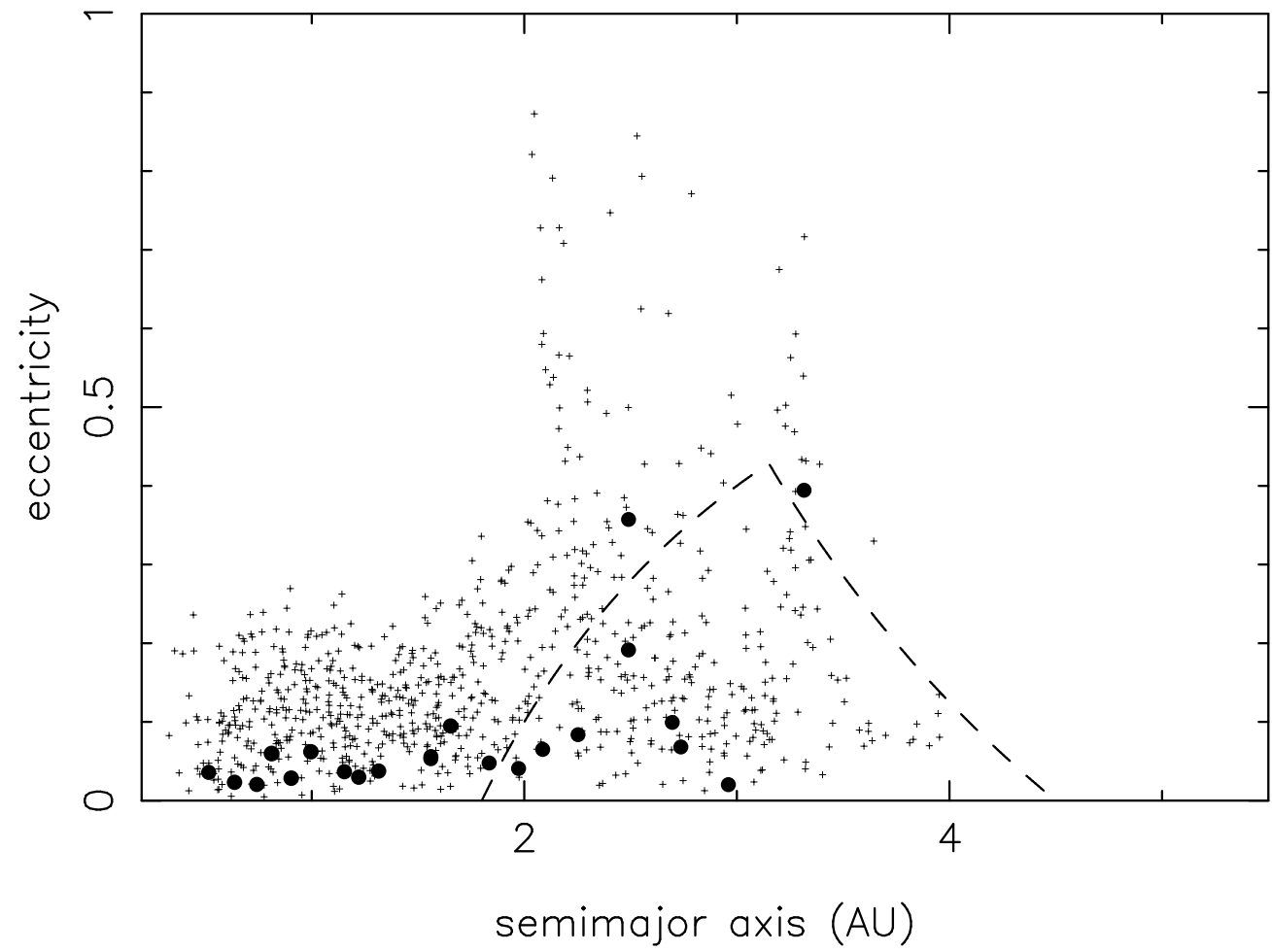
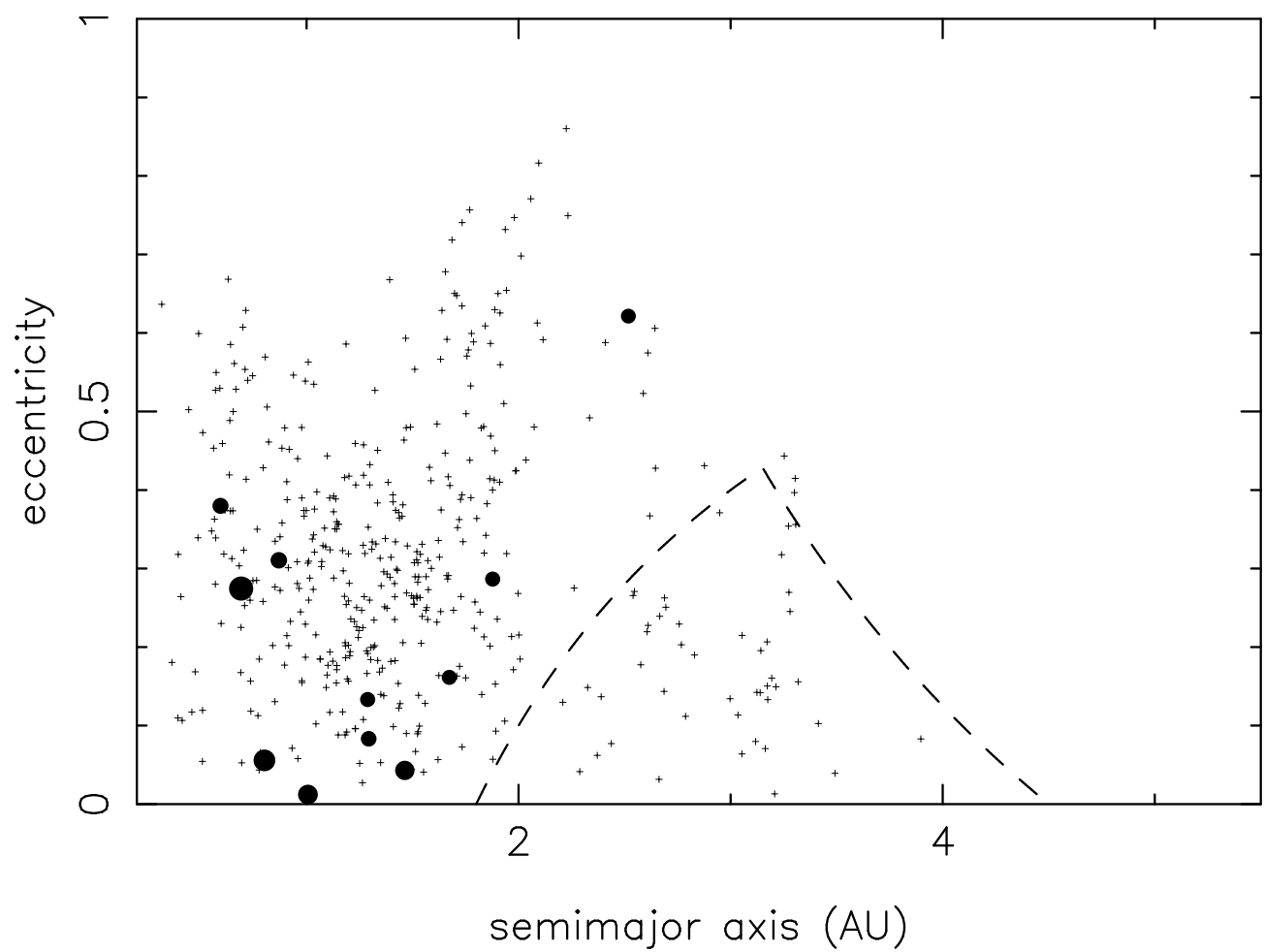
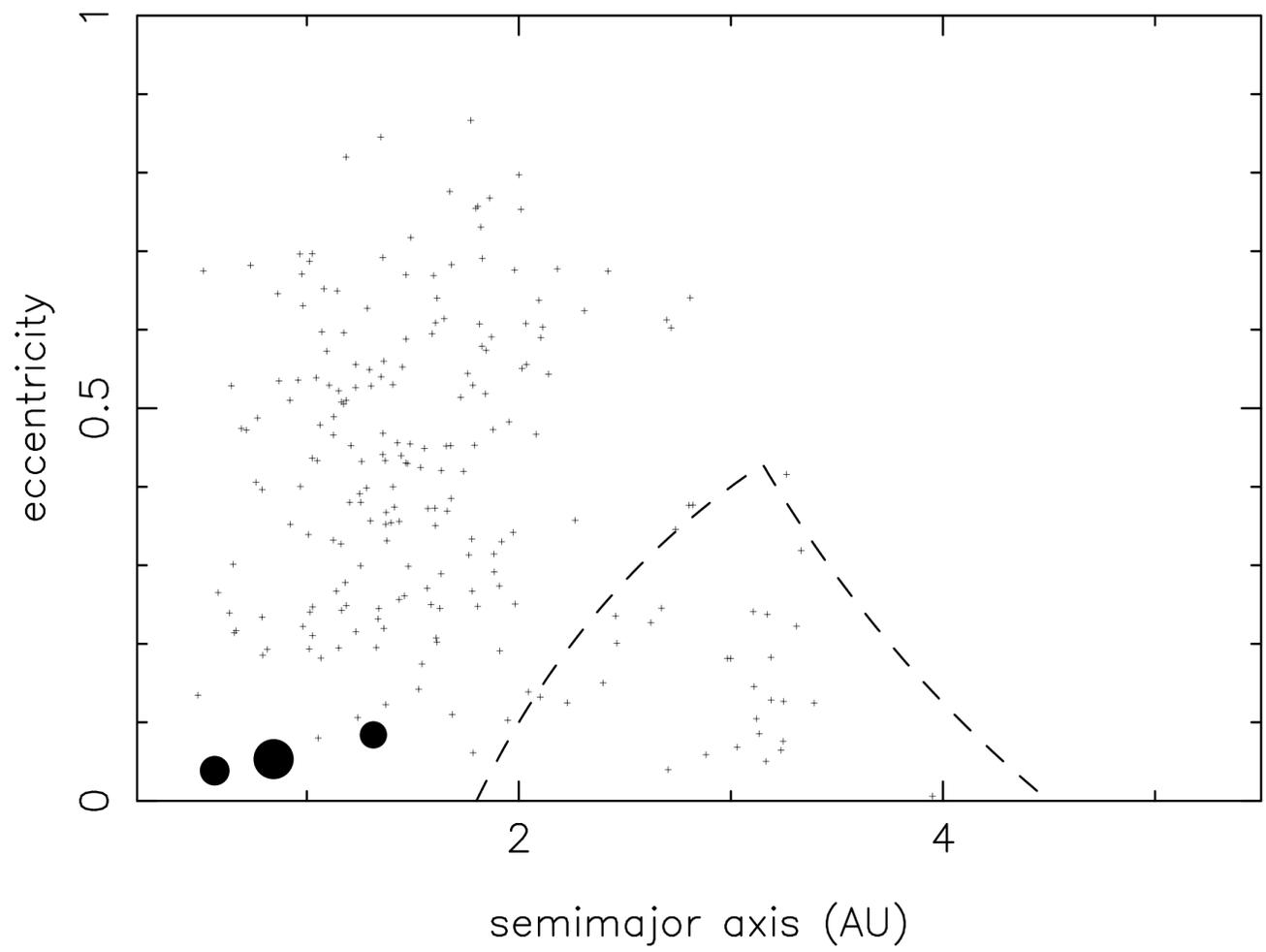

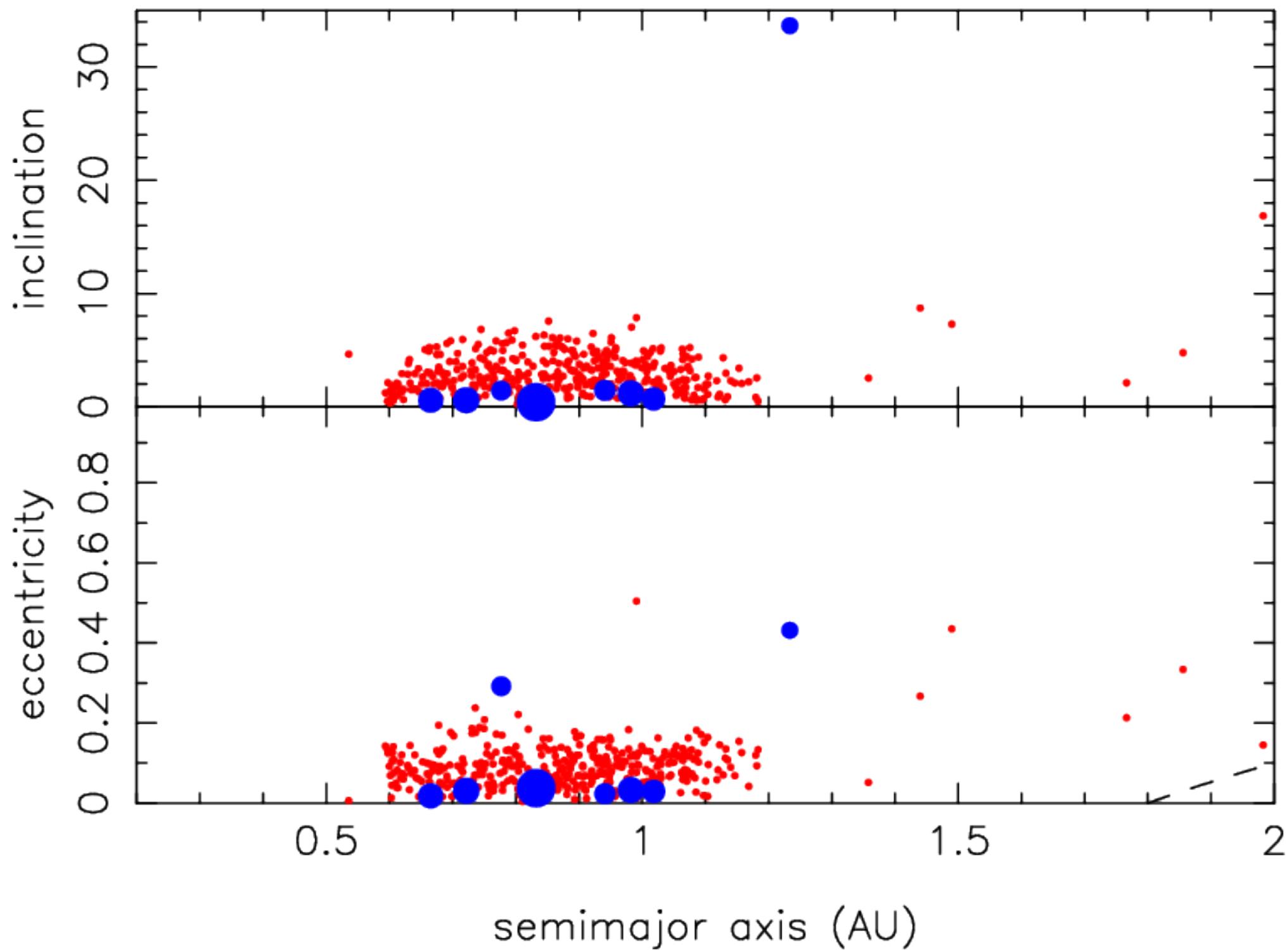

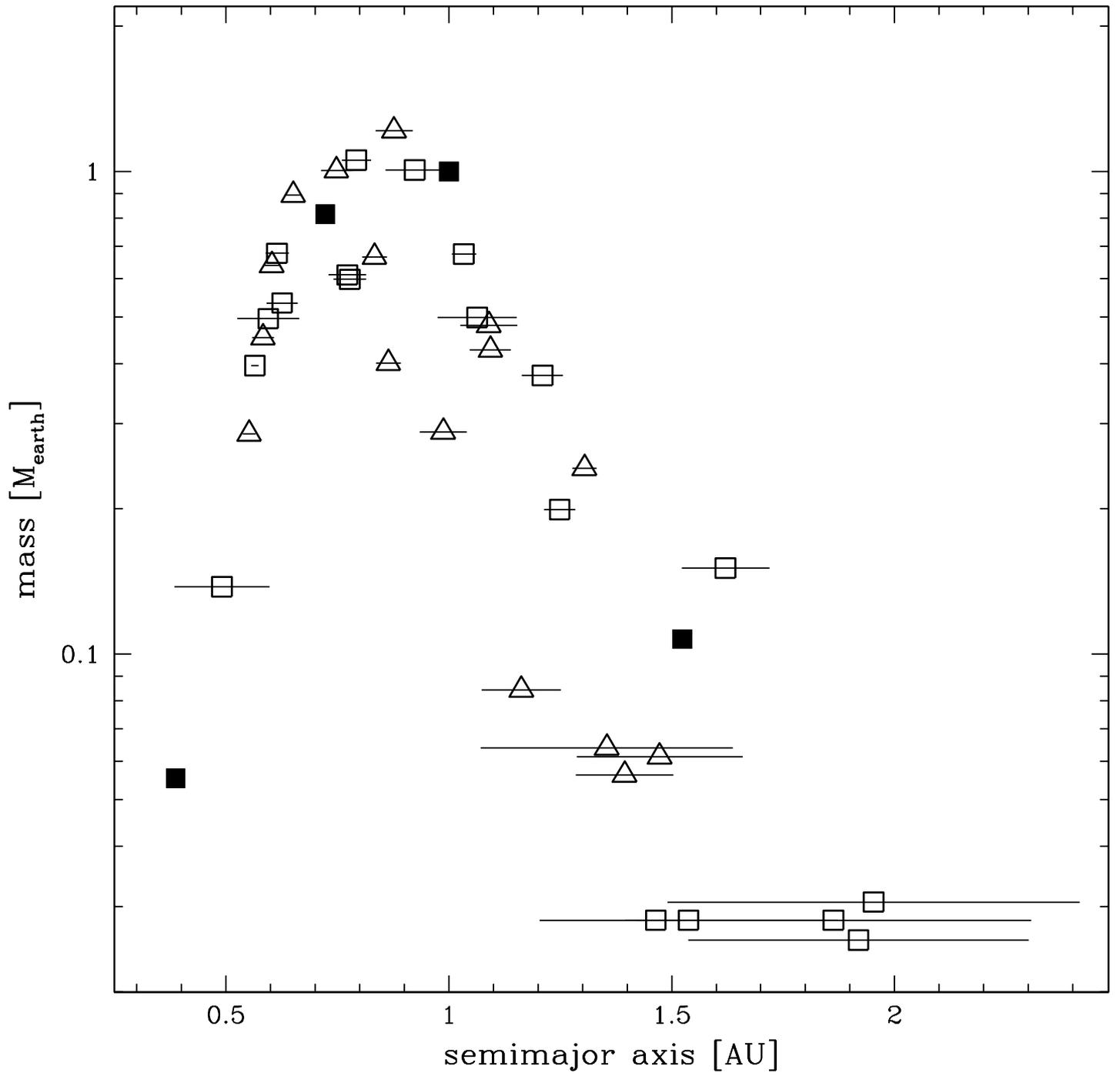

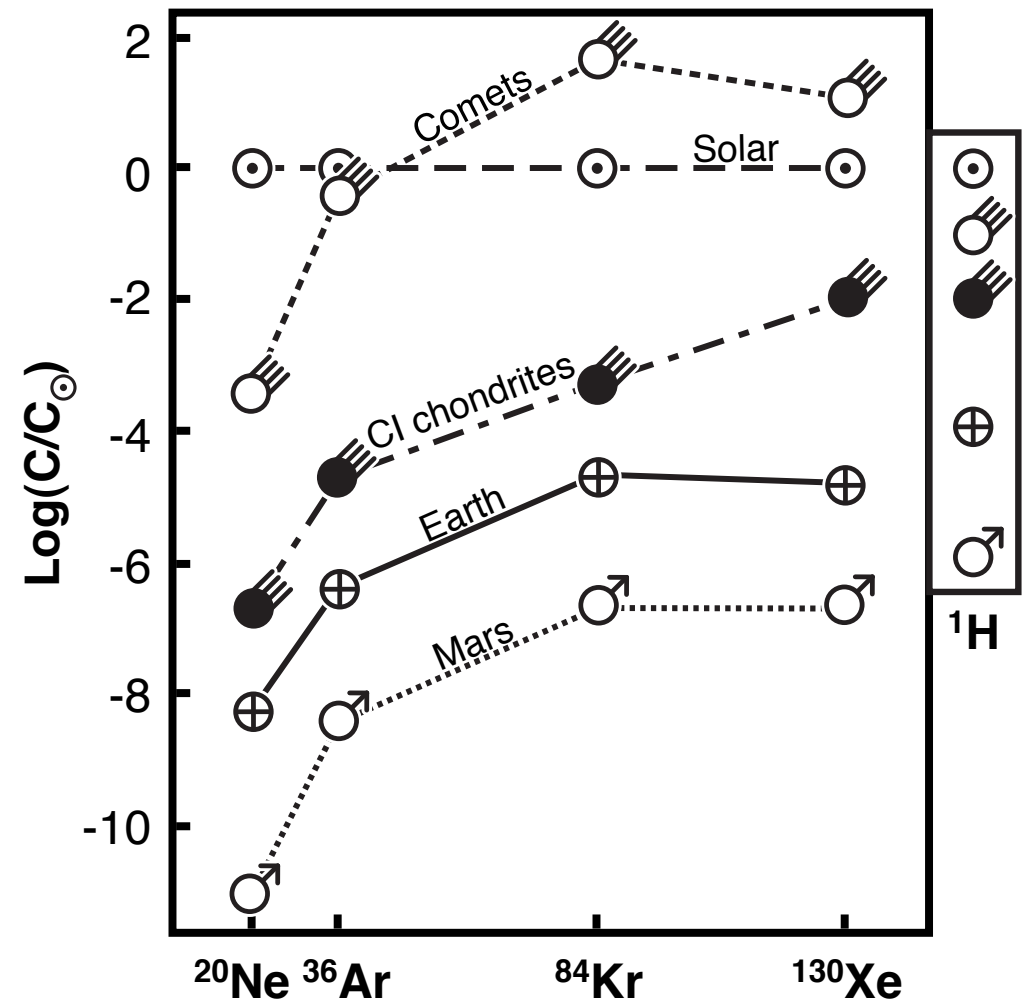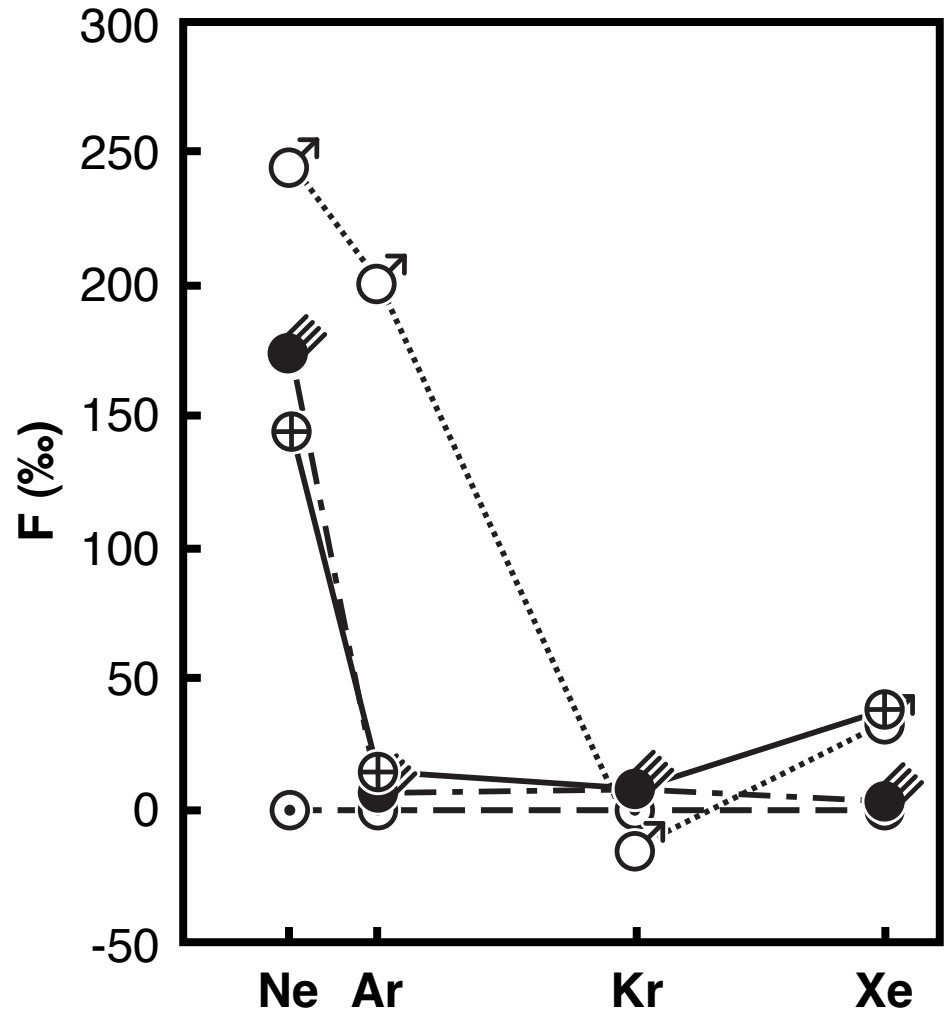

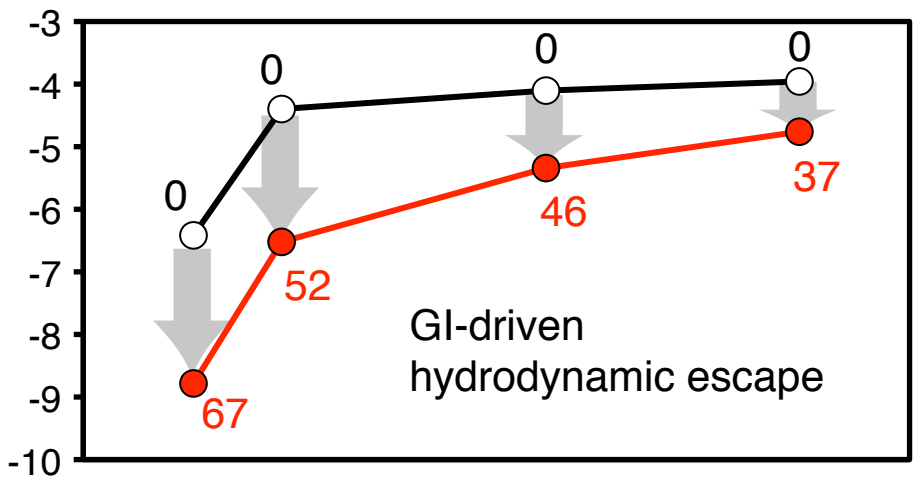
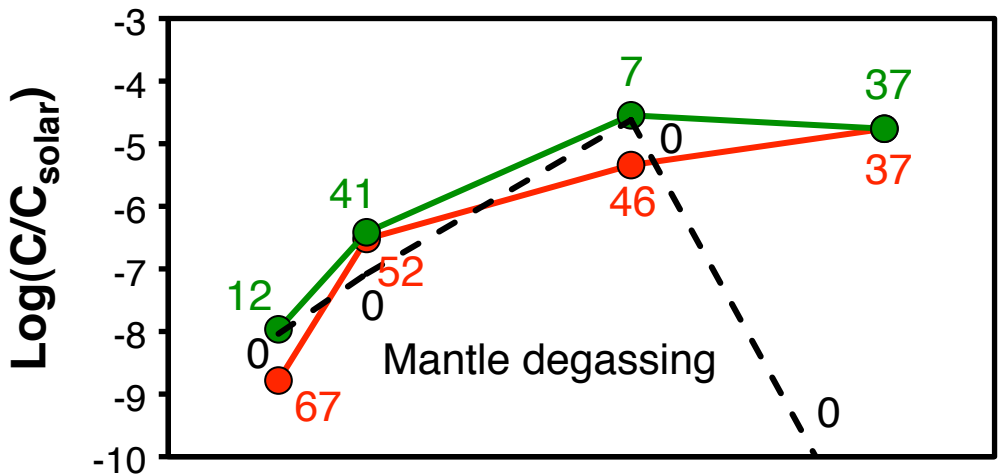
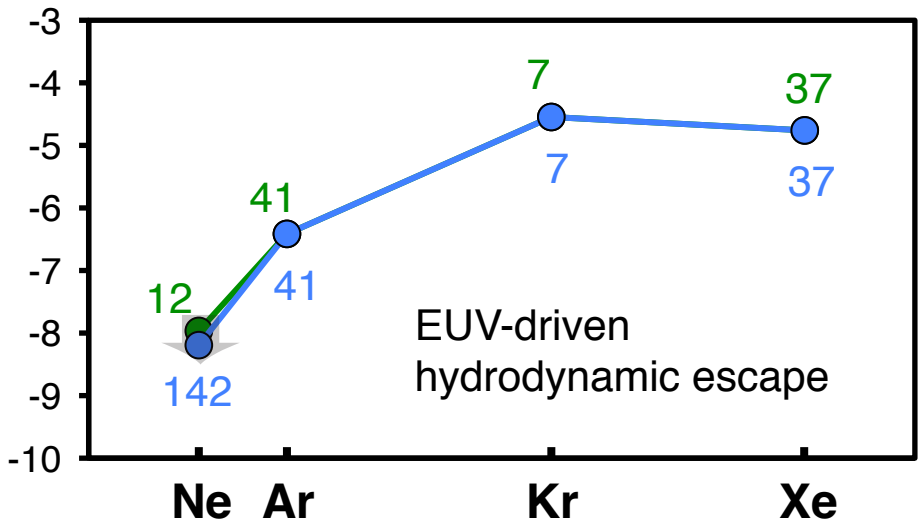

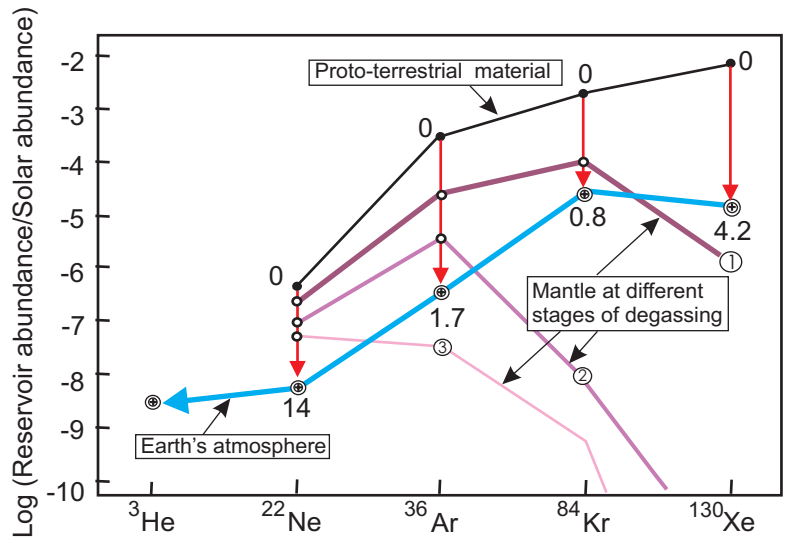

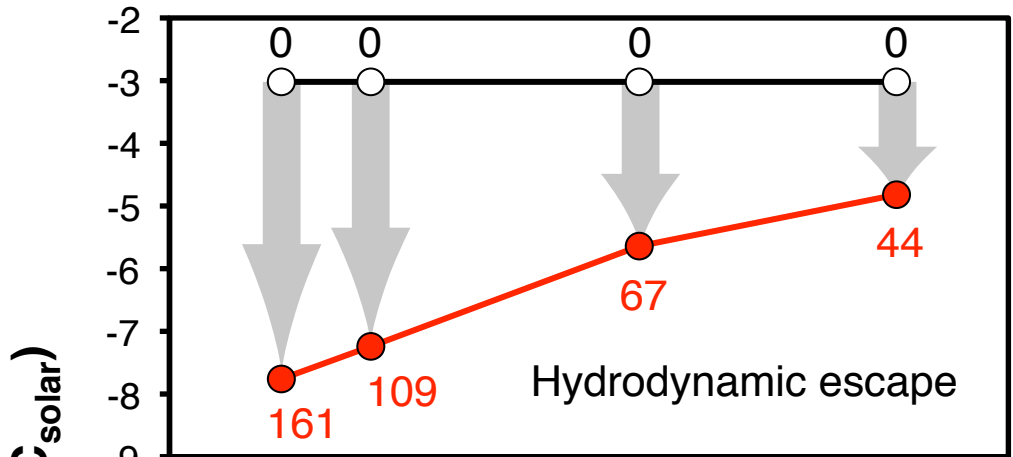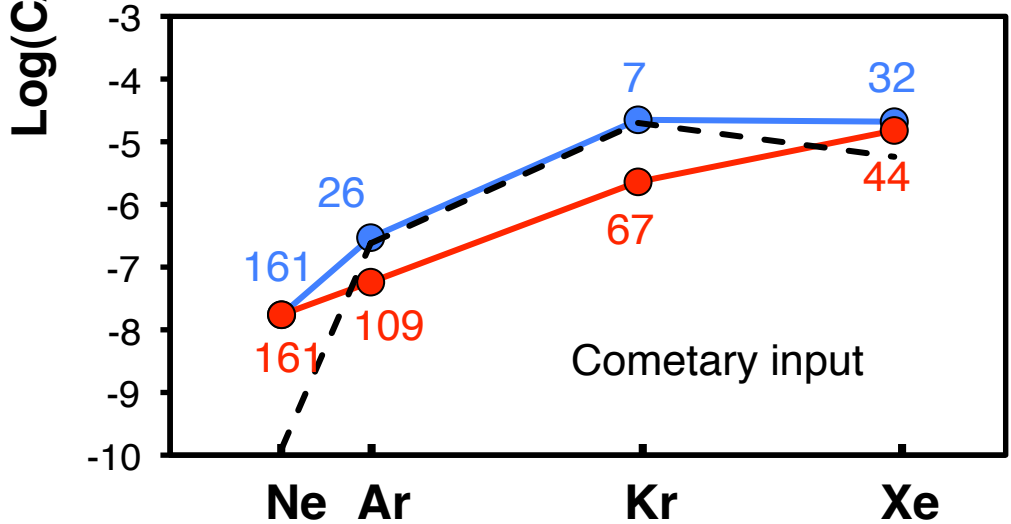

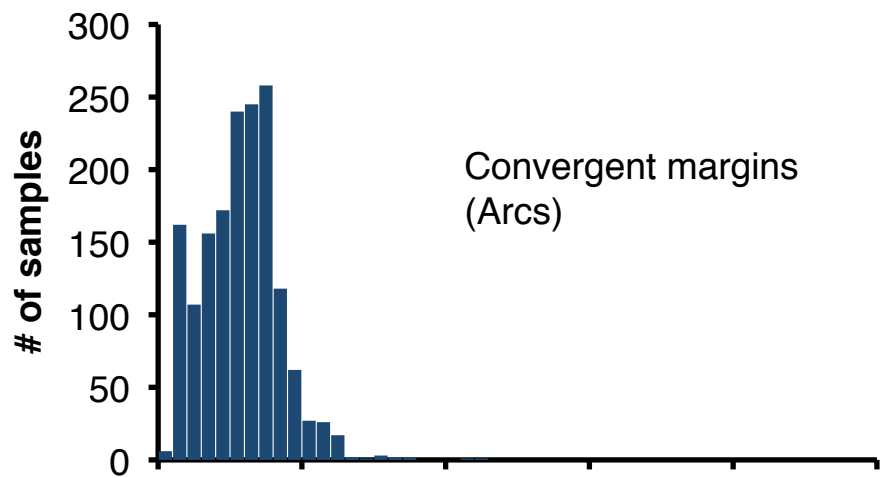
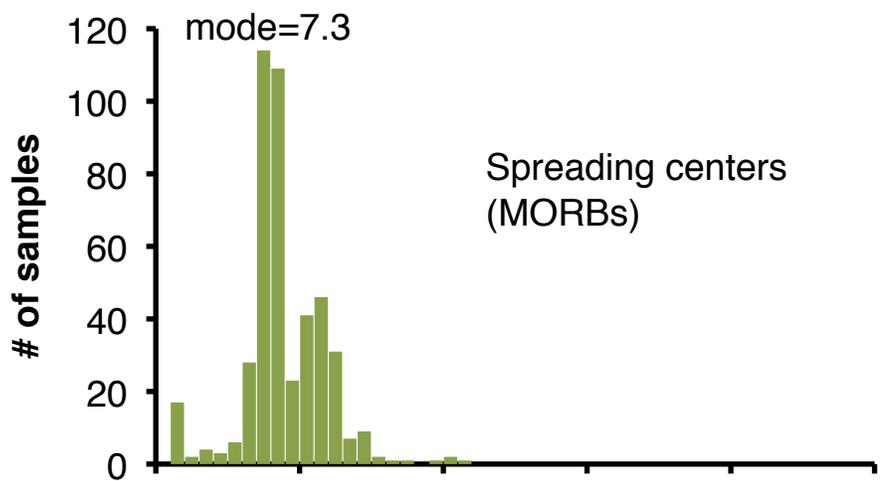
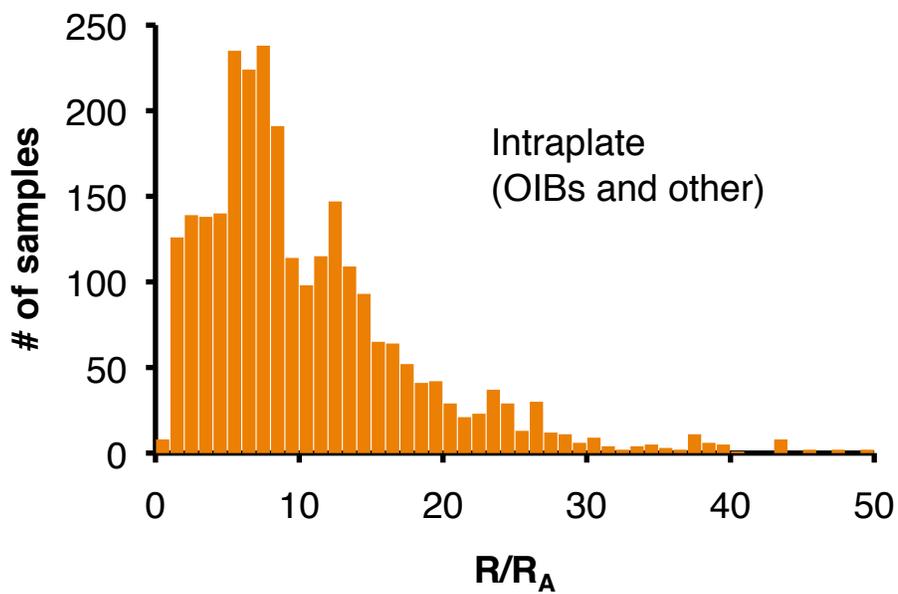

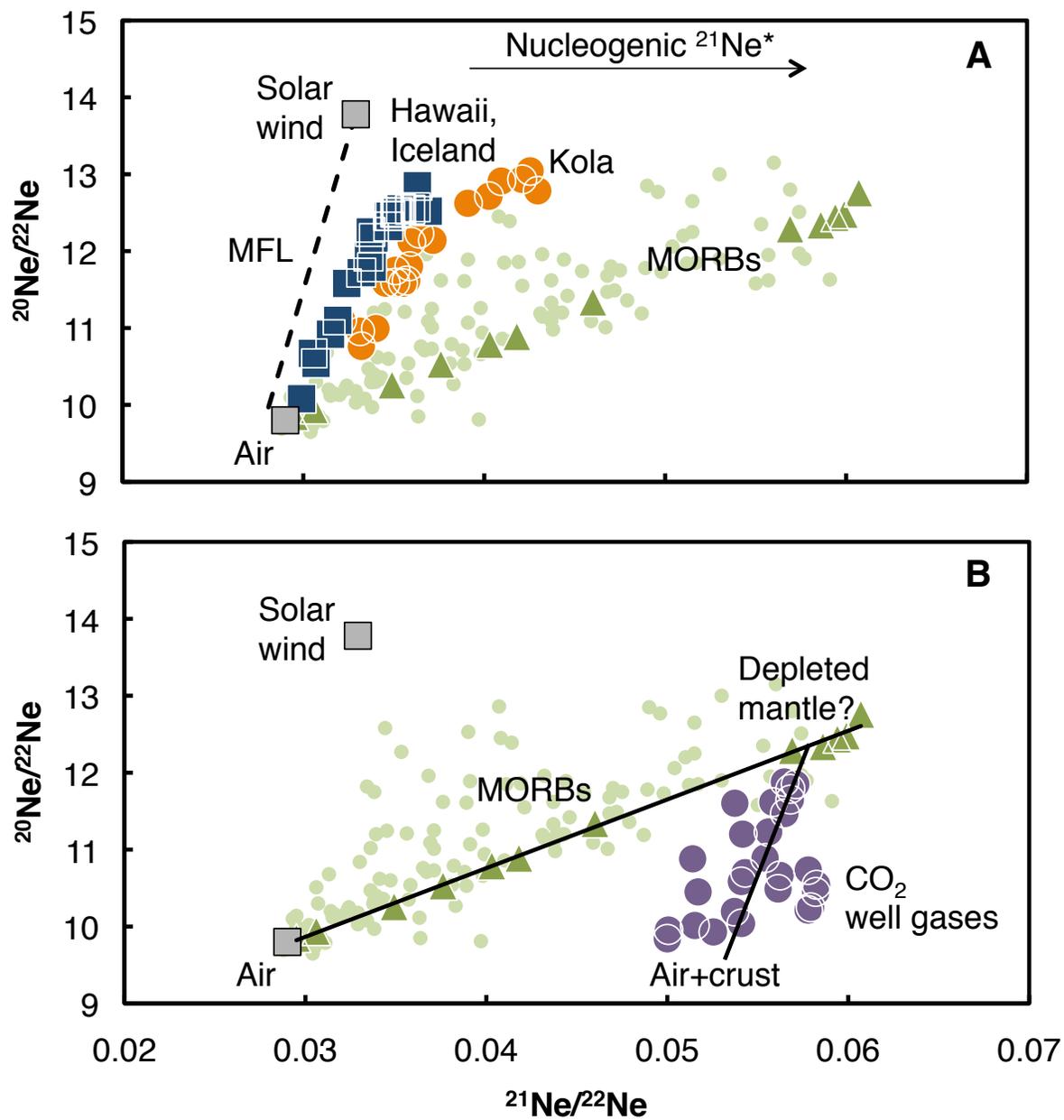

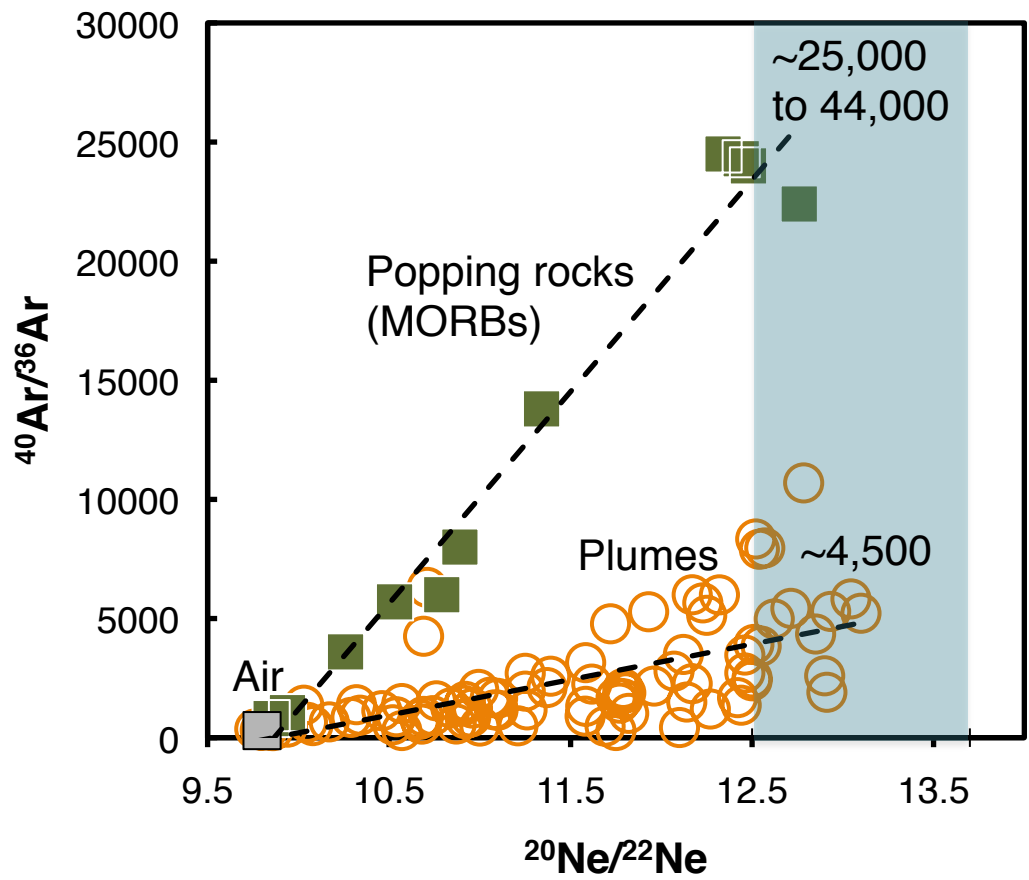

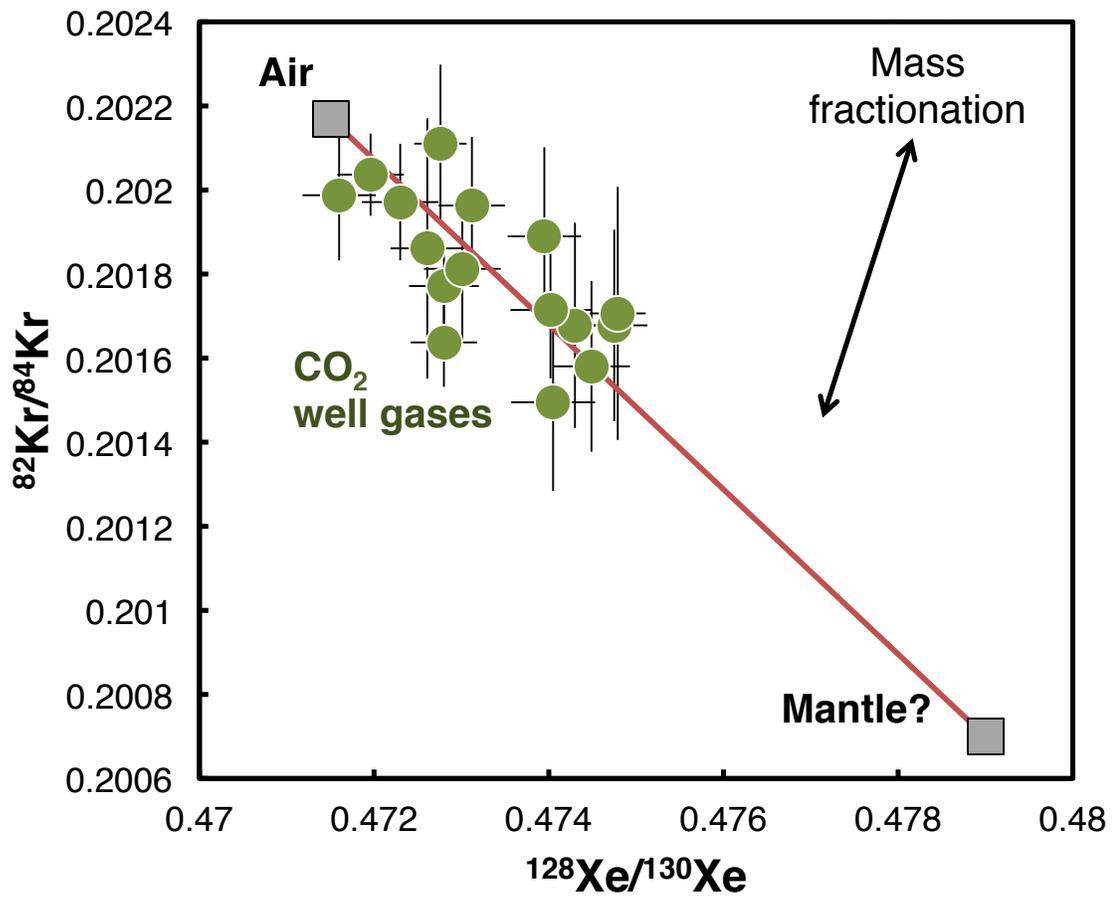

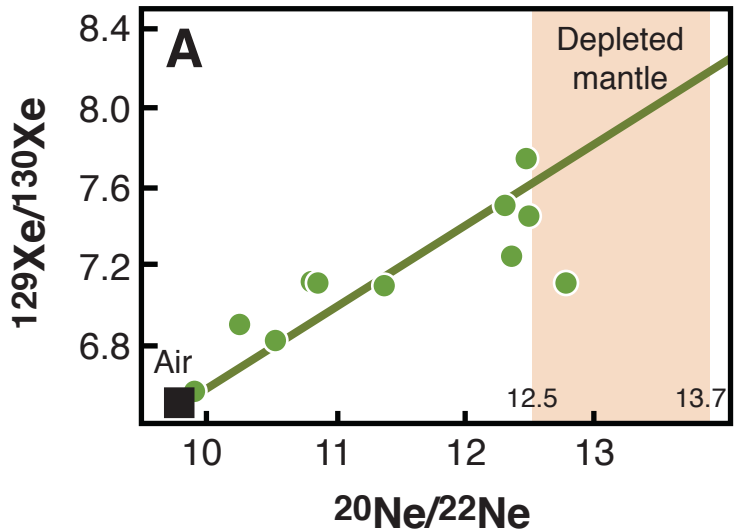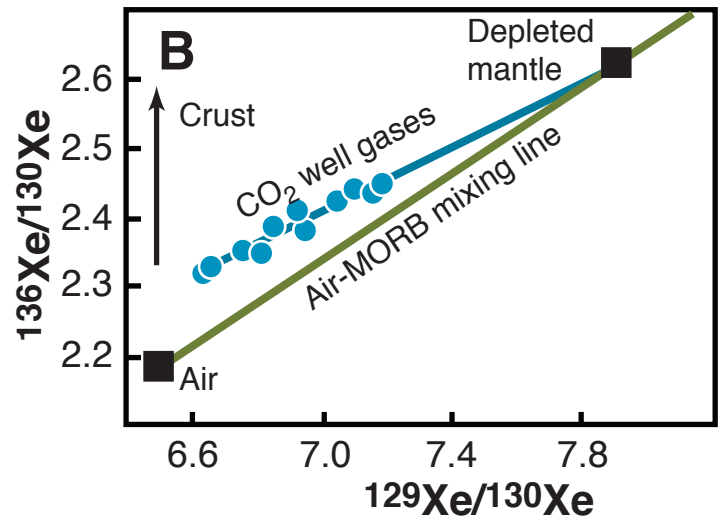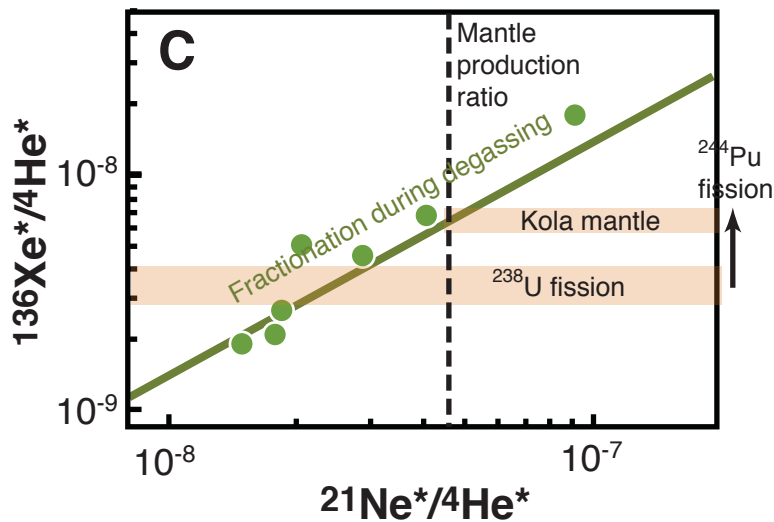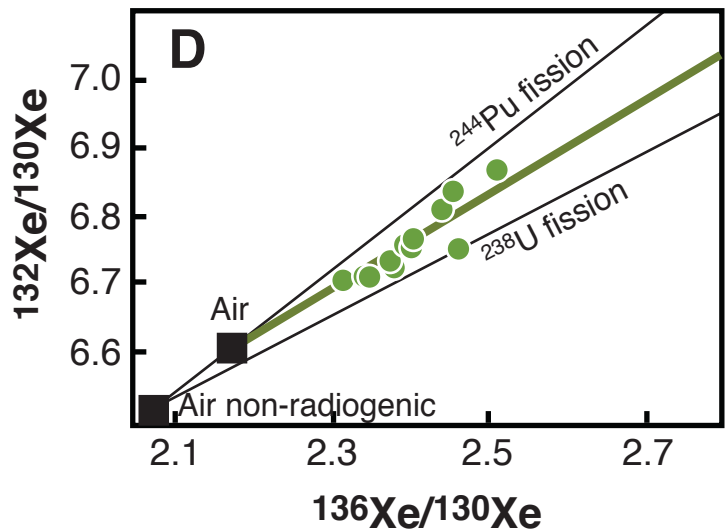

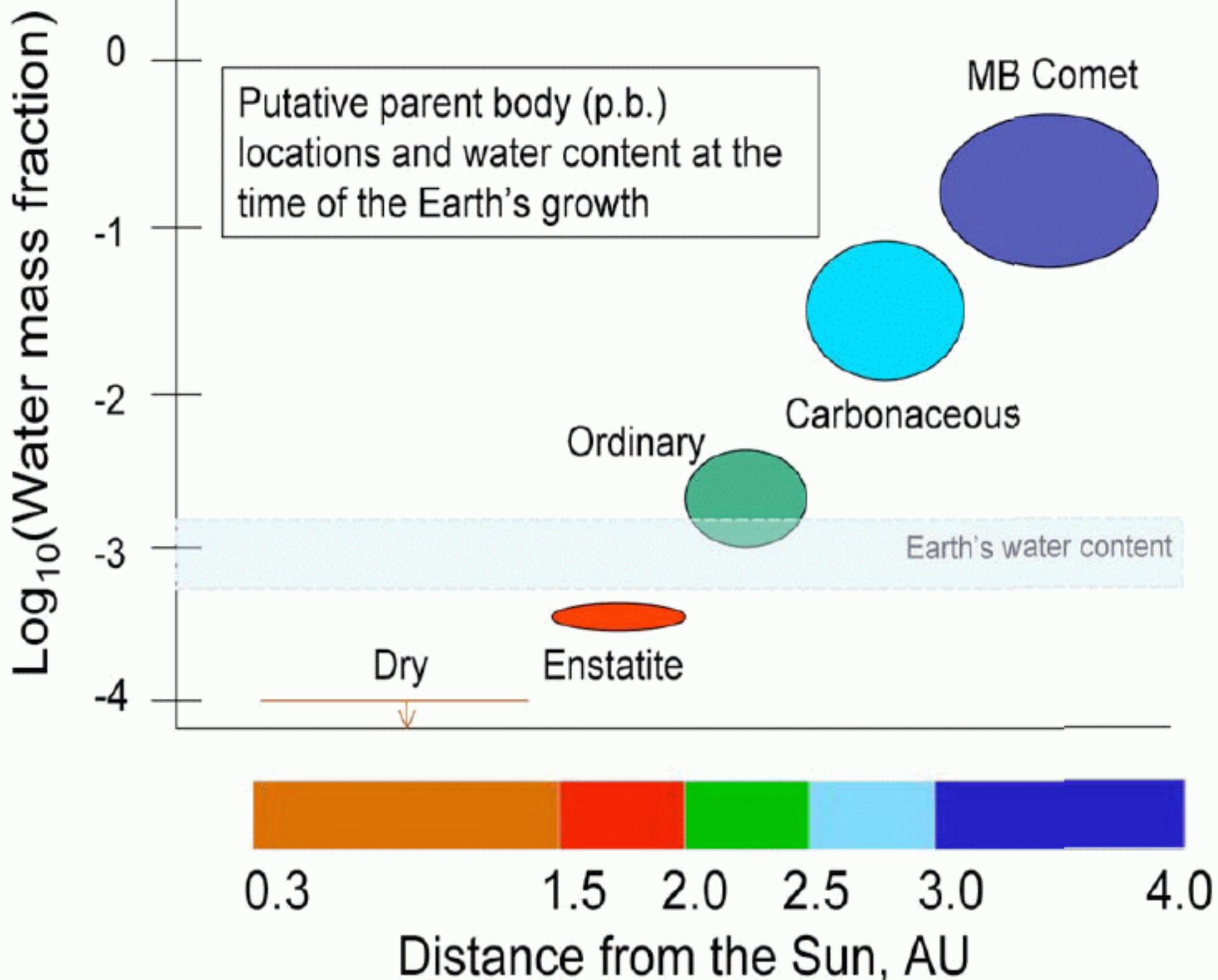

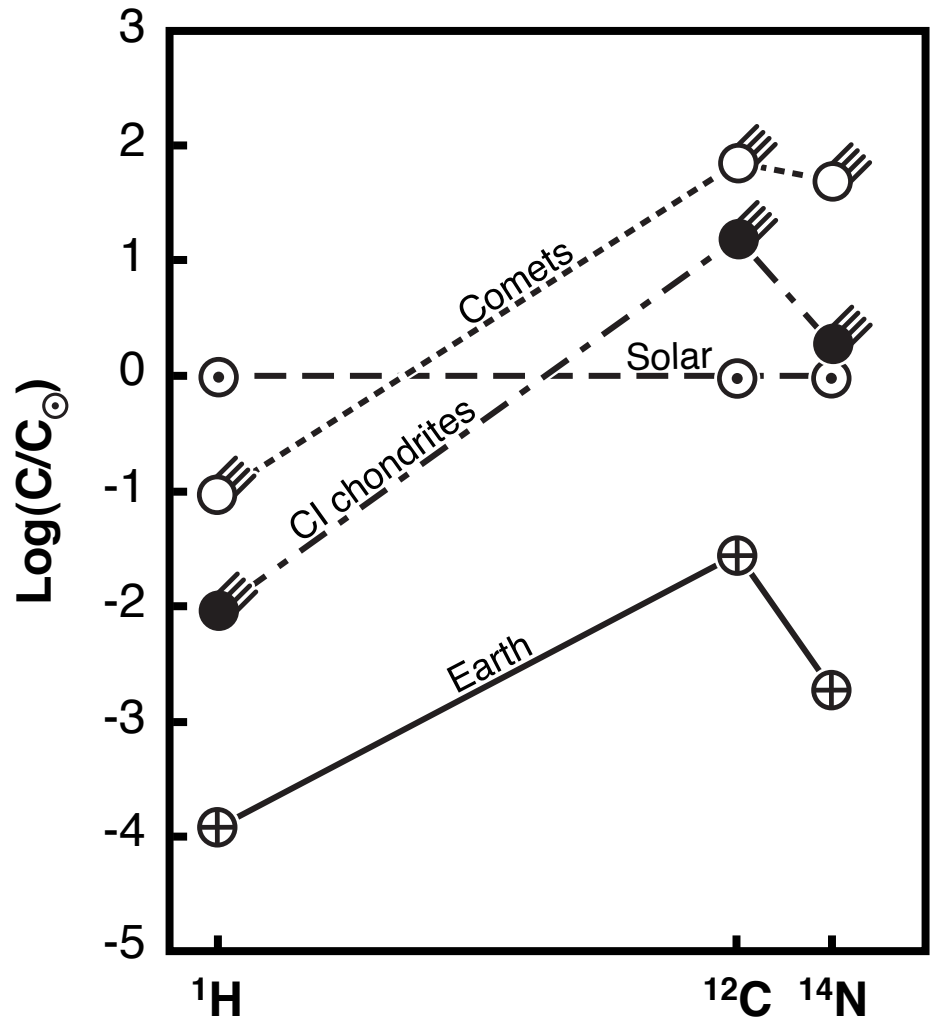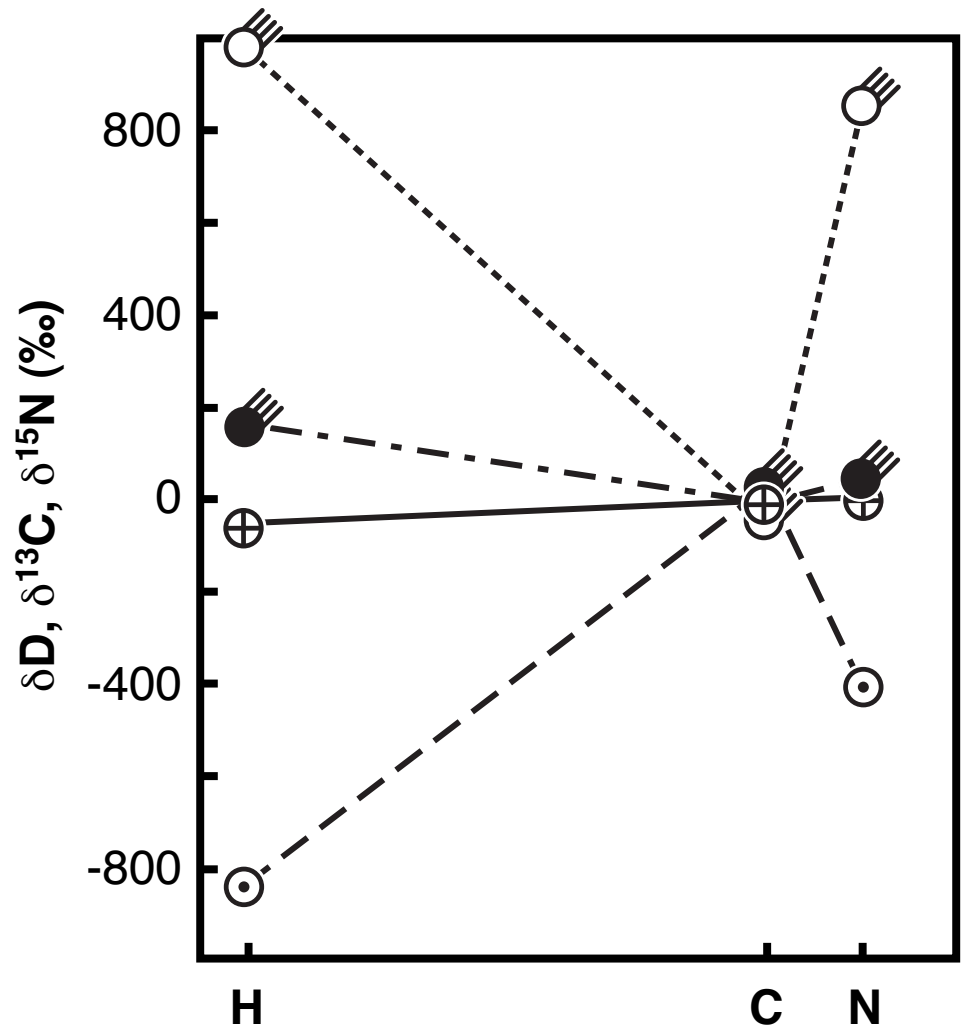

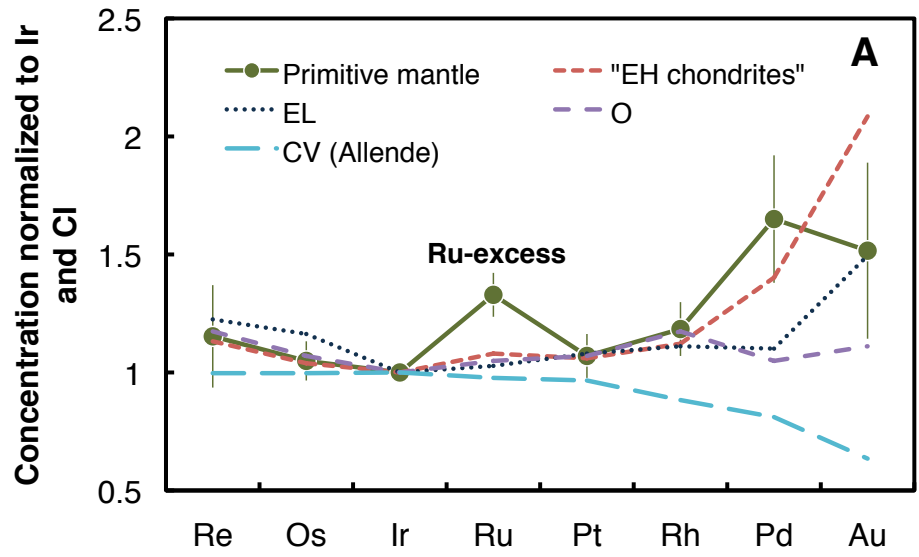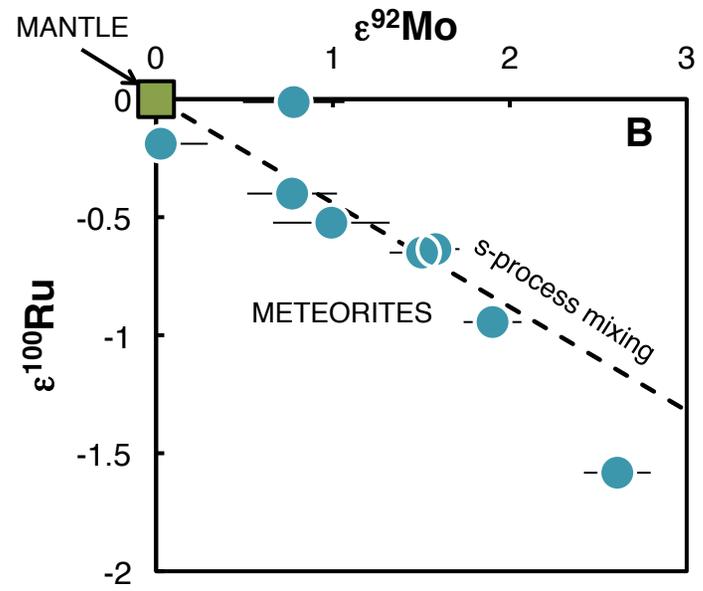

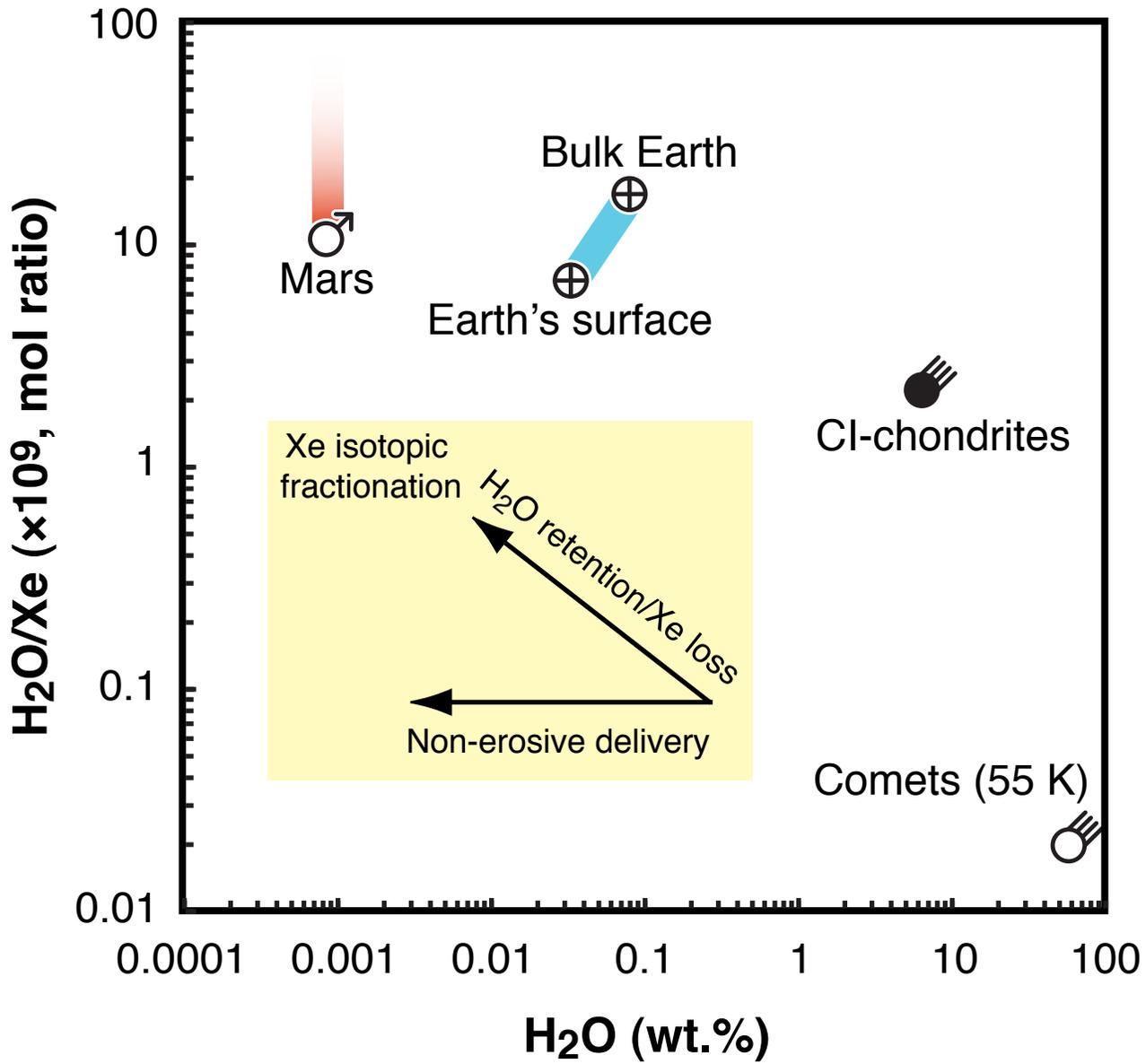

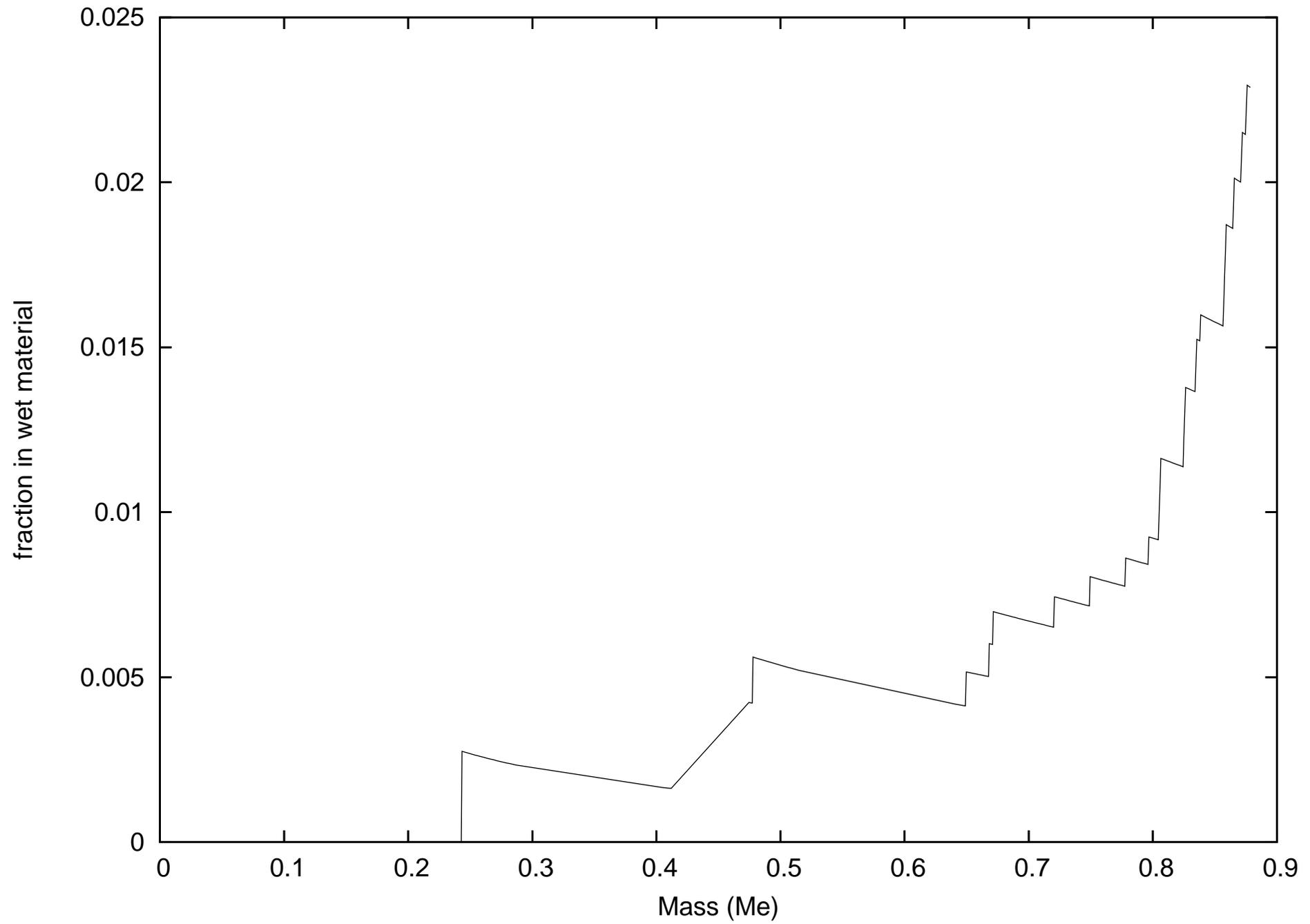

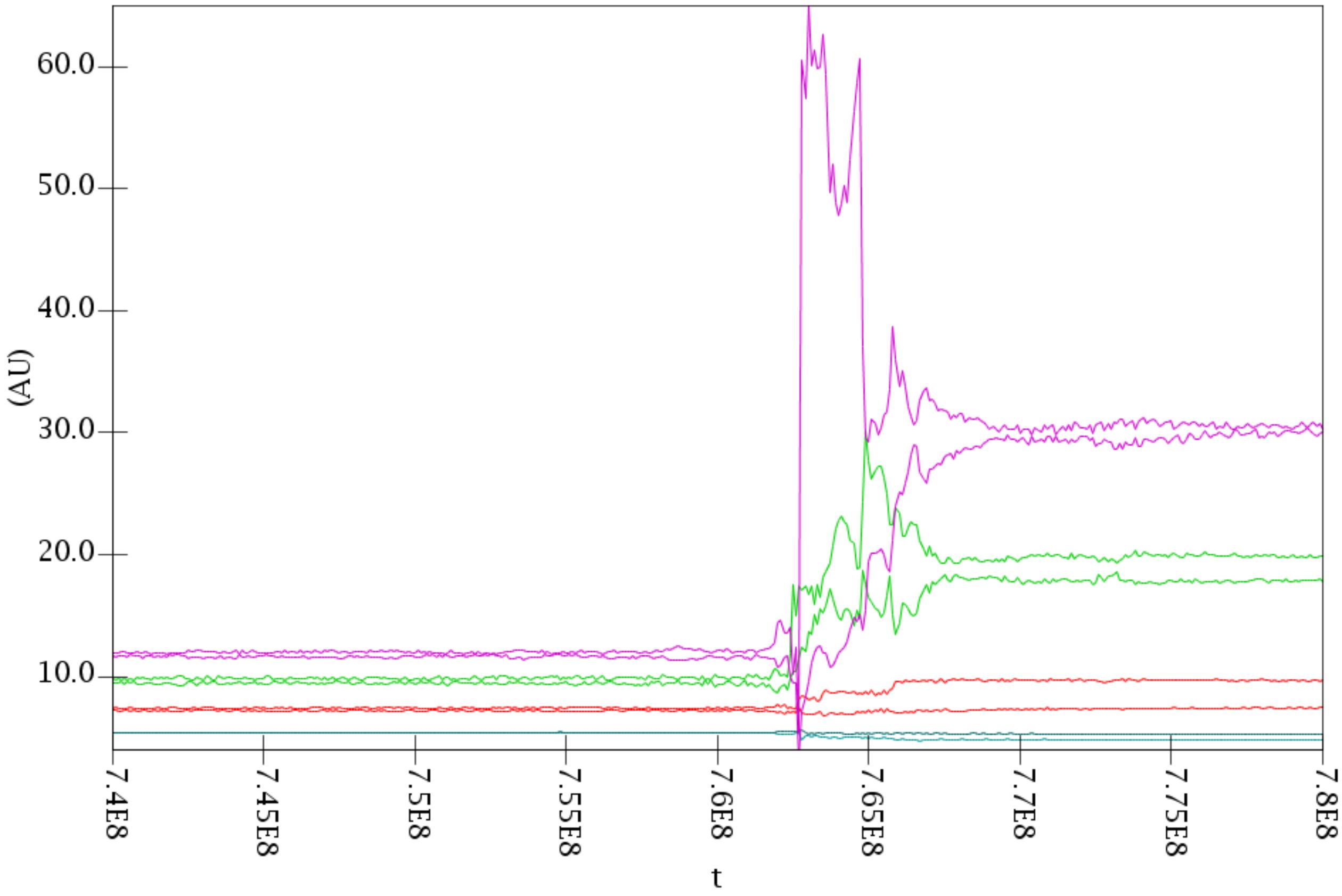